\documentclass[aps,twocolumn,amsmath,amssymb,preprintnumbers,superscriptaddress,prb,floatfix,nofootinbib]{revtex4-2}

\usepackage{comment}
\usepackage[version=4]{mhchem}
\usepackage[utf8]{inputenc}
\usepackage{newtxtext}
\usepackage[upint]{newtxmath}
\usepackage{microtype}
\usepackage{textcomp}
\usepackage{dsfont}
\usepackage{eucal}
\usepackage{siunitx}

\usepackage[normalem]{ulem}
\usepackage{enumerate}
\usepackage{amsfonts,amsmath,amssymb,amstext}
\usepackage{color}
\usepackage{soul}

\usepackage{todonotes}
\presetkeys%
    {todonotes}%
    {inline}{}

\usepackage{graphicx}
\usepackage{subfigure}

\usepackage[colorlinks,allcolors=blue]{hyperref}
\usepackage[capitalize]{cleveref} 
\usepackage{cleveref}

\newcommand{\ca}[2][]{c_{#2}^{\vphantom{\dagger}#1}} 
\newcommand{\cc}[2][]{c_{#2}^{{\dagger}#1}}          

\newcommand{\vk}{\boldsymbol{k}}

\newcommand{\RE}[1]{\,\mbox{Re}\left\{{#1}\right\}}

\newcommand{\df}[1]{\,\delta{\left(#1\right)}}

\definecolor{DarkBlue}{rgb}{0,0,0.80}
\definecolor{DarkRed}{rgb}{0.80,0,0}
\definecolor{Purple}{rgb}{0.55,0,0.55}
\definecolor{DarkGreen}{rgb}{0,0.80,0}

\let\epsilon\varepsilon

\begin{document}

\title{Interface-induced magnetization in altermagnets and antiferromagnets}

\author{Erik Wegner Hodt}
\thanks{These authors contributed equally to this work}
\email{erik.w.hodt@ntnu.no}
\author{Pavlo Sukhachov}
\thanks{These authors contributed equally to this work}
\email{pavlo.sukhachov@ntnu.no}
\affiliation{Center for Quantum Spintronics, Department of Physics, Norwegian \\ University of Science and Technology, NO-7491 Trondheim, Norway}

\author{Jacob Linder}
\email{jacob.linder@ntnu.no}
\affiliation{Center for Quantum Spintronics, Department of Physics, Norwegian \\ University of Science and Technology, NO-7491 Trondheim, Norway}

\date{\today}

\begin{abstract}
Altermagnets is a class of antiferromagnetic materials which has electron bands with lifted spin degeneracy in momentum space but vanishing net magnetization and no stray magnetic fields. Because of these properties, altermagnets have attracted much attention for potential use in spintronics. We here show that despite the absence of bulk magnetization, the itinerant electrons in altermagnets can generate a magnetization close to edges and vacuum interfaces. We find that surface-induced magnetization can also occur for conventional antiferromagnets with spin-degenerate bands, where the magnetization from the itinerant electrons originates from a subtle yet nonvanishing redistribution of the probability density on the unit-cell level. An intuitive explanation of this effect in a phenomenological model is provided. In the altermagnetic case, the induced magnetization has a different spatial dependence than in the antiferromagnetic case due to the anisotropy of the spin-polarized Fermi surfaces, causing the edge-induced Friedel oscillations of the spin-up and -down electron densities to have different periods. We employ both a low-energy effective continuum model and lattice tight-binding calculations. Our results have implications for the usage of altermagnets and antiferromagnets in nanoscale spintronic applications.
\end{abstract}

\maketitle

\section{Introduction}
\label{sec:intro}

The spin polarization of quasiparticles in solid-state materials is a key property for many quantum physical phenomena, not the least in the field of spintronics. Conventional antiferromagnetic materials feature spin-degenerate electron bands, as dictated by parity-time-reversal ($\mathcal{PT}$) symmetry. However, it has recently been understood that antiferromagnetic materials can display a momentum-dependent spin-splitting which is distinct from relativistically spin-orbit coupled systems. One such mechanism, originally envisioned in Ref.~\cite{Pekar-Rashba-CombinedResonanceCrystals-1965}, consists of a spin-lattice coupling due to an internal periodic magnetic field in the material. When $\mathcal{PT}$ symmetry is broken by the combined lattice geometry and spin ordering in a material, a large spin splitting can arise which depends on momentum $\vk$~\cite{Noda-Nakamura-MomentumdependentBandSpin-2016, Smejkal-Sinova:2020, Yuan-Zunger:2020, Hayami-Kusunose-BottomupDesignSpinsplit-2020, hayami2019momentum, Ahn-Kunes:2019}, 
while maintaining zero net magnetization upon integration over the entire first Brillouin zone.~\footnote{According to the extended classification in Ref.~\cite{Cheong-Huang:2024}, altermagnets with vanishing magnetization belong to types II and III altermagnets.} Such materials have been named altermagnets~\cite{Smejkal-Jungwirth:2022b, SmejkalPRX2022}, and \textit{ab initio} calculations have identified several possible material candidates that can host an altermagnetic state \cite{bai_arxiv_24}. 
This includes metals like RuO$_2$~\cite{Smejkal-Sinova:2020, Ahn-Kunes:2019, Gonzalez-Hernandez-Zelezny:2021} and Mn$_5$Si$_3$~\cite{Reichlova-Smejkal:2020} as well as semiconductors/insulators like MnTe~\cite{Gonzalez-Hernandez-Zelezny:2021,Smejkal-Jungwirth:2022b}, CrSb~\cite{Smejkal-Jungwirth:2022b}, MnF$_2$~\cite{Yuan-Zunger:2020,Egorov-Evarestov:2021}, and La$_2$CuO$_4$~\cite{Smejkal-Jungwirth:2022b}. 
Recent ARPES measurements in MnTe~\cite{Lee-Kim:2023, Krempasky-Jungwirth:2024, Osumi-Sato-ObservationGiantBand-2024}, RuO$_2$~\cite{Fedchenko-Elmers:2023, Li-Felser-Ma:2024}, and CrSb~\cite{Reimers-Jourdan:2023, Yang-Liu-ThreedimensionalMappingElectronic-2024, Zeng-Liu-ObservationSpinSplitting-2024} have corroborated several of these predictions~\footnote{The magnetic order in RuO$_2$ is debated~\cite{Hiraishi-Hiroi-NonmagneticGroundState-2024, Kessler-Moser-Ru02:2024}.}. Another piece of evidence supporting altermagnetism is provided by measuring the anomalous Hall effect~\cite{Feng-Liu:2022, Tschirner-Veyrat:2023} and spin-splitter effect and torques~\cite{Bose-Ralph:2022, Bai-Song:2022, Karube-Nitta:2022}. It is noteworthy that a Fermi-surface instability caused by quasiparticle interactions has been predicted to cause an altermagnetic spin-splitting in the band structure \cite{wu_prb_07} long before the recent surge of activity in the field. On the other hand, such an instability is interestingly restricted for $p$-wave magnetism \cite{kiselev_prb_17}.

A major part of the allure of altermagnets and their potential for spintronics~\cite{Naka-Seo-SpinCurrentGeneration-2019, Shao-Tsymbal:2021, Gonzalez-Hernandez-Zelezny:2021, Smejkal-Jungwirth:2022-TMR, Das-Soori:2023, sun_prb_23, hodt_prb_24} applications is the momentum-resolved spin polarization despite the absence of net magnetization. The latter part ensures a vanishing stray magnetic field, which is highly beneficial for miniaturizing altermagnet-based elements without causing disturbance to neighboring components in a device. However, an important question is to what extent the altermagnetic properties are retained when the dimensions of the material are scaled down.

In this work, we answer this question and show that altermagnets generically develop a nonvanishing magnetization at interfaces, except for very specific crystallographic orientations of the interface. This effect originates from the explicit breakdown of the symmetry between the spin-polarized Fermi surfaces of altermagnets. Then, the periods of the Friedel oscillations of the spin-up and -down electron densities become different leading to oscillating magnetization near the edges. The maximal breakdown is achieved if one of the spin-polarized Fermi surfaces is elongated along the confinement direction, i.e., $\epsilon_{\uparrow,\mathbf{k}_{\parallel}} \neq \epsilon_{\downarrow,-\mathbf{k}_{\parallel}}$, where $\mathbf{k}_{\parallel}$ is the momentum component along the interface and $\epsilon_{\sigma,\mathbf{k}}$ denotes the dispersion relation for spin $\sigma$. The symmetry between the Fermi surfaces and, as a result, the vanishing magnetization is preserved only for a specific orientation of the lobes such that $\epsilon_{\uparrow,\mathbf{k}_{\parallel}} = \epsilon_{\downarrow,-\mathbf{k}_{\parallel}}$. For any other orientation, a magnetization develops near the interface of the altermagnet, which means that it is a robust phenomenon.

In our work, we combine analysis via a low-energy effective continuum model with a tight-binding lattice calculation to address the contribution of itinerant electrons in the magnetization of metallic altermagnets and antiferromagnets. The results of the lattice calculations are more nuanced and contain information on the structure of the wave functions on the unit-cell level. While such resolution is not necessary to obtain an interface-induced magnetization in altermagnets, which has a strong contribution from the spin-split Fermi surface, it allows us to analyze the edge magnetization also in antiferromagnets and show that it too may demonstrate an oscillating behavior despite the spin-degenerate electron bands in the bulk. The latter originates from the redistribution of the electron wave functions near the uncompensated interfaces and is distinct from the edge magnetization in altermagnets which also contains the contribution from the spin-dependent Fermi momentum. The appearance of magnetization in antiferromagnets and altermagnets has important implications with regard to their potential role in nanoscale spintronic applications. 

The focus of our work on itinerant electrons distinguishes it from other very recent works on surface magnetization that was studied primarily in insulating antiferromagnets~\cite{Andreev:1996, Belashchenko-Belashchenko-EquilibriumMagnetizationBoundary-2010, Mitsui-Akai-MagneticFriedelOscillation-2020, Lapa-Schuller-DetectionUncompensatedMagnetization-2020, Weber-Spaldin-SurfaceMagnetizationAntiferromagnets-2023, Pylypovskyi-Makarov-SurfacesymmetrydrivenDzyaloshinskiiMoriya-2023}.
While the main attention in these works is devoted to the distortion of the localized spin moment texture of the lattice sites near the surface, in our work, we ignore any redistribution and dynamics of magnetic atoms. This allows us to distinguish the edge magnetization stemming from a different origin and identify universal features that are determined by the bulk properties rather than details of the interface. The treatment of the spins of magnetic atoms as a static background is justified in the limit of low temperatures, which we assume in our work.
Our model setup and results are also distinct from the recent proposal to realize magnetization in bent altermagnets~\cite{Yershov-Kravchuk-CurvatureInducedMagnetization-2024}, where the N\'{e}el order parameter becomes nonuniform.

The predicted magnetization can be realized in both two-dimensional (2D) and three-dimensional (3D) altermagnets and antiferromagnets. In 2D, it can be directly accessed via the spin-polarized tunneling microscopy~\cite{Wortmann-Blugel-ResolvingComplexAtomicScale-2001,Wiesendanger:review-2009,Schlenhoff-Wiesendanger-RealspaceImagingAtomicscale-2020}, nitrogen-vacancy magnetometry~\cite{Casola-Yacoby-ProbingCondensedMatter-2018,Hedrich-Maletinsky-NanoscaleMechanicsAntiferromagnetic-2021,Huxter-Degen-ScanningGradiometrySingle-2022}, and SQUID~\cite{Vasyukov-Zeldov-ScanningSuperconductingQuantum-2013,Persky-Kalisky-StudyingQuantumMaterials-2022}. The magnetization profiles in 3D samples could be tackled with M\"{o}ssbauer spectroscopy similar to the studies of magnetic Friedel oscillations~\cite{Wang-Freeman-SurfaceStatesSurface-1981, Freeman-Wimmer-MagnetismSurfacesInterfaces-1982, Mitsui-Akai-MagneticFriedelOscillation-2020}. 

Our paper is organized as follows. In Sec.~\ref{sec:analytics}, we introduce an effective low-energy model and provide an analytical calculation of the local density of state (LDOS) and magnetization in an altermagnetic ribbon; an antiferromagnetic ribbon is also discussed. Characterization of the interface-induced magnetization in a tight-binding model is provided in Sec.~\ref{sec:numerics}.
Technical details related to the derivation of the continuum models and the definition of the spectral function in the lattice model are presented in Appendixes~\ref{App:tight-binding} and \ref{App: A}, respectively.
The results are discussed and summarized in Sec.~\ref{sec:Summary}. Throughout this paper, we use $\hbar=k_{\rm B}=1$.

\section{Effective continuum model: theory and results}
\label{sec:analytics}

We start our investigation of edge-induced magnetization from the analysis in continuum low-energy models. Such models are derived from the lattice model, see Sec.~\ref{sec:numerics} and Appendix~\ref{App:tight-binding}, and are intended to capture the key aspects of materials. As we demonstrate in our work, while being successful in predicting the bulk properties and altermagnetic surface effects, the inherent coarse-graining of continuum models does not allow to capture a subtle interplay between boundaries and the structure of the unit-cell. Nevertheless, the use of continuum models is instructive and allows us to separate different mechanisms responsible for edge-induced magnetization.

\subsection{Antiferromagnets}
\label{sec:analytics-AFM}

As a warm-up, we consider an antiferromagnetic ribbon, see Fig.~\ref{fig:latticemodel}(a) for the setup geometry. Since an antiferromagnetic material has spin-degenerate energy bands, we expect no magnetization arising due to itinerant electrons unless a spin asymmetry is introduced via, e.g., boundary conditions. To illustrate this, in what follows, we consider a ribbon of an antiferromagnetic metal, and show that the corresponding LDOS remains spin-degenerate, i.e., no magnetization arises in a continuum model. However, this changes in the tight-binding model where the unit-cell geometry gives rise to phase shifts of the itinerant electron wave functions which turn out to produce a net interface magnetization, as we discuss later in Sec.~\ref{sec:numerics-AFM}. This suggests that the standard boundary conditions of a vanishing wave function at the vacuum edge are insufficient to capture the interfacial physics in antiferromagnetic systems.

\subsubsection{Model and eigenfunctions in antiferromagnetic ribbon}
\label{sec:analytics-AFM-model}
The simplest low-energy continuum Hamiltonian for an antiferromagnet is
\begin{equation}
\label{analytics-H-AF}
H_{\rm AFM}(\mathbf{k}) = \frac{\epsilon_0}{\sqrt{J_{\rm sd}^2+\epsilon_0^2}} \frac{k^2}{2m} -\left(\mu+\sqrt{J_{\rm sd}^2+\epsilon_0^2}\right),
\end{equation}
where $\mu$ is the chemical potential, $m$ is the effective mass, $\mathbf{k}$ is the momentum, $J_{\rm sd}$ is the strength of the \textit{s}-\textit{d} coupling, and $\epsilon_0$ is a constant with the dimension of energy. In what follows, we find it convenient to introduce dimensionless variables: $\tilde{\epsilon} = \epsilon/\mu$, $\tilde{k} = k/k_F$ with $k_F=\sqrt{2m\mu}$, and $\tilde{J}=J_{\rm sd}/\mu$.
The Hamiltonian (\ref{analytics-H-AF}) does not depend on the spin projection $s=\pm$ and describes two spin-degenerate energy bands.

The low-energy continuum model (\ref{analytics-H-AF}) can be straightforwardly derived from the lattice model defined in Sec.~\ref{sec:numerics-model} if one retains only the lowest-energy band and performs an expansion in momentum around the $\Gamma$ point, see Appendix~\ref{App:tight-binding} for details of the derivation. Being a continuum model, the model (\ref{analytics-H-AF}) loses information about the unit-cell structure. In particular, it becomes insensitive to features defined on the scale of the lattice constant where a more refined lattice treatment is required. Yet, as we will phenomenological show in Sec.~\ref{sec:numerics-AFM-phem}, the continuum model can still be useful in providing a qualitative picture of the lattice results.

The Hamiltonian (\ref{analytics-H-AF}) can be further simplified by introducing a new set of dimensionless variables: $\epsilon' = \epsilon/\left[\mu \left(1+\sqrt{\tilde{J}^2+\tilde{\epsilon}_0^2}\right)\right]$ and  $k' = k/q_F$ with $q_F = \sqrt{2m \mu \left(1+\sqrt{\tilde{J}^2+\tilde{\epsilon}_0^2}\right) \sqrt{\tilde{J}^2 +\tilde{\epsilon}_0^2}/\tilde{\epsilon}_0}$. Then, the dimensionless version of the Hamiltonian (\ref{analytics-H-AF}) reads
\begin{equation}
\label{analytics-H-AF-minus}
H_{\rm AFM}(\mathbf{q}) = \mu \left(1+\sqrt{\tilde{J}^2+\tilde{\epsilon}_0^2}\right) \left(\tilde{q}^2 -1\right).
\end{equation}

\begin{figure}[htb]
    \centering
    \includegraphics[width=\linewidth]{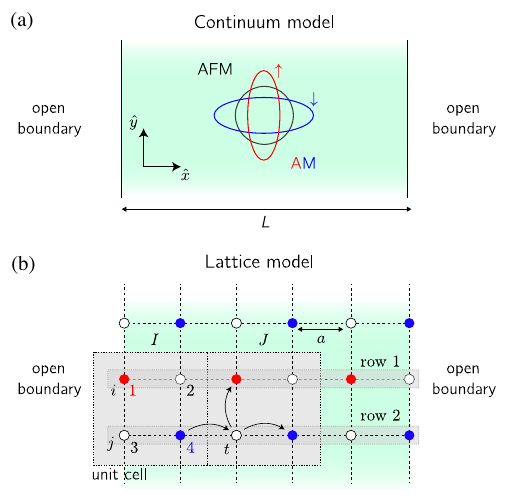}
    \caption{
    The antiferromagnetic (AFM) and altermagnetic (AM) ribbons are modeled both with a continuum model (a) and a tight-binding lattice model (b) with a four-site unit-cell which allows us to continuously tune the system from AFM to AM. In the continuum model, the ribbon is infinite in the \textit{y}-direction and of length \textit{L} in the \textit{x}-direction. In the lattice model, sites 1 and 4 are localized, anti-collinear magnetic moments, coupled to the itinerant electrons through an \textit{s-d} coupling. Sites 2 and 3 are nonmagnetic sites with an on-site energy $\epsilon_i$. The unit-cell indices are $(I,J)$, while the site indices are $(i,j)$. To model an AFM, we take $\epsilon_2=\epsilon_3=0$ while $\epsilon_2\neq\epsilon_3$ gives rise to AM order by introducing anisotropic spin-splitting mediated by the nonmagnetic sites. The nearest-neighbor hopping is described by $t$.
    }
    \label{fig:latticemodel}
\end{figure}

In the ribbon geometry shown in Fig.~\ref{fig:latticemodel}(a), only the momentum along the edges remains a good quantum number. For the sake of definiteness, we direct the $x$ axis normal to the ribbon and the $y$ axis along its edge. The antiferromagnetic ribbon is described by the Hamiltonian (\ref{analytics-H-AF-minus}) with $k_x \to -i\nabla_x$.

The wave function in the ribbon is 
\begin{equation}
\label{analytics-AF-sol}
\psi_{s,k'_y}(x') =  A_{+} \phi_{s,q'_{x},k'_y} e^{iq'_{x}x'+ik'_y y'} 
+A_{-} \phi_{s,-q'_{x},k'_y} e^{-iq'_{x}x'+ik'_y y'}
\end{equation}
with $x'= x q_F$ and $q'_{x} = \sqrt{\epsilon' +1 -(k'_y)^2}$.

We assume the following boundary conditions:
\begin{equation}
\label{analytics-AF-psi-BC}
\psi_{s,k_y}(x=0) = \psi_{s,k_y}(x=L) =0,
\end{equation}
which correspond to open boundaries. Since there is no spin mixing at the boundaries, each of the spin projections can be considered separately. 

The boundary conditions (\ref{analytics-AF-psi-BC}) lead to the following characteristic equation:
\begin{equation}
\label{analytics-AF-char-eq}
\sin{\left( L' q'_{x} \right)}=0,
\end{equation}
and allow us to determine the energy levels in the ribbon:
\begin{equation}
\label{analytics-AF-char-eq-sol}
\epsilon'_{n} = \left(\frac{\pi n}{L'}\right)^{2} +(k'_y)^2 -1
\end{equation}
with $n=0,1,2,\ldots$. Then, the normalized wave function in the ribbon is
\begin{equation}
\label{ribbon-AF-magnetism-v2-psi-norm}
\psi_{s,k_y',n}(x') = i\sqrt{\frac{2}{L}} \sin{\left(\frac{\pi n x'}{L'}\right)},
\end{equation}
where we used $\int_0^{L} dx |\psi_{s,k_y',n}(x')|^2 =1$.

\subsubsection{Density of states and magnetization}
\label{sec:analytics-AFM-DOS}

Interface effects are manifested in the local spin-resolved density of states (DOS) and magnetization. The LDOS in the ribbon is defined as: 
\begin{equation}
\label{analytics-AF-DOS-def}
\nu_s(\epsilon, x) = \sum_{n=0}^{\infty} \int \frac{dk_y}{2\pi} |\psi_{s,k_y,n}(x)|^2 \df{\epsilon -\epsilon_{n}}.
\end{equation}
By using the result (\ref{ribbon-AF-magnetism-v2-psi-norm}), we obtain
\begin{equation}
\label{analytics-AF-1}
\nu_s(\epsilon', x') = \frac{4\nu_0}{L'} \sum_{n=0}^{\infty} \RE{\frac{\sin^2{\left(\pi n x'/L'\right)}}{\sqrt{\epsilon' +1 -\left(\pi n/L'
\right)^{2}}}},
\end{equation}
where $\nu_0 = q_F^2 /\left[4\pi \mu \left(1+\sqrt{\tilde{J}^2+\tilde{\epsilon}_0^2}\right)\right]$ is the bulk DOS. As is evident from Eq.~(\ref{analytics-AF-1}), the LDOS in an antiferromagnetic ribbon is spin independent. Therefore, there is no edge-induced magnetization in this continuum model. To set up the stage for the comparison with altermagnets in Sec.~\ref{sec:analytics-AM-model}, we calculate the LDOS and the local charge density in a few approximations.

The local charge density is defined as
\begin{eqnarray}
\label{analytics-AF-n-def}
n_s(x') &=& -e\int d\epsilon\, \nu_s(\epsilon, x') n_F(\epsilon) \nonumber\\ 
&\stackrel{T\to0}{=}& 8n_0 \sum_{n=0}^{\infty}  \frac{\sin^2{\left(\frac{\pi n x'}{L'}\right)}}{L'} 
\RE{\sqrt{1 -\left(\frac{\pi n}{L'}
\right)^{2}}},
\end{eqnarray}
where $n_F(\epsilon)$ is the Fermi-Dirac distribution and $n_0=-e\nu_0 \mu \left(1+\sqrt{\tilde{J}^2+\tilde{\epsilon}_0^2}\right)$.

In the case of the wide ribbon, $L'\to\infty$, we replace $\pi n/ L' \to  p'_x$ and $\sum_{n=0}^{\infty} \to  L' \int_{-\infty}^{\infty} d p'_x/(2\pi)$ in Eqs.~(\ref{analytics-AF-1}) and (\ref{analytics-AF-n-def}). Then, the following compact formulas can be obtained
\begin{eqnarray}
\label{analytics-AF-large-L-DOS}
\lim_{L' \to \infty} \nu_s(\epsilon', x') &=& \nu_0 \left[1 - J_0{\left(2\sqrt{\epsilon'+1} x'\right)}\right],\\
\label{analytics-AF-large-L-n}
\lim_{L' \to \infty} n_s(x') &=& n_0 \left[1 - \frac{J_1{\left(2x'\right)}}{x'} \right]
\end{eqnarray}
with $J_n(x)$ being the Bessel functions of the first kind.

Away from the interfaces, $x'\gg1$, we extract the following asymptotics in Eqs.~(\ref{analytics-AF-large-L-DOS}) and (\ref{analytics-AF-large-L-n}):
\begin{eqnarray}
\label{analytics-AF-large-x-DOS}
\nu_s(\epsilon', x') &=& \nu_0 \!\left[1 -\frac{1}{\sqrt{\pi x'}} \frac{\sin{\left(2 \sqrt{\epsilon' +1} x'
 +\pi/4\right)}}{\left(\epsilon'+1\right)^{1/4}}
 \right] +\mathcal{O}{\left(\frac{1}{x'}\right)},\nonumber\\
 \\
\label{analytics-AF-large-x-n}
n_s(x') &=& n_0 \! \left[1 +\frac{1}{\sqrt{\pi} (x')^{3/2}} \cos{\left(2 x'
 +\frac{\pi}{4}\right)} \right] +\mathcal{O}{\left(\frac{1}{(x')^2}\right)}.\nonumber\\
\end{eqnarray}
Therefore, at $x' \gg 1$, the LDOS and the charge density approach their bulk values with the oscillating corrections decaying as $1/\sqrt{x'}$ and $1/x'^{3/2}$, respectively. Similar to regular Friedel oscillations of the electron density near defects in metals~\cite{Harrison:book, Mahan:book-2013}, the period of the oscillations is determined by the doubled Fermi momentum $q_F$.

Thus, since the energy bands in the AFM are spin-degenerate and the boundary conditions in the continuum model do not distinguish spin projections, the LDOS and the charge density in the continuum model of antiferromagnets remain spin-degenerate. Neither local nor net magnetizations occur in this case. As mentioned at the beginning of this section, we will see that this conclusion changes in a tight-binding model discussed in Sec.~\ref{sec:numerics}.

\subsection{Altermagnet}
\label{sec:analytics-AM-model}

While as in antiferromagnets, the net magnetization of bulk altermagnets vanishes, the momentum-dependent spin-splitting allows for qualitatively different behavior close to edges or interfaces. In this section, we show that, except for very specific orientations of the boundaries, finite-sized altermagnets allow for local and net magnetizations.

\subsubsection{Model and eigenfunctions in altermagnetic ribbon}
\label{sec:analytics-model}

To illustrate the generality of our results regarding the role of boundaries in altermagnets, we start with continuum model obtained from symmetry considerations~\cite{Smejkal-Jungwirth:2022}; such a model can also be derived from the lattice model defined in Sec.~\ref{sec:numerics-model}, see Appendix~\ref{App:tight-binding} for the details. We use the following Hamiltonian describing 2D $d$-wave altermagnets:
\begin{equation}
\label{ribbon-H-AM}
H(\mathbf{k}) = \frac{k^2}{2m} -\mu +\sigma_z \frac{1}{2m} \left[t_1 \left(k_x^2 -k_y^2\right) +2 t_2 k_x k_y\right].
\end{equation}
The anisotropy of the fully spin-polarized Fermi surfaces is described by the third term in Eq.~(\ref{ribbon-H-AM}) with $\sigma_z$ being the Pauli matrix in the spin space and dimensionless parameters $|t_1|<1$ and $|t_2|<1$ describing the strength of the altermagnetism; the corresponding band structure is schematically shown by the blue and red ellipses in Fig.~\ref{fig:latticemodel}(a) at $t_1\neq0$ and $t_2=0$. The case with $t_1=0$ and $t_2\neq0$ would correspond to a rotation of both the blue and red ellipses by $\pi/4$. In the latter case, the interface does not break the symmetry between the spin-polarized Fermi surfaces.
The ribbon is described by the Hamiltonian (\ref{ribbon-H-AM}) with $k_x \to -i\nabla_x$. We assume open boundary conditions that do not intermix spins, see Eq.~(\ref{analytics-AF-psi-BC}). This and the diagonal structure of the Hamiltonian (\ref{ribbon-H-AM}) allow us to treat each of the spin projections $s=\uparrow, \downarrow$ independently. 

The wave functions in the ribbon have the following form: 
\begin{equation}
\label{ribbon-psi}
\psi_{s,\tilde{k}_y}(\tilde{x}) = A_{s,+} e^{iq_{s,+} \tilde{x}} +A_{s,-} e^{iq_{s,-} \tilde{x}}
\end{equation}
cf. Eq.~(\ref{analytics-AF-sol}) and satisfy the boundary conditions (\ref{analytics-AF-psi-BC}). Here,
\begin{equation}
\label{ribbon-dwave-v1-lambda}
q_{s,\pm} = -\frac{s \tilde{k}_y t_2}{1+st_1} \pm \frac{1}{1+st_1} \sqrt{(1+st_1) (\tilde{\epsilon} +1) -\tilde{k}_y^2 \left(1 -t_1^2 -t_2^2\right)}
\end{equation}
and we used the dimensionless variables $\tilde{\epsilon} = \epsilon/\mu$, $\tilde{k}_y = k_y/k_F$, and $\tilde{x}=xk_F$ with $k_F=\sqrt{2m\mu}$.

Substituting the wave function (\ref{ribbon-psi}) into the boundary conditions (\ref{analytics-AF-psi-BC}), we find
\begin{eqnarray}
\label{ribbon-psi-n}
\psi_{s,\tilde{k}_y,n}(\tilde{x}) &=& i \sqrt{\frac{2}{L}} e^{-i\frac{s \tilde{k}_y t_2 \tilde{x}}{1+st_1}} \sin{\left(\frac{\pi n \tilde{x}}{\tilde{L}}\right)},\\
\label{ribbon-epsilon-n}
\tilde{\epsilon}_{n,s} &=& \tilde{k}_y^2 \frac{\left(1 -t_1^2 -t_2^2\right)}{1+st_1} + (1+st_1) \left(\frac{\pi n}{\tilde{L}}\right)^2 -1.
\end{eqnarray}

\subsubsection{Density of states and magnetization}
\label{sec:analytics-DOS}

The interplay of the altermagnetic spin-splitting and interface effects
is manifested in the local spin-resolved DOS and magnetization. By using the results in Eqs.~(\ref{ribbon-psi-n}) and (\ref{ribbon-epsilon-n}) in Eq.~(\ref{analytics-AF-DOS-def}), we obtain
\begin{eqnarray}
\label{ribbon-DOS-1}
\nu_s(\tilde{\epsilon}, \tilde{x}) &=& 4\nu_0  \sqrt{\frac{1+st_1}{1 -t_1^2 -t_2^2}} \sum_{n=0}^{\infty}  \frac{\sin^2{\left(\frac{\pi n \tilde{x}}{\tilde{L}}\right)}}{\tilde{L}} \nonumber\\
&\times& \frac{1}{\RE{\sqrt{\tilde{\epsilon} +1 -(1+st_1)\left(\frac{\pi n}{\tilde{L}}\right)^{2}}}},
\end{eqnarray}
where $\nu_0= m/(2\pi)$ is the bulk DOS in 2D. Unlike antiferromagnets, the LDOS of altermagnets retains the dependence on spin, cf. Eqs.~(\ref{analytics-AF-1}) and (\ref{ribbon-DOS-1}).

\begin{figure}[!ht]
\centering
\includegraphics[width=0.4\textwidth]{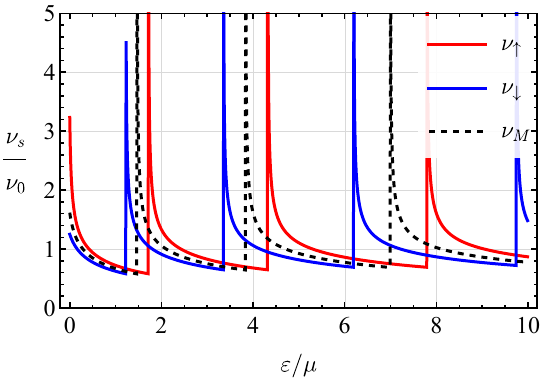}
\caption{
The spin-resolved LDOS at $x=L/2$. Solid red and blue lines correspond to the spin-up and spin-down LDOS, respectively, at $t_1 = 0.1$ and $t_2=0$. The dashed black line is a metallic LDOS $\nu_M$ at $t_1 = t_2=0$. We fixed $\tilde{L}=10$.
}
\label{fig:ribbon-DOS-eps}
\end{figure}

We present the spin-resolved LDOS in the middle of the ribbon, $x=L/2$, in Fig.~\ref{fig:ribbon-DOS-eps}. As expected from Eq.~(\ref{ribbon-DOS-1}), there is a well-pronounced splitting of the spin-polarized DOSs in altermagnets if $t_1\neq0$, i.e., when the altermagnetic lobes are perpendicular to the interface, see Fig.~\ref{fig:latticemodel}. No splitting is observed for $t_1=0$ and $t_2\neq0$ when the symmetry between the spin-polarized Fermi surfaces is preserved by the interface.  

The local charge density at vanishing temperature is
\begin{eqnarray}
\label{ribbon-n-def}
n_s(\tilde{x}) &=& 
8n_0 \sqrt{\frac{1+st_1}{1 -t_1^2 -t_2^2}} \sum_{n=0}^{\infty}  \frac{\sin^2{\left(\frac{\pi n \tilde{x}}{\tilde{L}}\right)}}{\tilde{L}} \nonumber\\
&\times&  \RE{\sqrt{1 -(1+st_1)\left(\frac{\pi n}{\tilde{L}}\right)^{2}}},
\end{eqnarray}
where we used the first line in Eq.~(\ref{analytics-AF-n-def}) and $n_0=-e\nu_0 \mu$. The local magnetization quantifies the difference between the spin-up and spin-down charge densities:
\begin{equation}
\label{ribbon-DOS-mz-def}
m_z(x) = -\frac{g\mu_B}{e} \left[ n_{\uparrow}(x) -n_{\downarrow}(x)\right],
\end{equation}
where $g$ is the Land\'{e} factor and $\mu_B$ is Bohr magneton.

While Eqs.~(\ref{ribbon-DOS-1}) and (\ref{ribbon-n-def}) are accessible for numerical calculations, it is instructive to consider a few simplifying assumptions. Assuming a wide ribbon with $\tilde{L} \gg 1$, the summation over the thickness modes can be replaced with the integration. The resulting expressions for the LDOS and the charge density read
\begin{eqnarray}
\label{ribbon-DOS-DOS-1}
\lim_{\tilde{L} \to \infty} \! \nu_s(\tilde{\epsilon}, \tilde{x}) &=& \frac{\nu_0}{\sqrt{1 -t_1^2 -t_2^2}} \left[1 - J_0{\left(2\sqrt{\frac{\tilde{\epsilon}+1}{1+st_1}} \tilde{x}\right)}\right],\\
\label{ribbon-DOS-n-1}
\lim_{\tilde{L} \to \infty} \! n_s(\tilde{x}) &=& \frac{n_0}{\sqrt{1 -t_1^2 -t_2^2}} \left[1 - \frac{\sqrt{1+st_1}}{\tilde{x}} J_1{\left(\frac{2\tilde{x}}{\sqrt{1+st_1}}\right)}\right], \nonumber\\
\end{eqnarray}
cf. Eqs.~(\ref{analytics-AF-large-L-DOS}) and (\ref{analytics-AF-large-L-n}).

Away from the interfaces, $\tilde{x}/\sqrt{1+st_1} \gg1$, we extract the following asymptotics in Eqs.~(\ref{ribbon-DOS-DOS-1}) and (\ref{ribbon-DOS-n-1}):
\begin{widetext}
\begin{eqnarray}
\label{ribbon-DOS-DOS-2}
\nu_s(\tilde{\epsilon}, \tilde{x}) &=& \frac{\nu_0}{\sqrt{1 -t_1^2 -t_2^2}} \left[1 -
\left(\frac{1 +st_1}{\tilde{\epsilon}+1}\right)^{1/4} \frac{1}{\sqrt{\pi \tilde{x}}} \sin{\left(2 \sqrt{\frac{\tilde{\epsilon} +1}{1+st_1}} \tilde{x} +\frac{\pi}{4}\right)} \right] +\mathcal{O}{\left(\frac{1}{\tilde{x}}\right)},\\
\label{ribbon-DOS-n-2}
n_s(\tilde{x}) &=& \frac{n_0}{\sqrt{1 -t_1^2 -t_2^2}} \left[1 +\frac{\left(1 +st_1\right)^{3/4}}{\sqrt{\pi} \tilde{x}^{3/2}} \cos{\left(\frac{2}{\sqrt{1+st_1}} \tilde{x}  +\frac{\pi}{4}\right)}  \right] +\mathcal{O}{\left(\frac{1}{\tilde{x}^2}\right)},
\end{eqnarray}
\end{widetext}
cf. Eqs.~(\ref{analytics-AF-large-x-DOS}) and (\ref{analytics-AF-large-x-n}).
While the decay with the coordinate is the same as in the antiferromagnet considered in Sec.~\ref{sec:analytics-AFM-DOS}, the period of the oscillations is determined by the Fermi momentum, which in the present altermagnetic case reads as $2k_F/\sqrt{1+st_1}$. Therefore, spin-up and spin-down charge densities are characterized by different oscillation frequencies. Hence, according to Eq.~(\ref{ribbon-DOS-mz-def}), {an} altermagnet is expected to develop oscillating magnetization.

Under the same approximations and assuming weak altermagnetic parameters, $t_1\ll1$ and $t_2\ll1$, the magnetization (\ref{ribbon-DOS-mz-def}) reads
\begin{equation}
\label{ribbon-DOS-mz-2}
m_z(\tilde{x}) \approx \frac{2t_1 m_0}{\sqrt{\pi \tilde{x}}} \sin{\left(2\tilde{x} +\frac{\pi}{4}\right)},
\end{equation}
where $m_0 = -g\mu_B n_0/e$. In writing Eq.~(\ref{ribbon-DOS-mz-2}), we expanded in $1/\tilde{x}$, $t_1$, and $t_2$ and kept only the leading-order nontrivial term.
Therefore, we expect a slower $\propto 1/\sqrt{\tilde{x}}$ decay compared to the charge density oscillations. In addition, the oscillations of the local charge density and the magnetization are shifted by $\pi/2$ at small $t_1$. 

While the LDOS and magnetization can be readily studied via local probes, it is also convenient to introduce the spatially averaged characteristics of the ribbon, which can be investigated without requiring high spatial resolution via, e.g., SQUID~\cite{Vasyukov-Zeldov-ScanningSuperconductingQuantum-2013, Granata-Vettoliere-NanoSuperconductingQuantum-2016, Buchner-Ney-TutorialBasicPrinciples-2018, Persky-Kalisky-StudyingQuantumMaterials-2022, Paulsen-Kiefer-UltralowFieldSQUID-2022}. 
For this, we integrate the magnetization over the ribbon width:
\begin{eqnarray}
\label{ribbon-Mz-def}
M_z &=& \int_0^{L}dx\, m_z(x)= 2 m_0\sum_{s} \frac{s
L}{\sqrt{1 -t_1^2 -t_2^2}} \nonumber\\
&\times& \left[1 - {}_{1}F_{2}\left(\frac{1}{2};\frac{3}{2},2; -\frac{\tilde{L}^2}{1+st_1}\right) \right] \nonumber\\
&\stackrel{\tilde{L} \gg1}{\approx}& 
\frac{2m_0}{k_F} \frac{1}{\sqrt{1 -t_1^2 -t_2^2}} \left(\sqrt{1-t_1} -\sqrt{1+t_1}\right) \nonumber\\
&\stackrel{t_1\ll1, t_2\ll1}{\approx}& -\frac{m_0}{k_F} t_1,
\end{eqnarray}
where ${}_{1}F_{2}\left(a;b_1,b_2; z\right)$ is the hypergeometric function. In the second expression, we used Eqs.~(\ref{ribbon-DOS-mz-def}) and (\ref{ribbon-DOS-n-1}); the factor $2$ stems from the contribution of two edges of the ribbon. Therefore, the ribbon becomes magnetized with the total magnetization depending linearly on the altermagnetic parameter $t_1\ll1$. 

As one can see from the second line of Eq.~(\ref{ribbon-Mz-def}), the net magnetization is determined by the asymmetry between the spin-polarized Fermi surfaces in the direction perpendicular to the edge of the ribbon; the asymmetry is quantified by the eccentricity of the ellipsoidal Fermi surfaces. The same asymmetry is responsible for the different frequencies of the Friedel oscillations for spin-up and spin-down electrons.

\begin{figure}[!ht]
\centering
\subfigure{\includegraphics[width=0.4\textwidth]{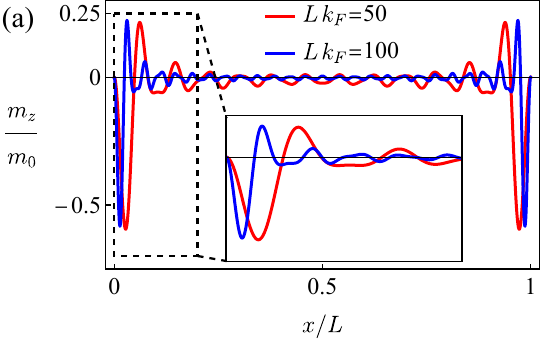}}
\subfigure{\includegraphics[width=0.4\textwidth]{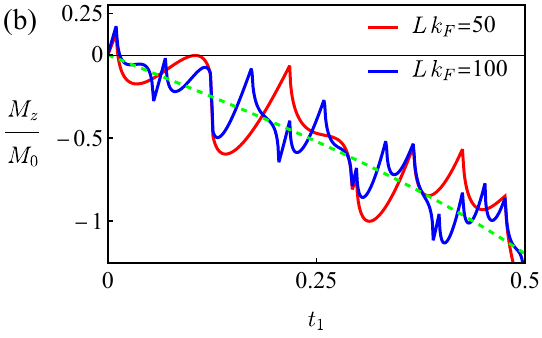}}
\caption{
(a) The spatial profile of the magnetization for a few widths of the ribbon at $t_1=0.5$. The periods of the oscillations are determined by two Fermi wave vectors $k_F/(1\pm t_1)$. The inset shows the enlarged region near the boundary.
Panel (b) The magnetization $M_z$ obtained by integrating over the ribbon width as a function of the altermagnetic parameter $t_1$. The green dashed line corresponds to the approximate expression in the last line in Eq.~(\ref{ribbon-Mz-def}). Here, $M_0 = m_0/k_F$. In both panels, we fix $t_2=0$. 
}
\label{fig:ribbon-DOS-mz}
\end{figure}

The local $m_z(x)$ and net $M_z$ magnetizations are shown in Figs.~\ref{fig:ribbon-DOS-mz}(a) and \ref{fig:ribbon-DOS-mz}(b), respectively. As expected, the oscillations of local magnetization contain two periods, see Eqs.~(\ref{ribbon-DOS-mz-def}) and (\ref{ribbon-DOS-n-2}), and decay into the bulk. For a larger width of the ribbon, the oscillations in Fig.~\ref{fig:ribbon-DOS-mz}(b) become indiscernible and approach the result given in the last line of Eq.~(\ref{ribbon-Mz-def}). 

To provide numerical estimates of the induced magnetization, we use the gyromagnetic ratio $g=2$ and estimate the effective mass from the tight-binding model expansion as $m=1/(2a^2t)$ (see, e.g., Eq.~(\ref{app-tb-afm})) with $a=5~\AA$ being the typical lattice constant and $t=1~\mbox{eV}$ being the hopping parameter. The characteristic magnetization $m_0$ and corresponding magnetic polarization $\mathcal{P}_{m_0}$ are
\begin{equation}
\label{estimate-m}
m_0 \approx 6\times 10^{-3} \mu_B \, \frac{\mu}{1~\mbox{eV}}, \quad \mathcal{P}_{m_0} \approx 0.1 \, \frac{\mu}{1~\mbox{eV}}~\mu\mbox{G}.
\end{equation}
The characteristic magnitude of the integrated over the ribbon width magnetization (i.e., the flux per unit length) $M_0=m_0/k_F$, see Eq.~(\ref{ribbon-Mz-def}), and the corresponding integrated magnetic polarization are estimated as
\begin{equation}
\label{estimate-M}
M_0 \approx 3\times 10^{-2}\, t_1 \mu_B\, \frac{\mu}{1~\mbox{eV}}\,\AA, \quad \mathcal{P}_{M_0} \approx 0.4 t_1 \, \frac{\mu}{1~\mbox{eV}}~\mu\mbox{G}\,\AA.
\end{equation}
For a ribbon with the length $1~\mbox{mm}$, $t_1=0.5$, and $\mu= 1~\mbox{eV}$, the induced magnetization corresponds to a flux almost 10 times larger than the magnetic field flux quantum. Therefore, the edge magnetization in altermagnets is measurable via SQUID techniques~\cite{Vasyukov-Zeldov-ScanningSuperconductingQuantum-2013, Granata-Vettoliere-NanoSuperconductingQuantum-2016, Buchner-Ney-TutorialBasicPrinciples-2018, Persky-Kalisky-StudyingQuantumMaterials-2022, Paulsen-Kiefer-UltralowFieldSQUID-2022}.

To summarize the results of this section, we found that the spin anisotropy of the Fermi surface in altermagnets is manifested in the spin-resolved LDOS and magnetization. By using a ribbon geometry as a representative example, we found the splitting of the spin-resolved LDOS and nonvanishing magnetization for generic orientations of the altermagnetic Fermi surfaces except for the single orientation that has combined mirror and spin-flip symmetry, corresponding to a compensated interface such that the spin-polarized lobes in Fig.~\ref{fig:latticemodel}(a) are rotated by $\pi/4$. The splitting of the spin-resolved LDOS and resulting magnetization arise from different length scales of the Friedel oscillations quantified by $2k_F/\sqrt{1+st_1}$. For a wide ribbon, we found that the magnetization decays as $1/\sqrt{k_F x}$ away from the edges of the ribbon and is shifted by $\pi/2$ in phase compared to the electric charge oscillations. 

The effective continuum model of antiferromagnets discussed in Sec.~\ref{sec:analytics-AFM} shows similar charge oscillations, albeit has spin-degenerate LDOS. Therefore, under the same conditions, antiferromagnets have no magnetization oscillations due to their itinerant electrons.

These analytical findings are confirmed in a lattice model for altermagnets in Sec.~\ref{sec:numerics}. The results of the lattice model, however, are more nuanced and provide access to additional information related to the unit-cell structure. In fact, we will see that the tight-binding model reveals a mechanism that induces an interface magnetization carried by the itinerant electrons which is not captured by the analytical model, such as the interface-induced magnetization in antiferromagnets.

\section{Lattice model: theory and results}
\label{sec:numerics}

\subsection{Tight-binding \textit{s}-\textit{d} model for antiferromagnet and altermagnet}
\label{sec:numerics-model}

We describe the AFM or AM ribbon on a lattice through a tight-binding model with an \textit{s-d} coupling to localized anti-collinear moments. The lattice is shown in Fig.~\ref{fig:latticemodel}(b). This model is inspired by the microscopic model for altermagnets introduced by Ref.~\cite{Brekke-Sudbo:2023}, but we consider the full four-site unit-cell. This allows us to continuously introduce the lattice asymmetry which takes the AFM order ($\epsilon_2 = \epsilon_3$) to the AM order ($\epsilon_2 \neq \epsilon_3$). Note in particular that we take the boundary conditions in the \textit{x}-direction to be open, while the \textit{y}-direction is periodic.

The tight-binding \textit{s-d} model is given by
\begin{multline}
    H=-t\sum_{\langle i,j\rangle, \sigma}\cc[]{i,\sigma}\ca[]{j,\sigma} 
    -J_\text{sd}\sum_{i,\sigma,\sigma'}(\boldsymbol{S}_i \cdot \boldsymbol{\sigma})_{\sigma\sigma'} \cc[]{i,\sigma}\ca[]{i,\sigma'} \\-\sum_{i,\sigma}(\mu - \delta_{i,i_\text{NM}}\epsilon_{\text{NM}})\cc[]{i,\sigma}\ca[]{i,\sigma},
    \label{eqn: model before unit}
\end{multline}
where \textit{i}, \textit{j} denote individual lattice sites, \textit{t} describes nearest-neighbor hopping, and $J_{\text{sd}}$ is the coupling between the itinerant electrons described by $\ca[]{}/\cc[]{}$ and the localized moments $\boldsymbol{S}_i$, the configuration of which is shown in Fig.~\ref{fig:latticemodel}(b).
Finally, $\mu$ is the chemical potential while an additional on-site energy $\epsilon_{\text{NM}} \in \{\epsilon_2, \epsilon_3\}$ is present if the site is nonmagnetic.

We now segment the lattice into four-site unit-cells shown in Fig.~\ref{fig:latticemodel}(b) using a band basis $\ca[]{I,\alpha,\sigma}/\cc[]{I,\alpha,\sigma}$, where \textit{I} is now a unit-cell index, $\alpha\in[1,2,3,4]$ denote the site index within the unit-cell and where $\sigma$ is the spin-index. To take advantage of the periodicity in \textit{y}, we also introduce Fourier-transformed band-operators in the \textit{y}-direction,
\begin{equation}
    \ca[]{(I_x,I_y),\alpha,\sigma}=\frac{1}{N_{y}^\text{u.c}}\sum_{k_y}\ca[]{I_x,k_y,\alpha,\sigma}e^{ik_y R_y},
\end{equation}
where $N_y^{\text{u.c.}}=N_y/2$ is the number of unit-cells in the \textit{y} direction, where $R_y$ is the \textit{y}-coordinate of the unit-cell and where the momentum $k_y\in(-\pi/2a,\pi/2a]$ runs over the reduced Brillouin zone corresponding to the $2\times 2$ real-space unit-cell and $a$ is the lattice constant. From now on, we omit \textit{x/y} subscripts on real- and momentum space indices, taking $I=I_x$ and $k=k_y$ for brevity of notation. 

The total Hamiltonian in the unit-cell basis is now given by 
\begin{align}
    H&=\sum_{k}\sum_{I,\sigma}\bigg\{
    \bigg[ \sum_{\Delta_x}\Tilde{c}_{I,k,\sigma}^\dagger h_{I,I+\Delta_x, k, \sigma}^{\vphantom{\dagger}}\Tilde{c}_{I+\Delta_x,k,\sigma}^{\vphantom{\dagger}} \bigg] \nonumber\\
    &+ \Tilde{c}_{I,k,\sigma}^\dagger h_{I,I, k, \sigma}^{\vphantom{\dagger}}\Tilde{c}_{I,k,\sigma}^{\vphantom{\dagger}} \bigg\},
    \label{eqn: total hamiltonian}
\end{align}
where $\Delta_x$ is the nearest-neighbor vector ($|\Delta_x|=|\Delta_y|=2a$) between two unit-cells in the \textit{x}-direction, the unit-cell basis is given as 
\begin{equation}
    \tilde{c}_{I,k,\sigma} = \begin{pmatrix}
        c_{I,k,1,\sigma} &
        c_{I,k,2,\sigma} &
        c_{I,k,3,\sigma} &
        c_{I,k,4,\sigma} 
    \end{pmatrix}^T,
\end{equation}
and where we defined the coefficient matrices 
\begin{widetext}
\begin{equation}
    h_{I,I,k,\sigma}= \begin{bmatrix}
        -J_\text{sd} \sigma -\mu & -t & -te^{2ika}-t & 0 \\
        -t & \epsilon_2 -\mu & 0 & -te^{2ika} - t \\
        -t\mathrm{e}^{-2ika} - t & 0 & \epsilon_3-\mu & -t \\
        0 & -te^{-2ika} -t & -t & J_\text{sd}\sigma -\mu 
    \end{bmatrix}, \qquad
    h_{I,I+\Delta_x,k,\sigma}=-t\begin{bmatrix}
        0 & \delta_{\Delta_x, -\hat{x}} & 0 & 0 \\
        \delta_{\Delta_x, \hat{x}} & 0  & 0 & 0 \\
        0 & 0 & 0 & \delta_{\Delta_x, -\hat{x}} \\
        0 & 0 & \delta_{\Delta_x, \hat{x}} & 0 
    \end{bmatrix}.
\end{equation}
\end{widetext}

\begin{figure}
    \includegraphics[width=1.0\linewidth]{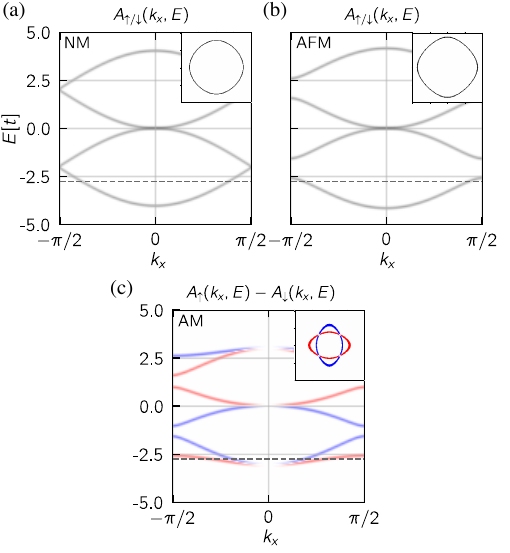}
    \caption{The spectral function $A_{\sigma}(k_x, E)$ (see Appendix \ref{App: A} for definition) is shown for $k_y=0$ in the lattice model with translational invariance in both \textit{x} and \textit{y}. The normal metal (NM) band structure is shown in (a) where the energy states between $E\sim-2t$ and $E\sim 2t$ are backfolded into the reduced Brillouin zone (RBZ) due to the unit-cell segmentation. Introducing a non-zero $J_\text{sd}=t$, gaps open at the AFM zone edge (b). If one additionally breaks the equivalence between the two nonmagnetic sites, a spin-split altermagnetic band structure emerges (c).
    By setting $\epsilon_3$ to a large value relative to other magnitudes in the system, $\epsilon_3=100t$, we make this site effectively unavailable. The insets show the Fermi surface at the chemical potential denoted by the dotted line.}
    \label{fig:bandstructure}
\end{figure}

We explicitly set $\boldsymbol{S}_i\propto\hat{\boldsymbol{z}}$. Moreover, we note that the antiferromagnetic state can be modeled with a simpler lattice, namely a two-site unit-cell that contains only magnetic atoms. We confirmed that the essential physical mechanisms underlying the interface-induced magnetization, which will be explained below, are also present in such a model. 

To elucidate the spectral information of this model, the spectral functions (see Appendix \ref{App: A} for definition) are plotted for the bulk model (translational invariant in both \textit{x} and \textit{y}) as the function of the momentum $k_x$ and the energy $E$ in Fig.~\ref{fig:bandstructure}. The latter shows the effects of a non-zero $J_\text{sd}$ and a broken equivalence between $\epsilon_2$ and $\epsilon_3$ on the spectral properties of the model.

Observables such as magnetization and charge density are now readily obtainable after diagonalization of Eq.~(\ref{eqn: total hamiltonian}). Depending on the interplay between spin-dependent ($J_\text{sd}$) and spin-independent ($\epsilon_2$, $\epsilon_3$) on-site couplings within the unit-cell, the effect of the interfaces on the system density can vary significantly. In the following, we begin by discussing how the introduction of nonmagnetic on-site potentials alters the system eigenfunctions and the subsequent consequences on the system density. Then, we discuss how these effects give rise to interface-induced magnetism in antiferromagnets and altermagnets.

\subsection{Wave function shift, magnitudes, and Friedel oscillations: illustration of concepts using a normal metal}

To provide the basis of our discussion of the magnetization and charge oscillations in ribbons of antiferromagnets and altermagnets, let us start by clarifying the nuances of the lattice treatment of a regular metal. The purpose of this section is to identify the key mechanisms at play in the microscopic model concerning spin-independent density modulations. With an understanding of the mechanisms that determine the oscillation pattern of the electron density near an interface, we can proceed to study the consequences of spin-dependent potentials for antiferromagnets and altermagnets.

The characterization of magnetization and edge effects revolves around the study of the spin-resolved electron density and how it varies across the lattice. The eigenfunctions of our lattice, $\psi_{n,k,\sigma}(r_I, \alpha)$, are obtained from the column vectors of the matrix $U$ diagonalizing the Hamiltonian, and are resolved at the unit-cell level. Each eigenfunction is described by three indices $k, n, \sigma$ and is a function of the \textit{x}-coordinate \textit{I} and the site index $\alpha$ within each unit-cell. The spin- and $\alpha$-dependent density at \textit{x}-position $r_I$ in the ribbon is defined as 
\begin{equation}
    n_{I,\alpha, \sigma} = \sum_{n,k}|\psi_{n, k, \sigma}({r}_I, \alpha)|^2n_F(\epsilon_{n, k,\sigma}),
    \label{eqn: density}
\end{equation}
where $\epsilon_{n,k, \sigma}$ is the energy eigenvalue of the eigenfunction $\psi_{n,k,\sigma}$.

\begin{figure}
    \centering
    \includegraphics{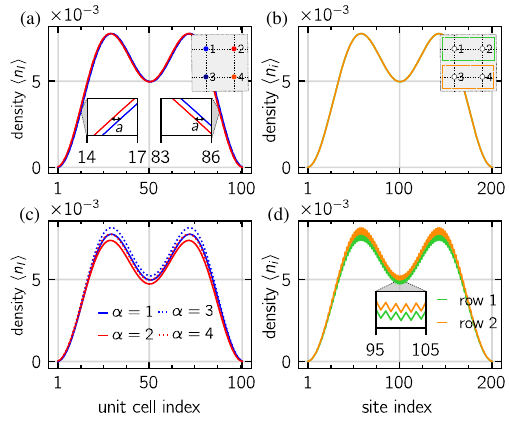}
    \caption{
    The NM densities obtained by summing over the two lowest energy harmonics for $k=0$ shown in the unit-cell-resolved (a) and row-resolved (b) case. Analogous plots in (c) and (d) are shown in the case of a non-zero on-site coupling $\epsilon_2=\delta$, $\epsilon_3=-\delta$ for $\delta \ll t$.  (a) In the NM, the densities are pair-wise equivalent with the ones corresponding to sites in the left column (blue) and right column (red) equal. The left and right column eigenfunctions are shifted by the lattice constant $a$, corresponding to their relative displacement within the unit-cell. (b) The row-resolved density can be illustrated by adding the densities for $\alpha=1$ and $\alpha=2$ for the row 1 and $\alpha=3$ and $\alpha=4$ for the row 2.  (c) A change in the on-site potentials changes the magnitude of the eigenfunction entries corresponding to the particular site, lifting the previous equivalence between the densities at different sites $\alpha$. The probability amplitude at $\alpha=2$ sites is suppressed due to $\epsilon_2=\delta$, while the negative potential $\epsilon_3=-\delta$ enhances the probability amplitude at the $\alpha=3$ sites. The row-resolved densities are given by the densities of unit-cell sites within the same row, shifted relatively by a lattice constant $a$. Due to the induced magnitude difference, the system density becomes a $\pi$-periodic density wave. 
    }
    \label{fig: 5}
\end{figure}

Before moving on, we briefly mention that in the following, we will use the terminology \textit{unit-cell-resolved} and \textit{row-resolved} extensively. \textit{Row-resolved} refers to quantities that are functions of individual lattice-site indices $(i,k)$ across one of the two distinct rows on our ribbon [first and second rows in the unit-cell, see Fig.~\ref{fig:latticemodel}(b)], while \textit{unit-cell-resolved} refers to quantities which are functions of unit-cell indices $(I,\alpha, k)$. 

\begin{figure*}[hbt]
    \centering
\includegraphics{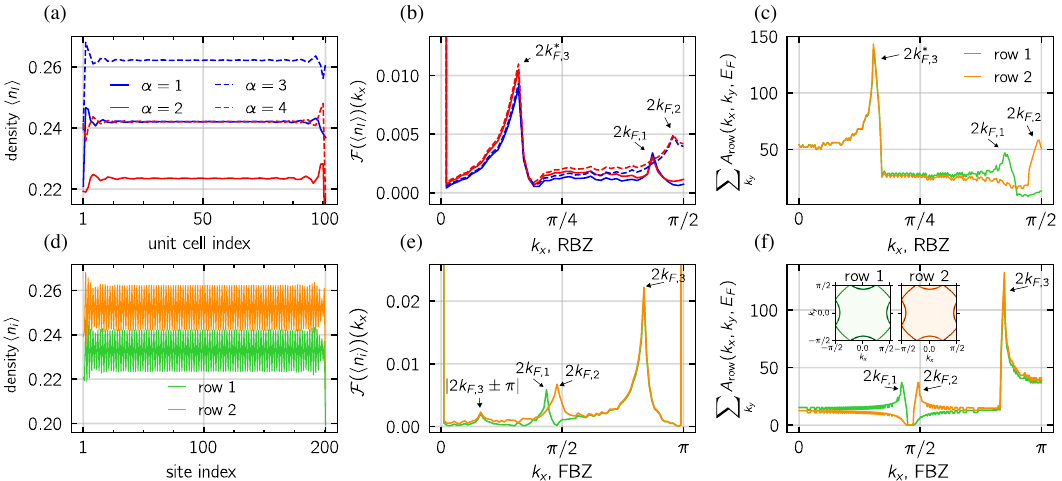}
\caption{
The density profiles for the split system $\epsilon_{2(3)}=\pm\delta$ showing Friedel oscillations at $\mu=-1.5t$. Both the unit-cell-resolved density (a) and row-resolved density (d) show clear Friedel oscillations, but the row-resolved density also shows the $\pi$-density wave due to the presence of non-zero on-site potentials. The Fourier transform of the unit-cell-resolved density is shown in (b) and the row-resolved in (e). Several peaks arise in both corresponding to the dominant frequencies in the density oscillations. {We label the dominant peaks with the notation $2k_{F,i}$ where \textit{i} is a simple index.} Due to the 2D geometry, we do not have one particular, but rather a distribution of $2k_F^x$ from states from different \textit{y}-modes. We quantify this distribution by summing the row-resolved bulk spectral function evaluated at the Fermi level across all \textit{y}-momenta, shown in (f) where $k_x$ runs over the full Brillouin zone (FBZ), obtaining a $2k_F^x$ distribution which matches well with the observed frequency content in the density oscillations. Note that values of $2k_F^x$ beyond $\pi/2$ appear in the reduced Brillouin zone (RBZ) in (c) due to aliasing and are distinguished by an asterisk ($2k_F^{*}$). (e) In the row-resolved density, an additional set of frequencies given by $|2k_F \pm \pi|$ arise due to the fact that the staggered $\pi$ order is modulated in magnitude by the underlying eigenfunctions. 
}
    \label{fig: 6}
\end{figure*}

Consider now a normal metal (NM) lattice with $J_\text{sd}=0$. We examine two distinct configurations of the nonmagnetic on-site potentials in order to elucidate the interplay between the unit-cell potentials and the lattice geometry. In Fig.~\ref{fig: 5}, we plot the unit-cell-resolved and row-resolved densities for a system with no on-site potentials in (a) and (b) and a split potential $\epsilon_{2(3)}=\pm\delta$ in (c) and (d) where $\delta/t\ll 1$ (see the insets for unit-cell layout). The densities are summed over the two first standing-wave harmonics in \textit{x} for the homogeneous \textit{y}-mode ($k = 0$). The first observation is that the densities on sites in the left and right column of the unit-cell are shifted  
by a lattice constant $a$ [see Fig. \ref{fig: 5}(a) and (c)]. In a unit-cell picture where both columns correspond to the same indices $(I,k)$, this shift originating from the relative \textit{x}-displacement between columns, manifests as a magnitude difference between the density profiles. The second observation is that the magnitudes of the $\alpha$-resolved densities are altered individually by the respective on-site potentials. In the case of $\epsilon_2=\epsilon_3=0$, all sites in the unit-cell are equivalent causing the only distinguishing factor to be the one lattice constant $a$ shift between densities in different columns. For the split configuration ($\epsilon_{2(3)}=\pm\delta$), this equivalence is lifted. In Fig.~\ref{fig: 5}(c), the $\alpha=2$ density is suppressed as $\epsilon_2=\delta$ represents an effective lowering of the chemical potential. Along the same lines, an enhancement is observed for $\alpha=3$ due to $\epsilon_3=-\delta$. The magnitudes of densities at sites $\alpha=1,4$ sites also change due to a changed unit-cell environment, but the effect is equal across the two sites and comparatively smaller. To understand the significance of these relative shifts in magnitude, the row-resolved picture is illustrative. The row-resolved densities are obtained by plotting the densities of unit-cell sites in the same row ($\alpha=1,2$ and $\alpha=3,4$) together, shifted relatively by a lattice constant $a$, shown in Figs.~\ref{fig: 5}(b) and \ref{fig: 5}(d). As is evident from Fig.~\ref{fig: 5}(b), the unit-cell description of the NM with $\epsilon_2=\epsilon_3=0$ does not essentially provide any new information not present in the original model. The advantage of the unit-cell picture arises when on-site potentials are introduced. While the unit-cell-resolved densities (Fig.~\ref{fig: 5}(c)) are modulated in magnitude due to the on-site potentials, they retain their smooth standing wave-like profiles as all sites of the same $\alpha$ see the same potential change. The row-resolved density, however, is given by alternating left and right column sites, and the row-resolved densities in Fig.~\ref{fig: 5}(d) are now distinctly different, where a $\pi$-periodic density wave emerges on each row due to the alternating magnitude modulation. When the sum in Eq.~(\ref{eqn: density}) is extended to a large number of eigenfunctions, the row-resolved density profiles can become intricate. As we have seen, however, the observed oscillations can be deconvoluted in the unit-cell picture as the combination of a one lattice constant $a$ shift and magnitude differences between densities in different columns, caused by the chosen on-site potential configuration and unit-cell geometry. 

In a system with broken translation invariance, the oscillations of the charge density known as Friedel oscillations~\cite{Harrison:book, Mahan:book-2013} arise in the vicinity of interfaces and defects. While the oscillations of eigenfunctions interfere and cancel out in the bulk of the ribbon, the open boundaries force all eigenfunctions to be roughly in phase close to the boundaries leading to the Friedel oscillations. The frequency of 1D Friedel oscillations is $2k_F$ where $k_F$ is the Fermi momentum. In 2D, the Fermi surface does in general give rise to a distribution of Fermi wave vectors, and the whole distribution of $k_F^x$, the \textit{x} component of the Fermi wave vectors, can leave a mark on the density oscillations. 

To exemplify how these density oscillations occur, the density for the split potential ($\epsilon_{2(3)}=\pm\delta$) is shown in Fig. \ref{fig: 6} for the unit-cell- \ref{fig: 6}(a) and row-resolved \ref{fig: 6}(d) representations. Both show clear oscillations in the vicinity of the interface while roughly homogeneous in the ribbon bulk. In Figs.~\ref{fig: 6}(b) and \ref{fig: 6}(e), the Fourier transform of the density in the two representations is shown, indicating the dominant frequencies in the density oscillations. These frequencies correspond to roughly twice the \textit{x}-component of the Fermi vector for the homogeneous \textit{y}-mode. Due to the 2D Fermi surface, however, eigenfunctions with a range of $k_F^x$ exist at the Fermi surface, giving rise to additional peaks in the frequency spectrum. These can be shown to correspond closely to the distribution of $k_F^x$ in the partially integrated spectral function evaluated at the Fermi surface, obtained by integrating over all \textit{y}-momenta, shown in Figs.~\ref{fig: 6}(c) and \ref{fig: 6}(f). Note in particular that the presence of the split potentials breaks the symmetry between the rows. As such, the row-resolved Fermi surfaces are identical in shape, but with a slight difference in the distribution of states on the Fermi surface. This is evident in Fig. \ref{fig: 6}(c) and \ref{fig: 6}(f) [see insets in \ref{fig: 6}(f)] where the distributions of \textit{x}-components are different on the two rows.

Finally, we point out that the density oscillations arising from the Fermi surface are largely independent of the specific on-site potentials as long as these are small relative to $t$. As long as the Fermi surface does not change significantly upon changing the potentials, the system eigenfunctions will exhibit similar Friedel oscillations, but with an additional staggered order on top. The non-zero potentials leave a more direct mark on the row-resolved density oscillations [Fig. \ref{fig: 6}(e)] through the emergence of a $\pi$ peak, due to the previously discussed mechanism, and through the appearance of additional peaks at frequencies $|2k_F \pm \pi|$, caused by the $2k_F$ modulation of the $\pi$-density wave magnitude. In effect, the presence of periodic potentials on the lattice gives rise to a magnitude modulation on top of the standing wave eigenfunctions. The magnitude of these modulations follows the magnitude of the eigenfunction as a whole, causing a coupling between the underlying $2k_F$ frequency of the initial eigenfunction, and the $\pi$ frequency arising from the lattice geometry. This gives rise to the additional peaks in the Fourier transform of the row-resolved density in Fig.~\ref{fig: 6}(d) close to $k_x=0$.

\subsection{Non-zero $J_\text{sd}$ coupling and AFM order}
\label{sec:numerics-AFM}

\begin{figure}
    \centering
    \includegraphics{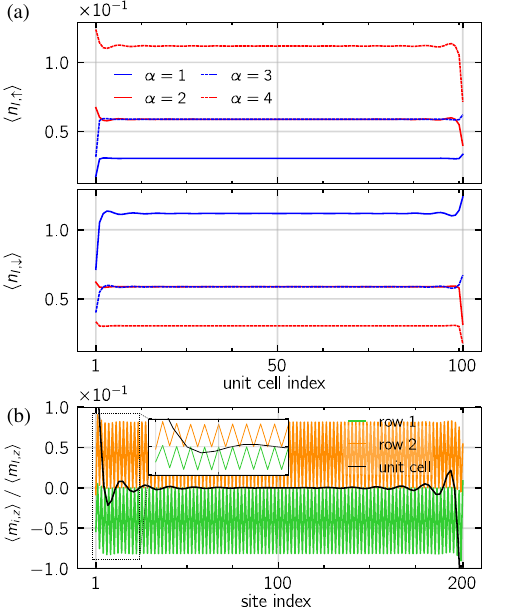}
    \caption{The spin-up and spin-down densities are shown in (a) for an AFM with $J_\text{sd}=t$, $\mu=-2.75t$. The spin-dependent $J_\text{sd}$ gives rise to highly spin-dependent occupations on the magnetic sites $\alpha=1,4$. (b) The magnitude differences in the underlying eigenfunctions give rise to a staggered magnetization (green and orange). The magnetization is out of phase between the two rows causing the total unit-cell magnetization (black) to approach zero in the ribbon center. Despite no spin dependence in the Friedel oscillations, we still observe uncompensated magnetization (black) in the vicinity of the interface. This is caused by an effective coupling between the staggered order and the spin-independent Friedel oscillations, not captured in the continuum model. Note that we here plot the magnitude of the unit-cell magnetization multiplied by a factor of 5 for clarity. Inset shows a zoom-in on the first 20 sites at the left interface.
    }
    \label{fig:magnetization-AFM-real}
\end{figure}

For a given spin projection, a non-zero $J_\text{sd}$ has a very similar effect as the split potentials discussed above and gives rise to a $\pi$-periodic density wave. The \textit{s-d} coupling is, however, spin-dependent and changes the sign between the two spin species. Due to the unit-cell geometry, the spin-up and -down density waves are out of phase, giving rise to a staggered magnetization persisting through the bulk of the ribbon. In our model, this is the mechanism responsible for the AFM staggered order in the itinerant electrons, a microscopic character which is absent in the effective model in Sec.~\ref{sec:analytics}. 

We now turn to the behavior close to the interface. The staggered magnetization arises due to magnitude differences between the spin-up and down densities. As the density waves for spin-up and down electrons are shifted relative to the interface, an effective coupling between Friedel oscillations and the staggered magnetization emerges, causing an interface-induced modulation of the bulk staggered order.

In Fig.~\ref{fig:magnetization-AFM-real}(a), we plot the spin-up and spin-down densities. In Fig.~\ref{fig:magnetization-AFM-real}(b), we plot the staggered (row-resolved) and unit-cell magnetizations, where the latter is averaged over all unit-cell sites. While the unit-cell magnetization practically vanishes in the center of the ribbon, significant deviations from the compensated bulk order arise in the vicinity of the interfaces. As the effective continuum model of AFM considered in Sec.~\ref{sec:analytics-AFM} does not capture the uncompensated nature of the interface nor the staggered order, it omits this type of coupling between Friedel oscillations and magnetization.

\begin{figure}[ht]
    \centering
    \includegraphics[width=1.0\linewidth]{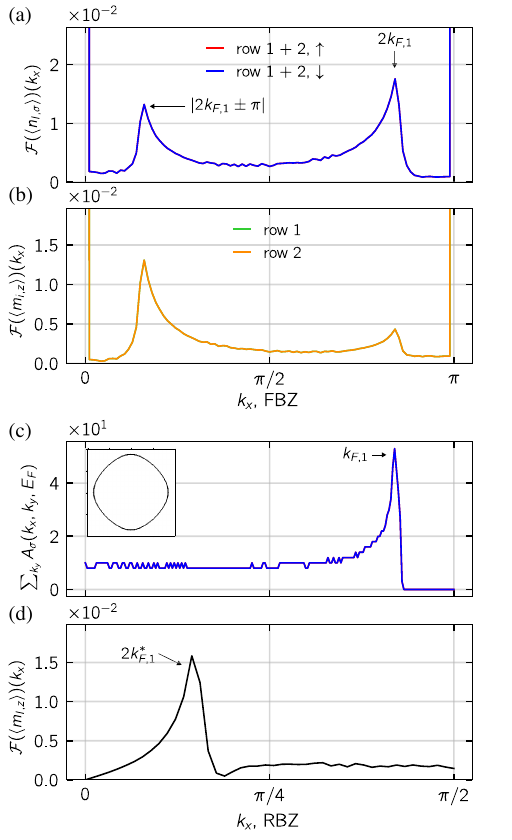}
    \caption{
    (a) The spin-up and -down row-resolved densities for the AFM in Fig. \ref{fig:magnetization-AFM-real} are Fourier transformed to obtain the frequencies of the density oscillations. The dominant frequency is denoted $2k_{F,1}$. By considering the spectral function integrated over all \textit{y} momenta and evaluated at the Fermi level (c), the dominant frequency can be seen to correspond closely to the peak in the distribution of $k_{F}^x$, i.e., \textit{x} components of the Fermi vector of states at the Fermi level. The additional peak in the spectrum in (a) arises because the magnitude of the $\pi$ staggered order itself is modulated by the density oscillations, giving rise to a $|2k_F\pm\pi|$ peak. (b) The row-resolved magnetization contains the same frequencies as those of the underlying spin-densities while the dominant frequency of the unit-cell magnetization (d) is $2k_{F,1}^*$, corresponding to the reduced Brillouin zone representation of $2k_{F,1}$, caused by aliasing.}
    \label{fig:magnetization-AFM-freq}
\end{figure}

We point out that the Fermi momentum itself is spin-independent in the AFM and one thus does not expect any magnetization contribution from Fermi surface mismatch, which the continuum model confirms. The frequency of the density and magnetization oscillations are shown in detail in Fig.~\ref{fig:magnetization-AFM-freq}. Notice in particular that we observe the unit-cell magnetization to oscillate at $2k_F$ [\ref{fig:magnetization-AFM-freq}(d)] exactly due to the effective coupling between the $\pi$-order and Friedel oscillations. Therefore, one of the ways to achieve surface-induced magnetization in continuum models would be to introduce spin-dependent boundary conditions, which would correspond to a different environment of spin-up and spin-down sites in the lattice model.

Following this reasoning, the emergence of uncompensated edge magnetization for itinerant electrons in an AFM depends critically on the unit-cell geometry and the nature of the interface. The interface-induced magnetization arises only if the up and down spins interact differently with the interface, creating a spin-dependent environment which does not vanish upon averaging over the interface. To achieve this, an uncompensated interface is typically needed. For a compensated interface, there will still be relative shifts between the spin-up and -down density waves, but the shifts will vanish when summing over a unit-cell as both spins interact with the interface symmetrically. Along these lines, we also expect this effect to be vulnerable to surface structure and roughness. As the magnetization is induced solely through the one lattice constant shift, a roughening of the surface could cancel out the spin-dependent interface environment. This is discussed in detail in Sec.~\ref{SEC:roughness}.

\subsection{Phenomenological model for magnetization oscillations in antiferromagnets}
\label{sec:numerics-AFM-phem}

Motivated by the results in the lattice model, where the spin-up and -down sites experience different environments and are shifted with respect to each other by a lattice constant, we propose a phenomenological model. We use the continuum model and the results of Sec.~\ref{sec:analytics-AFM}, however, we shift the coordinates for the spin-down electrons by a lattice constant $a$. Then the magnetization is determined by
\begin{equation}
\label{numerics-AFM-phem-mz-def}
m_z(x) = -\frac{g\mu_B}{e} \left[ n_{\uparrow}(x) -n_{\downarrow}(x+a)\right].
\end{equation}
While not being rigorous, the model captures the shift between the spin-up and -down populations crucial for the lattice model. Furthermore, one can view the shift by a lattice constant as modified boundary conditions where the wave functions of spin-up and -down electrons acquire different values at the interfaces.

For wide ribbons, $\tilde{L}\gg1$, the magnetization away from the edges $\tilde{x}\gg1$ reads as
\begin{eqnarray}
\label{numerics-AFM-phem-n-L-inf-diff-x-inf}
m_z(x) &=&
\frac{m_0 }{\sqrt{\pi}}\Bigg[\frac{\cos{\left(2\tilde{x} +\frac{\pi}{4}\right)}}{\tilde{x}^{3/2}} 
-\frac{\cos{\left(2\tilde{x} +2\tilde{a} +\frac{\pi}{4}\right)}}{(\tilde{x}+\tilde{a})^{3/2}} \Bigg]\! +\mathcal{O}{\left(\frac{1}{\tilde{x}^2}\right)} \nonumber\\
&=&
\frac{2m_0 \tilde{a}}{\sqrt{\pi}} \frac{1}{\tilde{x}^{3/2}} \sin{\left(2\tilde{x} +\frac{\pi}{4}\right)} +\mathcal{O}{\left(\tilde{a}^2,\frac{1}{\tilde{x}^2}\right)},
\end{eqnarray}
where we expanded in $\tilde{x}\gg1$ in the first line and in $\tilde{a}\ll1$ in the second line.
The oscillations of the magnetization in the phenomenological model of antiferromagnets are (i) suppressed as $\tilde{a} = ak_F \ll1$, (ii) shifted by $\pi/2$ in phase with respect to the particle density oscillations, and (iii) decay as $1/\tilde{x}^{3/2}$ at $\tilde{x}\to \infty$. In {addition,  
the} phenomenological model predicts vanishing net magnetization in antiferromagnets.

\subsection{Non-zero $J_\text{sd}$ coupling and AM order}
\label{sec:numerics-AM}

The advantage of the microscopic model is that it allows us to introduce the altermagnetic (AM) order on top of the underlying antiferromagnetic (AFM) lattice. By keeping the \textit{s-d} coupling and setting the on-site potentials $\epsilon_2 = 0$, $\epsilon_{3}=100t$, thus effectively making the $\alpha=3$ site unavailable,  $\mathcal{PT}$ symmetry is broken and the AFM becomes an AM with spin-split bands (see the bandstructure in Fig.~\ref{fig:bandstructure}).

\begin{figure}[ht]
    \centering    \includegraphics[width=1.0\linewidth]{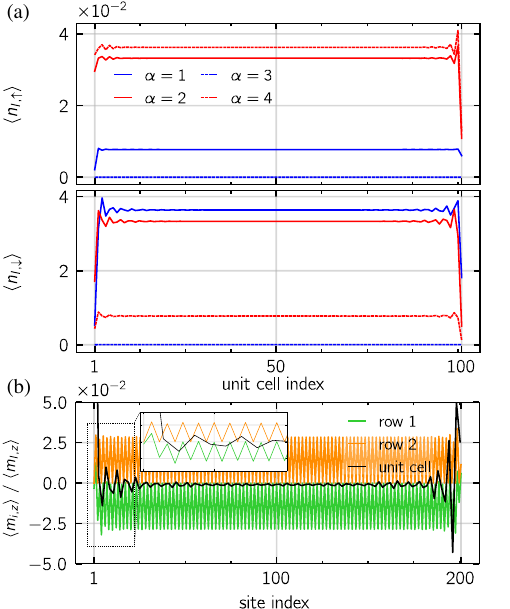}
    \caption{The spin-up and -down densities (a) for an AM with $J_\text{sd}=t$, $\epsilon_{2}=0$, $\epsilon_3 = 100t$ and $\mu=-2.75t$. Due to the split nonmagnetic potentials, the ribbon is no longer symmetric with respect to the combined action of time inversion and spatial inversion. The interface-induced magnetization shown in (b) is therefore not symmetric around the ribbon center, but depicts unique characteristics on the two interfaces. Note that we have plotted the unit-cell magnetization magnitude multiplied by a factor of 5 for clarity. Inset shows a zoom-in on the first 20 sites at the left interface. }
    \label{fig:magnetization-AM-real}
\end{figure}
\begin{figure}[ht!]
    \centering    
    \includegraphics[width=1.0\linewidth]{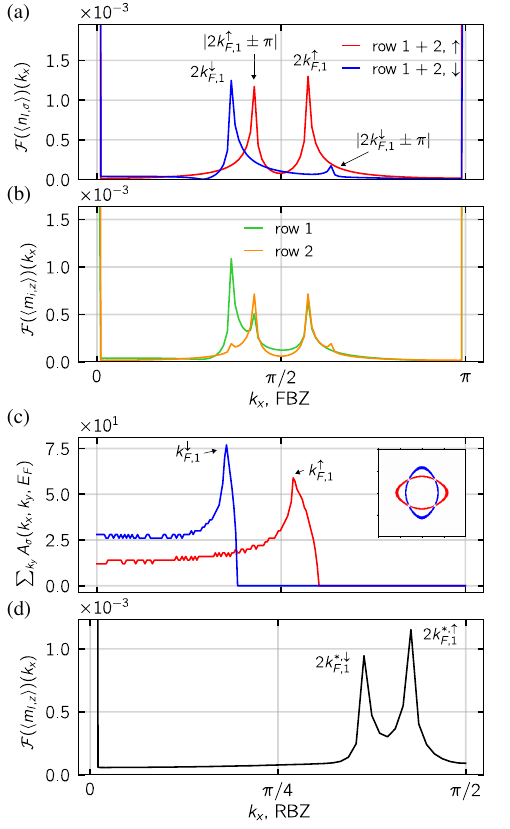}
    \caption{(a) The spin-up and -down row-resolved densities for the AM in Fig. \ref{fig:magnetization-AM-real} are Fourier transformed to obtain the frequencies of the density oscillations. In contrast with the AFM (Fig. \ref{fig:magnetization-AFM-freq}), the spin-up and -down densities depict different dominant frequencies. Note the presence of the additional peaks due to the magnitude modulation of the staggered order. The spin-asymmetry in the density frequencies can be attributed to the distribution of the spin-dependent $k_{F,\sigma}^x$ in the eigenfunctions at the Fermi level, obtained by a partial integration of the bulk spectral function (c). Notice also the inset in (c) showing the difference between the bulk spin-up and -down spectral function at the Fermi level. The frequencies present in the row-resolved magnetization are shown in (b) and a general observation is that the spin-dependence of the density oscillations gives rise to a magnetization that oscillates with all underlying frequencies. Finally, we plot the Fourier transform of the unit-cell magnetization in (d) and we observe two distinct peaks, ${k}_{F}^{*\uparrow}$, $k_{F}^{*\downarrow}$, corresponding to the dominant $k_{F}^{\uparrow}$, $k_{F}^{\downarrow}$ from (c) in the reduced Brillouin zone representation due to aliasing.}
    \label{fig:magnetization-AM-freq}
\end{figure}

The nature of the magnetization in the altermagnet is quite similar to the antiferromagnet discussed in Sec.~\ref{sec:numerics-AFM}. The \textit{s-d} coupling gives rise to $\pi$-periodic density waves, the ones for spin-up and -down species being out of phase [see Fig.~\ref{fig:magnetization-AM-real}(b)]. The presence of the $\epsilon_{3}$ potential only complicates the picture somewhat by lifting the equivalence between density oscillations on the two rows. As such, the interface-induced magnetization discussed in Sec.~\ref{sec:numerics-AFM} carries over to the AM as well. The uncompensated interface introduces an effective coupling between the staggered magnetization and Friedel oscillations, causing uncompensated unit-cell magnetization to emerge in the vicinity of the interface. 

However, a key distinguishing factor between the interface-induced magnetization in AFMs and AMs is the spin-dependence of the Fermi surface in the latter. The Friedel oscillations in AMs are themselves spin-dependent as the distribution of $k_F^x$ momenta on the Fermi surface is different for the two spin-species [see Figs.~\ref{fig:magnetization-AM-freq}(a) and \ref{fig:magnetization-AM-freq}(c)]. This gives rise to an additional source of magnetization as the spin-up and -down densities oscillate with different frequencies which also leaves an impact on the unit-cell magnetization in Fig.~\ref{fig:magnetization-AM-freq}(d). This is the effect found within the effective continuum model in Sec.~\ref{sec:analytics}. As these oscillations originate with the Fermi surface structure, we expect it to be more robust towards surface roughness and structure. Any source of Friedel oscillations in a system, be it an interface or an impurity,~\footnote{Both spinless and spinful impurities are expected to lead to the magnetization oscillations in altermagnets.}, 
is thus expected to be sources of uncompensated magnetism in altermagnets. The spin-dependent Friedel oscillations have also recently been shown to influence the interaction between impurities that are magnetic~\cite{lee_arxiv_23, amundsen_prb_24}. 

We point out that unit-cell geometry and the AM Fermi surface are intrinsically linked. In order to observe spin-dependent Friedel oscillations, the AM Fermi surface should be oriented in a way such that the distribution of Fermi vector components normal to the interface differs between the two spin species. Due to the link with lattice geometry, however, such an orientation would typically correspond to a unit-cell geometry which has an inherent asymmetry between spin-up and -down species leading to an uncompensated interface. As such, it is difficult to deconvolute the effects of the particular interface configuration which is highly important for the AFM order, from that of the Fermi-surface asymmetry expected to be important in AM. To disentangle magnetism arising from the interface from that arising from bulk properties, we turn to the role of the interface polarization and surface defects.

\subsection{The role of interface roughness} 
\label{SEC:roughness}

Our analysis so far has assumed atomically flat interfaces, but in real samples, the interfaces are commonly rough. The effect of such roughness can be investigated in our lattice model by making the interface jagged, which is achieved by adding a strong onsite potential on random lattice sites neighboring the vacuum interface. 
The question of the robustness of the induced magnetization with respect to such disorder is particularly important to make a connection to experiments.

\begin{figure}[t]
    \centering
    \includegraphics[width=0.9\linewidth]{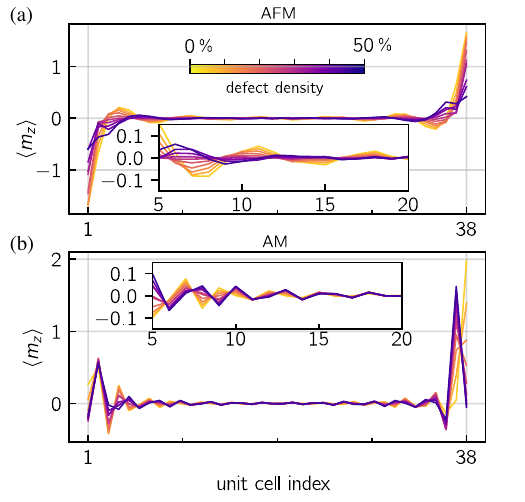}
    \caption{The unit-cell magnetization is shown for an AFM (a) and an AM (b) as a function of defect density in the first and last unit-cell on the ribbon. The system parameters are $N_x=76$, $N_y=100$, $J_{\text{sd}}=t$, $\epsilon_2=0$, and $\epsilon_3=100t$ (the two latter only relevant for the AM). The filling level is $\mu=-2.75t$. For each defect concentration, the specific percentage of sites in the first and last unit-cell is randomly made inaccessible by adding an on-site potential of $100t$. For each particular defect density, the unit-cell magnetization is averaged over 5 randomized defect configurations. The ribbon magnetization is observed to be affected by the density of defects and for the AFM, the oscillations of magnetization almost vanish when $30\%$ of interface cell sites are a vacancy. While defects have an effect on the AM as well, the suppression of the magnitude of oscillations is not as drastic as in the AFM.
    }
    \label{fig: AFM_vs_AM}
\end{figure}

In Fig. \ref{fig: AFM_vs_AM}, the unit-cell magnetization is shown for AFM (a) and AM (b) when interface roughness is included. The presence of defects breaks translational invariance in the \textit{y}-direction. We account for this by considering a ribbon of both finite width and length, with periodic boundary conditions between the first and last lattice site along the interface.
The rough interface is modeled by inserting vacancies, described by the $100t$ on-site potential, at random locations in the first and last layers of unit-cells on the ribbon. The resulting magnetization is averaged over several configurations of vacancies (in our case, we used 5 configurations).
We plot the magnetization averaged over all rows in the periodic direction as a function of the density of defects and observe that the magnetization is indeed affected by the roughness, both for the AM and AFM. For example, the magnetization oscillations in the AFM are almost completely suppressed for a defect density of approximately 30 \%. 
This is understood as destructive interference caused by the oscillating magnetization acquiring a row-dependent phase shift due to the local defect configuration. In the AM, the magnetization shown in the lower panel of Fig. \ref{fig: AFM_vs_AM} is not suppressed to the same extent. Based on our previous analysis, we would indeed expect the magnetization in the AFM to be more sensitive towards surface disorder, but it is difficult to draw a firm conclusion based on the magnetization profiles in the AFM and AM by themselves. 

\begin{figure}[h]
    \centering
    \includegraphics{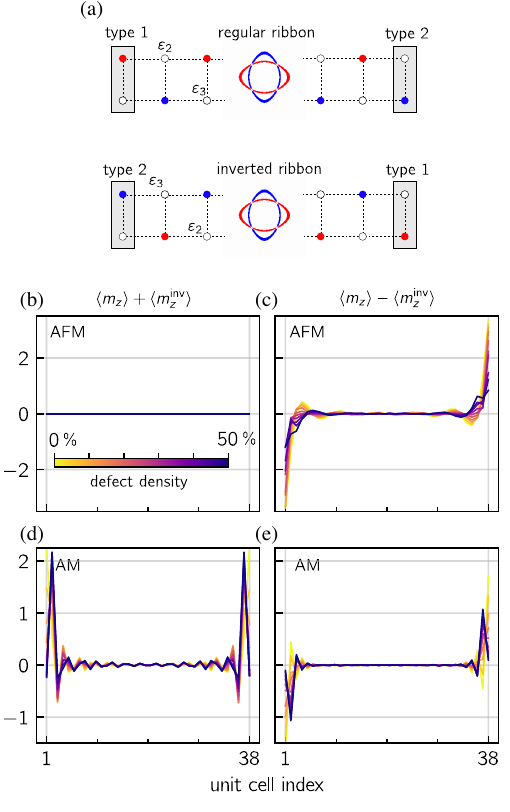}
    \caption{
    The relation between the regular and inverted ribbons is shown in (a). We plot the sum $\langle m_z \rangle + \langle m_z^\text{inv}\rangle$ (b) and (d), and difference $\langle m_z \rangle - \langle m_z^\text{inv}\rangle$ (c) and (e) for the same AFM (b) and (c) and AM (d) and (e) as in Fig. \ref{fig: AFM_vs_AM}. For the AFM, the sum $\langle m_z \rangle + \langle m_z^\text{inv}\rangle$ in (b) is zero, independent of defect density, indicating the surface origin of the magnetization.
    In the AM, the sum $\langle m_z \rangle + \langle m_z^\text{inv}\rangle$in (d) is non-zero and quite robust against surface defects, indicating that the magnetization is not purely initiated by the particular configuration of localized moments at the interface.
    }
    \label{fig: SUM_and_DIFFERENCE}
\end{figure}

Therefore, in order to demonstrate the difference between the AFM and AM responses to surface disorder, we introduce the concept of an inverted ribbon which has the same Fermi surface but a different lattice structure [see Fig. \ref{fig: SUM_and_DIFFERENCE}(a)]. Consider the ribbon as it is presented in Fig.~\ref{fig:latticemodel}(b). If one reverses the direction of the localized moments as well as switches $\epsilon_2$ and $\epsilon_3$ (the latter is only relevant for AM), a related ribbon can be obtained with the same ``bulk" environment (the same Fermi surface), but with different boundary conditions. In particular, the polarization of the boundaries has been reversed. In the following, we denote the magnetization of this inverted ribbon as $\langle m_z^\text{inv}\rangle$. The purpose of introducing the inverted ribbon is to separate the magnetization contribution arising from the interface and that arising due to intrinsic spin asymmetries in the Fermi surface. The former should be highly surface dependent, while the latter is expected to be more robust toward interface roughness. 

In Fig. \ref{fig: SUM_and_DIFFERENCE}, we plot the sum [(b) and (d)] and difference [(c) and (e)] between the magnetization of the original ribbon, $\langle m_z \rangle$, and the inverted ribbon $\langle m_z^\text{inv}\rangle$. By taking the sum, one should be left with the magnetization contribution which does not show a perfect inversion under a reversal of the interface polarization. Along the same lines, the difference should give exactly the surface-dependent contribution. In the case of the AFM, the sum is zero [see Fig.~\ref{fig: SUM_and_DIFFERENCE}(b)], independent of surface defects. Based on our previous analysis, the origin of the interface-induced magnetization in the AFM is solely due to the spin-asymmetric environment close to the interface. Upon a reversal of the interface polarization, it is reasonable that the ribbon magnetization inverts as well, giving rise to a perfect cancellation. This is exactly what we observe in Fig. \ref{fig: SUM_and_DIFFERENCE}(b). For the AFM, the difference $\langle m_z \rangle -\langle m_z^\text{inv}\rangle$ is highly sensitive to the surface disorder [see Fig.~\ref{fig: SUM_and_DIFFERENCE}(c)], supporting our statement that the magnetization in AFM is less robust toward surface roughness. 

For the AM, however, the magnetizations of the ribbon and its inverted counterpart do not cancel out, but leave a non zero remainder in the sum shown in Fig. \ref{fig: SUM_and_DIFFERENCE}(d). As has been discussed previously, altermagnets should have an additional contribution to the interface magnetization arising from the spin dependence of the Friedel oscillations. This asymmetry originates from the Fermi surface and should not be dependent on the exact realization of boundary conditions or surface defects. Indeed, as one can see in Figs.~\ref{fig: SUM_and_DIFFERENCE}(d) and \ref{fig: SUM_and_DIFFERENCE}(e), the sum $\langle m_z \rangle +\langle m_z^\text{inv}\rangle$ is less affected by the surface disorder than the difference $\langle m_z \rangle -\langle m_z^\text{inv}\rangle$. We argue that this analysis indicates that the interface-induced magnetization in altermagnets is more robust towards surface defects and the particular configuration of localized moments at the interface.

We note that the role of defects in an altermagnetic candidate MnTe was recently highlighted, albeit in a different context, in Ref.~\cite{Chilcote-Brahlek-StoichiometryinducedFerromagnetismAltermagnetic-2024}.

\section{Summary}
\label{sec:Summary}

In this work, we have characterized the nature of interface-induced magnetism carried by itinerant electrons in antiferromagnets and altermagnets, and quantified the decay of this magnetism into the bulk of the material. We find that altermagnets possess a unique contribution to the edge magnetization that is not present in conventional antiferromagnets. 

In a system with open boundaries, altermagnets quite generally acquire an uncompensated magnetization due to their spin-split Fermi surfaces [see Eqs.~(\ref{ribbon-DOS-mz-2}) and (\ref{ribbon-Mz-def}) as well as Fig.~\ref{fig:ribbon-DOS-mz} for the magnetization in a ribbon of an altermagnet]. In the effective continuum model, this magnetization originates from the nature of the spin-dependent Friedel oscillations whose frequencies are determined by different Fermi wave vectors, even for spin-independent boundary conditions. These oscillations are exemplified by the spin-resolved LDOS and the charge densities in Eqs.~(\ref{ribbon-DOS-1}), (\ref{ribbon-DOS-DOS-1}), and (\ref{ribbon-DOS-DOS-2}) and Eqs.~(\ref{ribbon-n-def}), (\ref{ribbon-DOS-n-1}), and (\ref{ribbon-DOS-n-2}), respectively; see also Fig.~\ref{fig:ribbon-DOS-eps}. Therefore, the combination of the momentum-dependent spin splitting and interfaces allows for oscillating magnetization in altermagnets. Since the Fermi surface in the continuum model of antiferromagnets used in Sec.~\ref{sec:analytics-AFM} is spin degenerate and the boundary conditions are spin insensitive, the model predicts no surface-induced magnetism.

In the lattice model, considered in Sec.~\ref{sec:numerics}, this picture is different. Being able to describe microscopic magnetic moments, the lattice model provides access to more details. We found that there is another contribution to interface-induced magnetization originating from an effective coupling between Friedel oscillations and staggered magnetic order; this contribution is qualitatively similar for antiferromagnets (see Figs.~\ref{fig:magnetization-AFM-real} and \ref{fig:magnetization-AFM-freq}), and altermagnets (see Figs.~\ref{fig:magnetization-AM-real} and \ref{fig:magnetization-AM-freq}). This coupling is the only source of edge magnetization for the antiferromagnet within our lattice model, and we argue that it corresponds to an effective spin-dependent boundary condition in the analytical continuum model. The corresponding phenomenological model is discussed in Sec.~\ref{sec:numerics-AFM-phem}.

The coupling between Friedel oscillations and the staggered order depends critically on the lattice geometry and yields a zero net interface magnetization for compensated AFM and AM interfaces. As we show in Sec.~\ref{SEC:roughness}, this contribution is expected to be sensitive to surface defects and roughness. The Fermi-surface contribution inherent to altermagnets requires only the spin-split structure of the Fermi surface and does not rely on the microscopic structure of the interface. Any nonmagnetic sources of density oscillations, caused by, e.g., interfaces or impurities, are therefore expected to give rise to local uncompensated magnetization, which is an important aspect to be aware of for the implementation altermagnet-based spintronics devices. The observation of robust magnetization induced by nonmagnetic defects could also provide an alternative way to probe altermagnets. Our order-of-magnitude estimate suggests that the edge-induced magnetization should be readily observable via, e.g., SQUID techniques~\cite{Vasyukov-Zeldov-ScanningSuperconductingQuantum-2013, Granata-Vettoliere-NanoSuperconductingQuantum-2016, Buchner-Ney-TutorialBasicPrinciples-2018, Persky-Kalisky-StudyingQuantumMaterials-2022, Paulsen-Kiefer-UltralowFieldSQUID-2022}.

In passing, we note that while we focused on 2D films of antiferromagnets and altermagnets, our study can be straightforwardly generalized to 3D materials and altermagnets with a more involved dispersion relation, such as $g$- or $i$-wave magnets.

\begin{acknowledgments}
We acknowledge useful communications with M.~Amundsen and B.~Brekke.
This work was supported by the Research Council of Norway through Grant No. 323766 and its Centres of Excellence funding scheme Grant No. 262633 “QuSpin.” Support from Sigma2 - the National Infrastructure for High-Performance Computing and Data Storage in Norway, project NN9577K, is acknowledged.
\end{acknowledgments}

\bibliography{library-short}

\begin{thebibliography}{64}%
\makeatletter
\providecommand \@ifxundefined [1]{%
 \@ifx{#1\undefined}
}%
\providecommand \@ifnum [1]{%
 \ifnum #1\expandafter \@firstoftwo
 \else \expandafter \@secondoftwo
 \fi
}%
\providecommand \@ifx [1]{%
 \ifx #1\expandafter \@firstoftwo
 \else \expandafter \@secondoftwo
 \fi
}%
\providecommand \natexlab [1]{#1}%
\providecommand \enquote  [1]{``#1''}%
\providecommand \bibnamefont  [1]{#1}%
\providecommand \bibfnamefont [1]{#1}%
\providecommand \citenamefont [1]{#1}%
\providecommand \href@noop [0]{\@secondoftwo}%
\providecommand \href [0]{\begingroup \@sanitize@url \@href}%
\providecommand \@href[1]{\@@startlink{#1}\@@href}%
\providecommand \@@href[1]{\endgroup#1\@@endlink}%
\providecommand \@sanitize@url [0]{\catcode `\\12\catcode `\$12\catcode `\&12\catcode `\#12\catcode `\^12\catcode `\_12\catcode `\%12\relax}%
\providecommand \@@startlink[1]{}%
\providecommand \@@endlink[0]{}%
\providecommand \url  [0]{\begingroup\@sanitize@url \@url }%
\providecommand \@url [1]{\endgroup\@href {#1}{\urlprefix }}%
\providecommand \urlprefix  [0]{URL }%
\providecommand \Eprint [0]{\href }%
\providecommand \doibase [0]{https://doi.org/}%
\providecommand \selectlanguage [0]{\@gobble}%
\providecommand \bibinfo  [0]{\@secondoftwo}%
\providecommand \bibfield  [0]{\@secondoftwo}%
\providecommand \translation [1]{[#1]}%
\providecommand \BibitemOpen [0]{}%
\providecommand \bibitemStop [0]{}%
\providecommand \bibitemNoStop [0]{.\EOS\space}%
\providecommand \EOS [0]{\spacefactor3000\relax}%
\providecommand \BibitemShut  [1]{\csname bibitem#1\endcsname}%
\let\auto@bib@innerbib\@empty
\bibitem [{\citenamefont {Pekar}\ and\ \citenamefont {Rashba}(1965)}]{Pekar-Rashba-CombinedResonanceCrystals-1965}%
  \BibitemOpen
  \bibfield  {author} {\bibinfo {author} {\bibfnamefont {S.}~\bibnamefont {Pekar}}\ and\ \bibinfo {author} {\bibfnamefont {G.}~\bibnamefont {Rashba}},\ }\bibfield  {title} {\bibinfo {title} {Combined {{Resonance}} in {{Crystals}} in {{Inhomogeneous Magnetic Fields}}},\ }\href {http://www.jetp.ras.ru/cgi-bin/e/index/e/20/5/p1295?a=list} {\bibfield  {journal} {\bibinfo  {journal} {JETP}\ }\textbf {\bibinfo {volume} {20}},\ \bibinfo {pages} {1295} (\bibinfo {year} {1965})}\BibitemShut {NoStop}%
\bibitem [{\citenamefont {Noda}\ \emph {et~al.}(2016)\citenamefont {Noda}, \citenamefont {Ohno},\ and\ \citenamefont {Nakamura}}]{Noda-Nakamura-MomentumdependentBandSpin-2016}%
  \BibitemOpen
  \bibfield  {author} {\bibinfo {author} {\bibfnamefont {Y.}~\bibnamefont {Noda}}, \bibinfo {author} {\bibfnamefont {K.}~\bibnamefont {Ohno}},\ and\ \bibinfo {author} {\bibfnamefont {S.}~\bibnamefont {Nakamura}},\ }\bibfield  {title} {\bibinfo {title} {Momentum-dependent band spin splitting in semiconducting {{MnO}}{$_2$} : A density functional calculation},\ }\href {https://doi.org/10.1039/C5CP07806G} {\bibfield  {journal} {\bibinfo  {journal} {Phys. Chem. Chem. Phys.}\ }\textbf {\bibinfo {volume} {18}},\ \bibinfo {pages} {13294} (\bibinfo {year} {2016})}\BibitemShut {NoStop}%
\bibitem [{\citenamefont {{\v{S}}mejkal}\ \emph {et~al.}(2020)\citenamefont {{\v{S}}mejkal}, \citenamefont {Gonz{\'{a}}lez-Hern{\'{a}}ndez}, \citenamefont {Jungwirth},\ and\ \citenamefont {Sinova}}]{Smejkal-Sinova:2020}%
  \BibitemOpen
  \bibfield  {author} {\bibinfo {author} {\bibfnamefont {L.}~\bibnamefont {{\v{S}}mejkal}}, \bibinfo {author} {\bibfnamefont {R.}~\bibnamefont {Gonz{\'{a}}lez-Hern{\'{a}}ndez}}, \bibinfo {author} {\bibfnamefont {T.}~\bibnamefont {Jungwirth}},\ and\ \bibinfo {author} {\bibfnamefont {J.}~\bibnamefont {Sinova}},\ }\bibfield  {title} {\bibinfo {title} {{Crystal time-reversal symmetry breaking and spontaneous Hall effect in collinear antiferromagnets}},\ }\href {https://doi.org/10.1126/sciadv.aaz8809} {\bibfield  {journal} {\bibinfo  {journal} {Sci. Adv.}\ }\textbf {\bibinfo {volume} {6}},\ \bibinfo {pages} {aaz8809} (\bibinfo {year} {2020})},\ \Eprint {https://arxiv.org/abs/1901.00445} {arXiv:1901.00445} \BibitemShut {NoStop}%
\bibitem [{\citenamefont {Yuan}\ \emph {et~al.}(2020)\citenamefont {Yuan}, \citenamefont {Wang}, \citenamefont {Luo}, \citenamefont {Rashba},\ and\ \citenamefont {Zunger}}]{Yuan-Zunger:2020}%
  \BibitemOpen
  \bibfield  {author} {\bibinfo {author} {\bibfnamefont {L.-D.}\ \bibnamefont {Yuan}}, \bibinfo {author} {\bibfnamefont {Z.}~\bibnamefont {Wang}}, \bibinfo {author} {\bibfnamefont {J.-W.}\ \bibnamefont {Luo}}, \bibinfo {author} {\bibfnamefont {E.~I.}\ \bibnamefont {Rashba}},\ and\ \bibinfo {author} {\bibfnamefont {A.}~\bibnamefont {Zunger}},\ }\bibfield  {title} {\bibinfo {title} {{Giant momentum-dependent spin splitting in centrosymmetric low- $\mathds{Z}$ antiferromagnets}},\ }\href {https://doi.org/10.1103/PhysRevB.102.014422} {\bibfield  {journal} {\bibinfo  {journal} {Phys. Rev. B}\ }\textbf {\bibinfo {volume} {102}},\ \bibinfo {pages} {014422} (\bibinfo {year} {2020})},\ \Eprint {https://arxiv.org/abs/1912.12689} {arXiv:1912.12689} \BibitemShut {NoStop}%
\bibitem [{\citenamefont {Hayami}\ \emph {et~al.}(2020)\citenamefont {Hayami}, \citenamefont {Yanagi},\ and\ \citenamefont {Kusunose}}]{Hayami-Kusunose-BottomupDesignSpinsplit-2020}%
  \BibitemOpen
  \bibfield  {author} {\bibinfo {author} {\bibfnamefont {S.}~\bibnamefont {Hayami}}, \bibinfo {author} {\bibfnamefont {Y.}~\bibnamefont {Yanagi}},\ and\ \bibinfo {author} {\bibfnamefont {H.}~\bibnamefont {Kusunose}},\ }\bibfield  {title} {\bibinfo {title} {Bottom-up design of spin-split and reshaped electronic band structures in spin-orbit-coupling free antiferromagnets: {{Procedure}} on the basis of augmented multipoles},\ }\href {https://doi.org/10.1103/PhysRevB.102.144441} {\bibfield  {journal} {\bibinfo  {journal} {Phys. Rev. B}\ }\textbf {\bibinfo {volume} {102}},\ \bibinfo {pages} {144441} (\bibinfo {year} {2020})},\ \Eprint {https://arxiv.org/abs/2008.10815} {arxiv:2008.10815 [cond-mat]} \BibitemShut {NoStop}%
\bibitem [{\citenamefont {Hayami}\ \emph {et~al.}(2019)\citenamefont {Hayami}, \citenamefont {Yanagi},\ and\ \citenamefont {Kusunose}}]{hayami2019momentum}%
  \BibitemOpen
  \bibfield  {author} {\bibinfo {author} {\bibfnamefont {S.}~\bibnamefont {Hayami}}, \bibinfo {author} {\bibfnamefont {Y.}~\bibnamefont {Yanagi}},\ and\ \bibinfo {author} {\bibfnamefont {H.}~\bibnamefont {Kusunose}},\ }\bibfield  {title} {\bibinfo {title} {Momentum-{{Dependent Spin Splitting}} by {{Collinear Antiferromagnetic Ordering}}},\ }\href {https://doi.org/10.7566/JPSJ.88.123702} {\bibfield  {journal} {\bibinfo  {journal} {J. Phys. Soc. Jpn.}\ }\textbf {\bibinfo {volume} {88}},\ \bibinfo {pages} {123702} (\bibinfo {year} {2019})}\BibitemShut {NoStop}%
\bibitem [{\citenamefont {Ahn}\ \emph {et~al.}(2019)\citenamefont {Ahn}, \citenamefont {Hariki}, \citenamefont {Lee},\ and\ \citenamefont {Kune{\v{s}}}}]{Ahn-Kunes:2019}%
  \BibitemOpen
  \bibfield  {author} {\bibinfo {author} {\bibfnamefont {K.-H.}\ \bibnamefont {Ahn}}, \bibinfo {author} {\bibfnamefont {A.}~\bibnamefont {Hariki}}, \bibinfo {author} {\bibfnamefont {K.-W.}\ \bibnamefont {Lee}},\ and\ \bibinfo {author} {\bibfnamefont {J.}~\bibnamefont {Kune{\v{s}}}},\ }\bibfield  {title} {\bibinfo {title} {{Antiferromagnetism in RuO$_2$ as d-wave Pomeranchuk instability}},\ }\href {https://doi.org/10.1103/PhysRevB.99.184432} {\bibfield  {journal} {\bibinfo  {journal} {Phys. Rev. B}\ }\textbf {\bibinfo {volume} {99}},\ \bibinfo {pages} {184432} (\bibinfo {year} {2019})},\ \Eprint {https://arxiv.org/abs/1902.04436} {arXiv:1902.04436} \BibitemShut {NoStop}%
\bibitem [{\citenamefont {Cheong}\ and\ \citenamefont {Huang}(2024)}]{Cheong-Huang:2024}%
  \BibitemOpen
  \bibfield  {author} {\bibinfo {author} {\bibfnamefont {S.-W.}\ \bibnamefont {Cheong}}\ and\ \bibinfo {author} {\bibfnamefont {F.-T.}\ \bibnamefont {Huang}},\ }\bibfield  {title} {\bibinfo {title} {{Altermagnetism with non-collinear spins}},\ }\href {https://doi.org/10.1038/s41535-024-00626-6} {\bibfield  {journal} {\bibinfo  {journal} {npj Quantum Mater.}\ }\textbf {\bibinfo {volume} {9}},\ \bibinfo {pages} {13} (\bibinfo {year} {2024})},\ \Eprint {https://arxiv.org/abs/2401.13069} {arXiv:2401.13069} \BibitemShut {NoStop}%
\bibitem [{\citenamefont {\v{S}mejkal}\ \emph {et~al.}(2022)\citenamefont {\v{S}mejkal}, \citenamefont {Sinova},\ and\ \citenamefont {Jungwirth}}]{Smejkal-Jungwirth:2022b}%
  \BibitemOpen
  \bibfield  {author} {\bibinfo {author} {\bibfnamefont {L.}~\bibnamefont {\v{S}mejkal}}, \bibinfo {author} {\bibfnamefont {J.}~\bibnamefont {Sinova}},\ and\ \bibinfo {author} {\bibfnamefont {T.}~\bibnamefont {Jungwirth}},\ }\bibfield  {title} {\bibinfo {title} {{Beyond Conventional Ferromagnetism and Antiferromagnetism: A Phase with Nonrelativistic Spin and Crystal Rotation Symmetry}},\ }\href {https://doi.org/10.1103/PhysRevX.12.031042} {\bibfield  {journal} {\bibinfo  {journal} {Phys. Rev. X}\ }\textbf {\bibinfo {volume} {12}},\ \bibinfo {pages} {031042} (\bibinfo {year} {2022})},\ \Eprint {https://arxiv.org/abs/2105.05820} {arXiv:2105.05820} \BibitemShut {NoStop}%
\bibitem [{\citenamefont {\ifmmode~\check{S}\else \v{S}\fi{}mejkal}\ \emph {et~al.}(2022)\citenamefont {\ifmmode~\check{S}\else \v{S}\fi{}mejkal}, \citenamefont {Sinova},\ and\ \citenamefont {Jungwirth}}]{SmejkalPRX2022}%
  \BibitemOpen
  \bibfield  {author} {\bibinfo {author} {\bibfnamefont {L.}~\bibnamefont {\ifmmode~\check{S}\else \v{S}\fi{}mejkal}}, \bibinfo {author} {\bibfnamefont {J.}~\bibnamefont {Sinova}},\ and\ \bibinfo {author} {\bibfnamefont {T.}~\bibnamefont {Jungwirth}},\ }\bibfield  {title} {\bibinfo {title} {{Emerging Research Landscape of Altermagnetism}},\ }\href {https://doi.org/10.1103/PhysRevX.12.040501} {\bibfield  {journal} {\bibinfo  {journal} {Phys. Rev. X}\ }\textbf {\bibinfo {volume} {12}},\ \bibinfo {pages} {040501} (\bibinfo {year} {2022})}\BibitemShut {NoStop}%
\bibitem [{\citenamefont {Bai}\ \emph {et~al.}(2024)\citenamefont {Bai}, \citenamefont {Feng}, \citenamefont {Liu}, \citenamefont {{\v S}mejkal}, \citenamefont {Mokrousov},\ and\ \citenamefont {Yao}}]{bai_arxiv_24}%
  \BibitemOpen
  \bibfield  {author} {\bibinfo {author} {\bibfnamefont {L.}~\bibnamefont {Bai}}, \bibinfo {author} {\bibfnamefont {W.}~\bibnamefont {Feng}}, \bibinfo {author} {\bibfnamefont {S.}~\bibnamefont {Liu}}, \bibinfo {author} {\bibfnamefont {L.}~\bibnamefont {{\v S}mejkal}}, \bibinfo {author} {\bibfnamefont {Y.}~\bibnamefont {Mokrousov}},\ and\ \bibinfo {author} {\bibfnamefont {Y.}~\bibnamefont {Yao}},\ }\href {https://arxiv.org/abs/2406.02123} {\bibinfo {title} {Altermagnetism: Exploring new frontiers in magnetism and spintronics}} (\bibinfo {year} {2024}),\ \Eprint {https://arxiv.org/abs/2406.02123} {arxiv:2406.02123 [cond-mat]} \BibitemShut {NoStop}%
\bibitem [{\citenamefont {Gonz{\'{a}}lez-Hern{\'{a}}ndez}\ \emph {et~al.}(2021)\citenamefont {Gonz{\'{a}}lez-Hern{\'{a}}ndez}, \citenamefont {{\v{S}}mejkal}, \citenamefont {V{\'{y}}born{\'{y}}}, \citenamefont {Yahagi}, \citenamefont {Sinova}, \citenamefont {Jungwirth},\ and\ \citenamefont {{\v{Z}}elezn{\'{y}}}}]{Gonzalez-Hernandez-Zelezny:2021}%
  \BibitemOpen
  \bibfield  {author} {\bibinfo {author} {\bibfnamefont {R.}~\bibnamefont {Gonz{\'{a}}lez-Hern{\'{a}}ndez}}, \bibinfo {author} {\bibfnamefont {L.}~\bibnamefont {{\v{S}}mejkal}}, \bibinfo {author} {\bibfnamefont {K.}~\bibnamefont {V{\'{y}}born{\'{y}}}}, \bibinfo {author} {\bibfnamefont {Y.}~\bibnamefont {Yahagi}}, \bibinfo {author} {\bibfnamefont {J.}~\bibnamefont {Sinova}}, \bibinfo {author} {\bibfnamefont {T.}~\bibnamefont {Jungwirth}},\ and\ \bibinfo {author} {\bibfnamefont {J.}~\bibnamefont {{\v{Z}}elezn{\'{y}}}},\ }\bibfield  {title} {\bibinfo {title} {{Efficient Electrical Spin Splitter Based on Nonrelativistic Collinear Antiferromagnetism}},\ }\href {https://doi.org/10.1103/PhysRevLett.126.127701} {\bibfield  {journal} {\bibinfo  {journal} {Phys. Rev. Lett.}\ }\textbf {\bibinfo {volume} {126}},\ \bibinfo {pages} {127701} (\bibinfo {year} {2021})},\ \Eprint {https://arxiv.org/abs/2002.07073} {arXiv:2002.07073} \BibitemShut {NoStop}%
\bibitem [{\citenamefont {Reichlov{\'{a}}}\ \emph {et~al.}(2020)\citenamefont {Reichlov{\'{a}}}, \citenamefont {Seeger}, \citenamefont {Gonz{\'{a}}lez-Hern{\'{a}}ndez}, \citenamefont {Kounta}, \citenamefont {Schlitz}, \citenamefont {Kriegner}, \citenamefont {Ritzinger}, \citenamefont {Lammel}, \citenamefont {Leivisk{\"{a}}}, \citenamefont {Pet{\v{r}}{\'{i}}{\v{c}}ek}, \citenamefont {Dole{\v{z}}al}, \citenamefont {Schmoranzerov{\'{a}}}, \citenamefont {Bad'ura}, \citenamefont {Thomas}, \citenamefont {Baltz}, \citenamefont {Michez}, \citenamefont {Sinova}, \citenamefont {Goennenwein}, \citenamefont {Jungwirth},\ and\ \citenamefont {{\v{S}}mejkal}}]{Reichlova-Smejkal:2020}%
  \BibitemOpen
  \bibfield  {author} {\bibinfo {author} {\bibfnamefont {H.}~\bibnamefont {Reichlov{\'{a}}}}, \bibinfo {author} {\bibfnamefont {R.~L.}\ \bibnamefont {Seeger}}, \bibinfo {author} {\bibfnamefont {R.}~\bibnamefont {Gonz{\'{a}}lez-Hern{\'{a}}ndez}}, \bibinfo {author} {\bibfnamefont {I.}~\bibnamefont {Kounta}}, \bibinfo {author} {\bibfnamefont {R.}~\bibnamefont {Schlitz}}, \bibinfo {author} {\bibfnamefont {D.}~\bibnamefont {Kriegner}}, \bibinfo {author} {\bibfnamefont {P.}~\bibnamefont {Ritzinger}}, \bibinfo {author} {\bibfnamefont {M.}~\bibnamefont {Lammel}}, \bibinfo {author} {\bibfnamefont {M.}~\bibnamefont {Leivisk{\"{a}}}}, \bibinfo {author} {\bibfnamefont {V.}~\bibnamefont {Pet{\v{r}}{\'{i}}{\v{c}}ek}}, \bibinfo {author} {\bibfnamefont {P.}~\bibnamefont {Dole{\v{z}}al}}, \bibinfo {author} {\bibfnamefont {E.}~\bibnamefont {Schmoranzerov{\'{a}}}}, \bibinfo {author} {\bibfnamefont {A.}~\bibnamefont {Bad'ura}}, \bibinfo {author} {\bibfnamefont {A.}~\bibnamefont {Thomas}}, \bibinfo {author} {\bibfnamefont
  {V.}~\bibnamefont {Baltz}}, \bibinfo {author} {\bibfnamefont {L.}~\bibnamefont {Michez}}, \bibinfo {author} {\bibfnamefont {J.}~\bibnamefont {Sinova}}, \bibinfo {author} {\bibfnamefont {S.~T.~B.}\ \bibnamefont {Goennenwein}}, \bibinfo {author} {\bibfnamefont {T.}~\bibnamefont {Jungwirth}},\ and\ \bibinfo {author} {\bibfnamefont {L.}~\bibnamefont {{\v{S}}mejkal}},\ }\href@noop {} {\bibinfo {title} {{Macroscopic time reversal symmetry breaking by staggered spin-momentum interaction}}} (\bibinfo {year} {2020}),\ \Eprint {https://arxiv.org/abs/2012.15651} {arXiv:2012.15651} \BibitemShut {NoStop}%
\bibitem [{\citenamefont {Egorov}\ and\ \citenamefont {Evarestov}(2021)}]{Egorov-Evarestov:2021}%
  \BibitemOpen
  \bibfield  {author} {\bibinfo {author} {\bibfnamefont {S.~A.}\ \bibnamefont {Egorov}}\ and\ \bibinfo {author} {\bibfnamefont {R.~A.}\ \bibnamefont {Evarestov}},\ }\bibfield  {title} {\bibinfo {title} {{Colossal Spin Splitting in the Monolayer of the Collinear Antiferromagnet MnF$_2$}},\ }\href {https://doi.org/10.1021/acs.jpclett.1c00282} {\bibfield  {journal} {\bibinfo  {journal} {J. Phys. Chem. Lett.}\ }\textbf {\bibinfo {volume} {12}},\ \bibinfo {pages} {2363} (\bibinfo {year} {2021})}\BibitemShut {NoStop}%
\bibitem [{\citenamefont {Lee}\ \emph {et~al.}(2024)\citenamefont {Lee}, \citenamefont {Lee}, \citenamefont {Jung}, \citenamefont {Jung}, \citenamefont {Kim}, \citenamefont {Lee}, \citenamefont {Seok}, \citenamefont {Kim}, \citenamefont {Park}, \citenamefont {{\v{S}}mejkal}, \citenamefont {Kang},\ and\ \citenamefont {Kim}}]{Lee-Kim:2023}%
  \BibitemOpen
  \bibfield  {author} {\bibinfo {author} {\bibfnamefont {S.}~\bibnamefont {Lee}}, \bibinfo {author} {\bibfnamefont {S.}~\bibnamefont {Lee}}, \bibinfo {author} {\bibfnamefont {S.}~\bibnamefont {Jung}}, \bibinfo {author} {\bibfnamefont {J.}~\bibnamefont {Jung}}, \bibinfo {author} {\bibfnamefont {D.}~\bibnamefont {Kim}}, \bibinfo {author} {\bibfnamefont {Y.}~\bibnamefont {Lee}}, \bibinfo {author} {\bibfnamefont {B.}~\bibnamefont {Seok}}, \bibinfo {author} {\bibfnamefont {J.}~\bibnamefont {Kim}}, \bibinfo {author} {\bibfnamefont {B.~G.}\ \bibnamefont {Park}}, \bibinfo {author} {\bibfnamefont {L.}~\bibnamefont {{\v{S}}mejkal}}, \bibinfo {author} {\bibfnamefont {C.-J.}\ \bibnamefont {Kang}},\ and\ \bibinfo {author} {\bibfnamefont {C.}~\bibnamefont {Kim}},\ }\bibfield  {title} {\bibinfo {title} {{Broken Kramers Degeneracy in Altermagnetic MnTe}},\ }\href {https://doi.org/10.1103/PhysRevLett.132.036702} {\bibfield  {journal} {\bibinfo  {journal} {Phys. Rev. Lett.}\ }\textbf {\bibinfo {volume} {132}},\ \bibinfo
  {pages} {036702} (\bibinfo {year} {2024})},\ \Eprint {https://arxiv.org/abs/2308.11180} {arXiv:2308.11180} \BibitemShut {NoStop}%
\bibitem [{\citenamefont {Krempask{\'{y}}}\ \emph {et~al.}(2024)\citenamefont {Krempask{\'{y}}}, \citenamefont {{\v{S}}mejkal}, \citenamefont {D'Souza}, \citenamefont {Hajlaoui}, \citenamefont {Springholz}, \citenamefont {Uhl{\'{i}}řov{\'{a}}}, \citenamefont {Alarab}, \citenamefont {Constantinou}, \citenamefont {Strocov}, \citenamefont {Usanov}, \citenamefont {Pudelko}, \citenamefont {Gonz{\'{a}}lez-Hern{\'{a}}ndez}, \citenamefont {{Birk Hellenes}}, \citenamefont {Jansa}, \citenamefont {Reichlov{\'{a}}}, \citenamefont {{\v{S}}ob{\'{a}}ň}, \citenamefont {{Gonzalez Betancourt}}, \citenamefont {Wadley}, \citenamefont {Sinova}, \citenamefont {Kriegner}, \citenamefont {Min{\'{a}}r}, \citenamefont {Dil},\ and\ \citenamefont {Jungwirth}}]{Krempasky-Jungwirth:2024}%
  \BibitemOpen
  \bibfield  {author} {\bibinfo {author} {\bibfnamefont {J.}~\bibnamefont {Krempask{\'{y}}}}, \bibinfo {author} {\bibfnamefont {L.}~\bibnamefont {{\v{S}}mejkal}}, \bibinfo {author} {\bibfnamefont {S.~W.}\ \bibnamefont {D'Souza}}, \bibinfo {author} {\bibfnamefont {M.}~\bibnamefont {Hajlaoui}}, \bibinfo {author} {\bibfnamefont {G.}~\bibnamefont {Springholz}}, \bibinfo {author} {\bibfnamefont {K.}~\bibnamefont {Uhl{\'{i}}řov{\'{a}}}}, \bibinfo {author} {\bibfnamefont {F.}~\bibnamefont {Alarab}}, \bibinfo {author} {\bibfnamefont {P.~C.}\ \bibnamefont {Constantinou}}, \bibinfo {author} {\bibfnamefont {V.}~\bibnamefont {Strocov}}, \bibinfo {author} {\bibfnamefont {D.}~\bibnamefont {Usanov}}, \bibinfo {author} {\bibfnamefont {W.~R.}\ \bibnamefont {Pudelko}}, \bibinfo {author} {\bibfnamefont {R.}~\bibnamefont {Gonz{\'{a}}lez-Hern{\'{a}}ndez}}, \bibinfo {author} {\bibfnamefont {A.}~\bibnamefont {{Birk Hellenes}}}, \bibinfo {author} {\bibfnamefont {Z.}~\bibnamefont {Jansa}}, \bibinfo {author} {\bibfnamefont
  {H.}~\bibnamefont {Reichlov{\'{a}}}}, \bibinfo {author} {\bibfnamefont {Z.}~\bibnamefont {{\v{S}}ob{\'{a}}ň}}, \bibinfo {author} {\bibfnamefont {R.~D.}\ \bibnamefont {{Gonzalez Betancourt}}}, \bibinfo {author} {\bibfnamefont {P.}~\bibnamefont {Wadley}}, \bibinfo {author} {\bibfnamefont {J.}~\bibnamefont {Sinova}}, \bibinfo {author} {\bibfnamefont {D.}~\bibnamefont {Kriegner}}, \bibinfo {author} {\bibfnamefont {J.}~\bibnamefont {Min{\'{a}}r}}, \bibinfo {author} {\bibfnamefont {J.~H.}\ \bibnamefont {Dil}},\ and\ \bibinfo {author} {\bibfnamefont {T.}~\bibnamefont {Jungwirth}},\ }\bibfield  {title} {\bibinfo {title} {{Altermagnetic lifting of Kramers spin degeneracy}},\ }\href {https://doi.org/10.1038/s41586-023-06907-7} {\bibfield  {journal} {\bibinfo  {journal} {Nature}\ }\textbf {\bibinfo {volume} {626}},\ \bibinfo {pages} {517} (\bibinfo {year} {2024})}\BibitemShut {NoStop}%
\bibitem [{\citenamefont {Osumi}\ \emph {et~al.}(2024)\citenamefont {Osumi}, \citenamefont {Souma}, \citenamefont {Aoyama}, \citenamefont {Yamauchi}, \citenamefont {Honma}, \citenamefont {Nakayama}, \citenamefont {Takahashi}, \citenamefont {Ohgushi},\ and\ \citenamefont {Sato}}]{Osumi-Sato-ObservationGiantBand-2024}%
  \BibitemOpen
  \bibfield  {author} {\bibinfo {author} {\bibfnamefont {T.}~\bibnamefont {Osumi}}, \bibinfo {author} {\bibfnamefont {S.}~\bibnamefont {Souma}}, \bibinfo {author} {\bibfnamefont {T.}~\bibnamefont {Aoyama}}, \bibinfo {author} {\bibfnamefont {K.}~\bibnamefont {Yamauchi}}, \bibinfo {author} {\bibfnamefont {A.}~\bibnamefont {Honma}}, \bibinfo {author} {\bibfnamefont {K.}~\bibnamefont {Nakayama}}, \bibinfo {author} {\bibfnamefont {T.}~\bibnamefont {Takahashi}}, \bibinfo {author} {\bibfnamefont {K.}~\bibnamefont {Ohgushi}},\ and\ \bibinfo {author} {\bibfnamefont {T.}~\bibnamefont {Sato}},\ }\bibfield  {title} {\bibinfo {title} {Observation of a giant band splitting in altermagnetic {{MnTe}}},\ }\href {https://doi.org/10.1103/PhysRevB.109.115102} {\bibfield  {journal} {\bibinfo  {journal} {Phys. Rev. B}\ }\textbf {\bibinfo {volume} {109}},\ \bibinfo {pages} {115102} (\bibinfo {year} {2024})}\BibitemShut {NoStop}%
\bibitem [{\citenamefont {Fedchenko}\ \emph {et~al.}(2024)\citenamefont {Fedchenko}, \citenamefont {Min{\'a}r}, \citenamefont {Akashdeep}, \citenamefont {D'Souza}, \citenamefont {Vasilyev}, \citenamefont {Tkach}, \citenamefont {Odenbreit}, \citenamefont {Nguyen}, \citenamefont {Kutnyakhov}, \citenamefont {Wind}, \citenamefont {Wenthaus}, \citenamefont {Scholz}, \citenamefont {Rossnagel}, \citenamefont {Hoesch}, \citenamefont {Aeschlimann}, \citenamefont {Stadtm{\"u}ller}, \citenamefont {Kl{\"a}ui}, \citenamefont {Sch{\"o}nhense}, \citenamefont {Jungwirth}, \citenamefont {Hellenes}, \citenamefont {Jakob}, \citenamefont {{\v S}mejkal}, \citenamefont {Sinova},\ and\ \citenamefont {Elmers}}]{Fedchenko-Elmers:2023}%
  \BibitemOpen
  \bibfield  {author} {\bibinfo {author} {\bibfnamefont {O.}~\bibnamefont {Fedchenko}}, \bibinfo {author} {\bibfnamefont {J.}~\bibnamefont {Min{\'a}r}}, \bibinfo {author} {\bibfnamefont {A.}~\bibnamefont {Akashdeep}}, \bibinfo {author} {\bibfnamefont {S.~W.}\ \bibnamefont {D'Souza}}, \bibinfo {author} {\bibfnamefont {D.}~\bibnamefont {Vasilyev}}, \bibinfo {author} {\bibfnamefont {O.}~\bibnamefont {Tkach}}, \bibinfo {author} {\bibfnamefont {L.}~\bibnamefont {Odenbreit}}, \bibinfo {author} {\bibfnamefont {Q.}~\bibnamefont {Nguyen}}, \bibinfo {author} {\bibfnamefont {D.}~\bibnamefont {Kutnyakhov}}, \bibinfo {author} {\bibfnamefont {N.}~\bibnamefont {Wind}}, \bibinfo {author} {\bibfnamefont {L.}~\bibnamefont {Wenthaus}}, \bibinfo {author} {\bibfnamefont {M.}~\bibnamefont {Scholz}}, \bibinfo {author} {\bibfnamefont {K.}~\bibnamefont {Rossnagel}}, \bibinfo {author} {\bibfnamefont {M.}~\bibnamefont {Hoesch}}, \bibinfo {author} {\bibfnamefont {M.}~\bibnamefont {Aeschlimann}}, \bibinfo {author} {\bibfnamefont
  {B.}~\bibnamefont {Stadtm{\"u}ller}}, \bibinfo {author} {\bibfnamefont {M.}~\bibnamefont {Kl{\"a}ui}}, \bibinfo {author} {\bibfnamefont {G.}~\bibnamefont {Sch{\"o}nhense}}, \bibinfo {author} {\bibfnamefont {T.}~\bibnamefont {Jungwirth}}, \bibinfo {author} {\bibfnamefont {A.~B.}\ \bibnamefont {Hellenes}}, \bibinfo {author} {\bibfnamefont {G.}~\bibnamefont {Jakob}}, \bibinfo {author} {\bibfnamefont {L.}~\bibnamefont {{\v S}mejkal}}, \bibinfo {author} {\bibfnamefont {J.}~\bibnamefont {Sinova}},\ and\ \bibinfo {author} {\bibfnamefont {H.-J.}\ \bibnamefont {Elmers}},\ }\bibfield  {title} {\bibinfo {title} {{Observation of time-reversal symmetry breaking in the band structure of altermagnetic RuO$_2$}},\ }\href {https://doi.org/10.1126/sciadv.adj4883} {\bibfield  {journal} {\bibinfo  {journal} {Science Advances}\ }\textbf {\bibinfo {volume} {10}},\ \bibinfo {pages} {eadj4883} (\bibinfo {year} {2024})}\BibitemShut {NoStop}%
\bibitem [{\citenamefont {Lin}\ \emph {et~al.}(2024)\citenamefont {Lin}, \citenamefont {Chen}, \citenamefont {Lu}, \citenamefont {Liang}, \citenamefont {Feng}, \citenamefont {Yamagami}, \citenamefont {Osiecki}, \citenamefont {Leandersson}, \citenamefont {Thiagarajan}, \citenamefont {Liu}, \citenamefont {Felser},\ and\ \citenamefont {Ma}}]{Li-Felser-Ma:2024}%
  \BibitemOpen
  \bibfield  {author} {\bibinfo {author} {\bibfnamefont {Z.}~\bibnamefont {Lin}}, \bibinfo {author} {\bibfnamefont {D.}~\bibnamefont {Chen}}, \bibinfo {author} {\bibfnamefont {W.}~\bibnamefont {Lu}}, \bibinfo {author} {\bibfnamefont {X.}~\bibnamefont {Liang}}, \bibinfo {author} {\bibfnamefont {S.}~\bibnamefont {Feng}}, \bibinfo {author} {\bibfnamefont {K.}~\bibnamefont {Yamagami}}, \bibinfo {author} {\bibfnamefont {J.}~\bibnamefont {Osiecki}}, \bibinfo {author} {\bibfnamefont {M.}~\bibnamefont {Leandersson}}, \bibinfo {author} {\bibfnamefont {B.}~\bibnamefont {Thiagarajan}}, \bibinfo {author} {\bibfnamefont {J.}~\bibnamefont {Liu}}, \bibinfo {author} {\bibfnamefont {C.}~\bibnamefont {Felser}},\ and\ \bibinfo {author} {\bibfnamefont {J.}~\bibnamefont {Ma}},\ }\href@noop {} {\bibinfo {title} {{Observation of Giant Spin Splitting and d-wave Spin Texture in Room Temperature Altermagnet RuO$_2$}}} (\bibinfo {year} {2024}),\ \Eprint {https://arxiv.org/abs/2402.04995} {arXiv:2402.04995} \BibitemShut {NoStop}%
\bibitem [{\citenamefont {Reimers}\ \emph {et~al.}(2024)\citenamefont {Reimers}, \citenamefont {Odenbreit}, \citenamefont {{\v{S}}mejkal}, \citenamefont {Strocov}, \citenamefont {Constantinou}, \citenamefont {Hellenes}, \citenamefont {{Jaeschke Ubiergo}}, \citenamefont {Campos}, \citenamefont {Bharadwaj}, \citenamefont {Chakraborty}, \citenamefont {Denneulin}, \citenamefont {Shi}, \citenamefont {Dunin-Borkowski}, \citenamefont {Das}, \citenamefont {Kl{\"{a}}ui}, \citenamefont {Sinova},\ and\ \citenamefont {Jourdan}}]{Reimers-Jourdan:2023}%
  \BibitemOpen
  \bibfield  {author} {\bibinfo {author} {\bibfnamefont {S.}~\bibnamefont {Reimers}}, \bibinfo {author} {\bibfnamefont {L.}~\bibnamefont {Odenbreit}}, \bibinfo {author} {\bibfnamefont {L.}~\bibnamefont {{\v{S}}mejkal}}, \bibinfo {author} {\bibfnamefont {V.~N.}\ \bibnamefont {Strocov}}, \bibinfo {author} {\bibfnamefont {P.}~\bibnamefont {Constantinou}}, \bibinfo {author} {\bibfnamefont {A.~B.}\ \bibnamefont {Hellenes}}, \bibinfo {author} {\bibfnamefont {R.}~\bibnamefont {{Jaeschke Ubiergo}}}, \bibinfo {author} {\bibfnamefont {W.~H.}\ \bibnamefont {Campos}}, \bibinfo {author} {\bibfnamefont {V.~K.}\ \bibnamefont {Bharadwaj}}, \bibinfo {author} {\bibfnamefont {A.}~\bibnamefont {Chakraborty}}, \bibinfo {author} {\bibfnamefont {T.}~\bibnamefont {Denneulin}}, \bibinfo {author} {\bibfnamefont {W.}~\bibnamefont {Shi}}, \bibinfo {author} {\bibfnamefont {R.~E.}\ \bibnamefont {Dunin-Borkowski}}, \bibinfo {author} {\bibfnamefont {S.}~\bibnamefont {Das}}, \bibinfo {author} {\bibfnamefont {M.}~\bibnamefont {Kl{\"{a}}ui}},
  \bibinfo {author} {\bibfnamefont {J.}~\bibnamefont {Sinova}},\ and\ \bibinfo {author} {\bibfnamefont {M.}~\bibnamefont {Jourdan}},\ }\bibfield  {title} {\bibinfo {title} {{Direct observation of altermagnetic band splitting in CrSb thin films}},\ }\href {https://doi.org/10.1038/s41467-024-46476-5} {\bibfield  {journal} {\bibinfo  {journal} {Nat. Commun.}\ }\textbf {\bibinfo {volume} {15}},\ \bibinfo {pages} {2116} (\bibinfo {year} {2024})},\ \Eprint {https://arxiv.org/abs/2310.17280} {arXiv:2310.17280} \BibitemShut {NoStop}%
\bibitem [{\citenamefont {Yang}\ \emph {et~al.}(2024)\citenamefont {Yang}, \citenamefont {Li}, \citenamefont {Yang}, \citenamefont {Li}, \citenamefont {Zheng}, \citenamefont {Zhu}, \citenamefont {Cao}, \citenamefont {Zhao}, \citenamefont {Zhang}, \citenamefont {Ye}, \citenamefont {Song}, \citenamefont {Hu}, \citenamefont {Yang}, \citenamefont {Shi}, \citenamefont {Yuan}, \citenamefont {Zhang}, \citenamefont {Xu},\ and\ \citenamefont {Liu}}]{Yang-Liu-ThreedimensionalMappingElectronic-2024}%
  \BibitemOpen
  \bibfield  {author} {\bibinfo {author} {\bibfnamefont {G.}~\bibnamefont {Yang}}, \bibinfo {author} {\bibfnamefont {Z.}~\bibnamefont {Li}}, \bibinfo {author} {\bibfnamefont {S.}~\bibnamefont {Yang}}, \bibinfo {author} {\bibfnamefont {J.}~\bibnamefont {Li}}, \bibinfo {author} {\bibfnamefont {H.}~\bibnamefont {Zheng}}, \bibinfo {author} {\bibfnamefont {W.}~\bibnamefont {Zhu}}, \bibinfo {author} {\bibfnamefont {S.}~\bibnamefont {Cao}}, \bibinfo {author} {\bibfnamefont {W.}~\bibnamefont {Zhao}}, \bibinfo {author} {\bibfnamefont {J.}~\bibnamefont {Zhang}}, \bibinfo {author} {\bibfnamefont {M.}~\bibnamefont {Ye}}, \bibinfo {author} {\bibfnamefont {Y.}~\bibnamefont {Song}}, \bibinfo {author} {\bibfnamefont {L.-H.}\ \bibnamefont {Hu}}, \bibinfo {author} {\bibfnamefont {L.}~\bibnamefont {Yang}}, \bibinfo {author} {\bibfnamefont {M.}~\bibnamefont {Shi}}, \bibinfo {author} {\bibfnamefont {H.}~\bibnamefont {Yuan}}, \bibinfo {author} {\bibfnamefont {Y.}~\bibnamefont {Zhang}}, \bibinfo {author} {\bibfnamefont
  {Y.}~\bibnamefont {Xu}},\ and\ \bibinfo {author} {\bibfnamefont {Y.}~\bibnamefont {Liu}},\ }\href {http://arxiv.org/abs/2405.12575} {\bibinfo {title} {Three-dimensional mapping and electronic origin of large altermagnetic splitting near {{Fermi}} level in {{CrSb}}}} (\bibinfo {year} {2024}),\ \Eprint {https://arxiv.org/abs/2405.12575} {arxiv:2405.12575 [cond-mat]} \BibitemShut {NoStop}%
\bibitem [{\citenamefont {Zeng}\ \emph {et~al.}(2024)\citenamefont {Zeng}, \citenamefont {Zhu}, \citenamefont {Zhu}, \citenamefont {Liu}, \citenamefont {Ma}, \citenamefont {Hao}, \citenamefont {Liu}, \citenamefont {Qu}, \citenamefont {Yang}, \citenamefont {Jiang}, \citenamefont {Yamagami}, \citenamefont {Arita}, \citenamefont {Zhang}, \citenamefont {Shao}, \citenamefont {Dai}, \citenamefont {Shimada}, \citenamefont {Liu}, \citenamefont {Ye}, \citenamefont {Huang}, \citenamefont {Liu},\ and\ \citenamefont {Liu}}]{Zeng-Liu-ObservationSpinSplitting-2024}%
  \BibitemOpen
  \bibfield  {author} {\bibinfo {author} {\bibfnamefont {M.}~\bibnamefont {Zeng}}, \bibinfo {author} {\bibfnamefont {M.-Y.}\ \bibnamefont {Zhu}}, \bibinfo {author} {\bibfnamefont {Y.-P.}\ \bibnamefont {Zhu}}, \bibinfo {author} {\bibfnamefont {X.-R.}\ \bibnamefont {Liu}}, \bibinfo {author} {\bibfnamefont {X.-M.}\ \bibnamefont {Ma}}, \bibinfo {author} {\bibfnamefont {Y.-J.}\ \bibnamefont {Hao}}, \bibinfo {author} {\bibfnamefont {P.}~\bibnamefont {Liu}}, \bibinfo {author} {\bibfnamefont {G.}~\bibnamefont {Qu}}, \bibinfo {author} {\bibfnamefont {Y.}~\bibnamefont {Yang}}, \bibinfo {author} {\bibfnamefont {Z.}~\bibnamefont {Jiang}}, \bibinfo {author} {\bibfnamefont {K.}~\bibnamefont {Yamagami}}, \bibinfo {author} {\bibfnamefont {M.}~\bibnamefont {Arita}}, \bibinfo {author} {\bibfnamefont {X.}~\bibnamefont {Zhang}}, \bibinfo {author} {\bibfnamefont {T.-H.}\ \bibnamefont {Shao}}, \bibinfo {author} {\bibfnamefont {Y.}~\bibnamefont {Dai}}, \bibinfo {author} {\bibfnamefont {K.}~\bibnamefont {Shimada}}, \bibinfo {author}
  {\bibfnamefont {Z.}~\bibnamefont {Liu}}, \bibinfo {author} {\bibfnamefont {M.}~\bibnamefont {Ye}}, \bibinfo {author} {\bibfnamefont {Y.}~\bibnamefont {Huang}}, \bibinfo {author} {\bibfnamefont {Q.}~\bibnamefont {Liu}},\ and\ \bibinfo {author} {\bibfnamefont {C.}~\bibnamefont {Liu}},\ }\href {http://arxiv.org/abs/2405.12679} {\bibinfo {title} {Observation of {{Spin Splitting}} in {{Room-Temperature Metallic Antiferromagnet CrSb}}}} (\bibinfo {year} {2024}),\ \Eprint {https://arxiv.org/abs/2405.12679} {arxiv:2405.12679 [cond-mat]} \BibitemShut {NoStop}%
\bibitem [{\citenamefont {Hiraishi}\ \emph {et~al.}(2024)\citenamefont {Hiraishi}, \citenamefont {Okabe}, \citenamefont {Koda}, \citenamefont {Kadono}, \citenamefont {Muroi}, \citenamefont {Hirai},\ and\ \citenamefont {Hiroi}}]{Hiraishi-Hiroi-NonmagneticGroundState-2024}%
  \BibitemOpen
  \bibfield  {author} {\bibinfo {author} {\bibfnamefont {M.}~\bibnamefont {Hiraishi}}, \bibinfo {author} {\bibfnamefont {H.}~\bibnamefont {Okabe}}, \bibinfo {author} {\bibfnamefont {A.}~\bibnamefont {Koda}}, \bibinfo {author} {\bibfnamefont {R.}~\bibnamefont {Kadono}}, \bibinfo {author} {\bibfnamefont {T.}~\bibnamefont {Muroi}}, \bibinfo {author} {\bibfnamefont {D.}~\bibnamefont {Hirai}},\ and\ \bibinfo {author} {\bibfnamefont {Z.}~\bibnamefont {Hiroi}},\ }\bibfield  {title} {\bibinfo {title} {Nonmagnetic {{Ground State}} in {{RuO}}{$_2$} {{Revealed}} by {{Muon Spin Rotation}}},\ }\href {https://doi.org/10.1103/PhysRevLett.132.166702} {\bibfield  {journal} {\bibinfo  {journal} {Phys. Rev. Lett.}\ }\textbf {\bibinfo {volume} {132}},\ \bibinfo {pages} {166702} (\bibinfo {year} {2024})},\ \Eprint {https://arxiv.org/abs/2403.10028} {arxiv:2403.10028 [cond-mat]} \BibitemShut {NoStop}%
\bibitem [{\citenamefont {Ke{\ss}ler}\ \emph {et~al.}(2024)\citenamefont {Ke{\ss}ler}, \citenamefont {{Garcia-Gassull}}, \citenamefont {Suter}, \citenamefont {Prokscha}, \citenamefont {Salman}, \citenamefont {Khalyavin}, \citenamefont {Manuel}, \citenamefont {Orlandi}, \citenamefont {Mazin}, \citenamefont {Valent{\i}},\ and\ \citenamefont {Moser}}]{Kessler-Moser-Ru02:2024}%
  \BibitemOpen
  \bibfield  {author} {\bibinfo {author} {\bibfnamefont {P.}~\bibnamefont {Ke{\ss}ler}}, \bibinfo {author} {\bibfnamefont {L.}~\bibnamefont {{Garcia-Gassull}}}, \bibinfo {author} {\bibfnamefont {A.}~\bibnamefont {Suter}}, \bibinfo {author} {\bibfnamefont {T.}~\bibnamefont {Prokscha}}, \bibinfo {author} {\bibfnamefont {Z.}~\bibnamefont {Salman}}, \bibinfo {author} {\bibfnamefont {D.}~\bibnamefont {Khalyavin}}, \bibinfo {author} {\bibfnamefont {P.}~\bibnamefont {Manuel}}, \bibinfo {author} {\bibfnamefont {F.}~\bibnamefont {Orlandi}}, \bibinfo {author} {\bibfnamefont {I.~I.}\ \bibnamefont {Mazin}}, \bibinfo {author} {\bibfnamefont {R.}~\bibnamefont {Valent{\i}}},\ and\ \bibinfo {author} {\bibfnamefont {S.}~\bibnamefont {Moser}},\ }\href {http://arxiv.org/abs/2405.10820} {\bibinfo {title} {Absence of magnetic order in {{RuO}}{$_2$}: Insights from {$\mu$}{{SR}} spectroscopy and neutron diffraction}} (\bibinfo {year} {2024}),\ \Eprint {https://arxiv.org/abs/2405.10820} {arxiv:2405.10820 [cond-mat]} \BibitemShut
  {NoStop}%
\bibitem [{\citenamefont {Feng}\ \emph {et~al.}(2022)\citenamefont {Feng}, \citenamefont {Zhou}, \citenamefont {{\v{S}}mejkal}, \citenamefont {Wu}, \citenamefont {Zhu}, \citenamefont {Guo}, \citenamefont {Gonz{\'{a}}lez-Hern{\'{a}}ndez}, \citenamefont {Wang}, \citenamefont {Yan}, \citenamefont {Qin}, \citenamefont {Zhang}, \citenamefont {Wu}, \citenamefont {Chen}, \citenamefont {Meng}, \citenamefont {Liu}, \citenamefont {Xia}, \citenamefont {Sinova}, \citenamefont {Jungwirth},\ and\ \citenamefont {Liu}}]{Feng-Liu:2022}%
  \BibitemOpen
  \bibfield  {author} {\bibinfo {author} {\bibfnamefont {Z.}~\bibnamefont {Feng}}, \bibinfo {author} {\bibfnamefont {X.}~\bibnamefont {Zhou}}, \bibinfo {author} {\bibfnamefont {L.}~\bibnamefont {{\v{S}}mejkal}}, \bibinfo {author} {\bibfnamefont {L.}~\bibnamefont {Wu}}, \bibinfo {author} {\bibfnamefont {Z.}~\bibnamefont {Zhu}}, \bibinfo {author} {\bibfnamefont {H.}~\bibnamefont {Guo}}, \bibinfo {author} {\bibfnamefont {R.}~\bibnamefont {Gonz{\'{a}}lez-Hern{\'{a}}ndez}}, \bibinfo {author} {\bibfnamefont {X.}~\bibnamefont {Wang}}, \bibinfo {author} {\bibfnamefont {H.}~\bibnamefont {Yan}}, \bibinfo {author} {\bibfnamefont {P.}~\bibnamefont {Qin}}, \bibinfo {author} {\bibfnamefont {X.}~\bibnamefont {Zhang}}, \bibinfo {author} {\bibfnamefont {H.}~\bibnamefont {Wu}}, \bibinfo {author} {\bibfnamefont {H.}~\bibnamefont {Chen}}, \bibinfo {author} {\bibfnamefont {Z.}~\bibnamefont {Meng}}, \bibinfo {author} {\bibfnamefont {L.}~\bibnamefont {Liu}}, \bibinfo {author} {\bibfnamefont {Z.}~\bibnamefont {Xia}}, \bibinfo
  {author} {\bibfnamefont {J.}~\bibnamefont {Sinova}}, \bibinfo {author} {\bibfnamefont {T.}~\bibnamefont {Jungwirth}},\ and\ \bibinfo {author} {\bibfnamefont {Z.}~\bibnamefont {Liu}},\ }\bibfield  {title} {\bibinfo {title} {{An anomalous Hall effect in altermagnetic ruthenium dioxide}},\ }\href {https://doi.org/10.1038/s41928-022-00866-z} {\bibfield  {journal} {\bibinfo  {journal} {Nat. Electron.}\ }\textbf {\bibinfo {volume} {5}},\ \bibinfo {pages} {735} (\bibinfo {year} {2022})},\ \Eprint {https://arxiv.org/abs/2002.08712} {arXiv:2002.08712} \BibitemShut {NoStop}%
\bibitem [{\citenamefont {Tschirner}\ \emph {et~al.}(2023)\citenamefont {Tschirner}, \citenamefont {Ke{\ss}ler}, \citenamefont {Gonzalez~Betancourt}, \citenamefont {Kotte}, \citenamefont {Kriegner}, \citenamefont {B\"{u}chner}, \citenamefont {Dufouleur}, \citenamefont {Kamp}, \citenamefont {Jovic}, \citenamefont {Smejkal}, \citenamefont {Sinova}, \citenamefont {Claessen}, \citenamefont {Jungwirth}, \citenamefont {Moser}, \citenamefont {Reichlova},\ and\ \citenamefont {Veyrat}}]{Tschirner-Veyrat:2023}%
  \BibitemOpen
  \bibfield  {author} {\bibinfo {author} {\bibfnamefont {T.}~\bibnamefont {Tschirner}}, \bibinfo {author} {\bibfnamefont {P.}~\bibnamefont {Ke{\ss}ler}}, \bibinfo {author} {\bibfnamefont {R.~D.}\ \bibnamefont {Gonzalez~Betancourt}}, \bibinfo {author} {\bibfnamefont {T.}~\bibnamefont {Kotte}}, \bibinfo {author} {\bibfnamefont {D.}~\bibnamefont {Kriegner}}, \bibinfo {author} {\bibfnamefont {B.}~\bibnamefont {B\"{u}chner}}, \bibinfo {author} {\bibfnamefont {J.}~\bibnamefont {Dufouleur}}, \bibinfo {author} {\bibfnamefont {M.}~\bibnamefont {Kamp}}, \bibinfo {author} {\bibfnamefont {V.}~\bibnamefont {Jovic}}, \bibinfo {author} {\bibfnamefont {L.}~\bibnamefont {Smejkal}}, \bibinfo {author} {\bibfnamefont {J.}~\bibnamefont {Sinova}}, \bibinfo {author} {\bibfnamefont {R.}~\bibnamefont {Claessen}}, \bibinfo {author} {\bibfnamefont {T.}~\bibnamefont {Jungwirth}}, \bibinfo {author} {\bibfnamefont {S.}~\bibnamefont {Moser}}, \bibinfo {author} {\bibfnamefont {H.}~\bibnamefont {Reichlova}},\ and\ \bibinfo {author}
  {\bibfnamefont {L.}~\bibnamefont {Veyrat}},\ }\bibfield  {title} {\bibinfo {title} {{Saturation of the anomalous Hall effect at high magnetic fields in altermagnetic RuO$_2$}},\ }\href {https://doi.org/10.1063/5.0160335} {\bibfield  {journal} {\bibinfo  {journal} {APL Materials}\ }\textbf {\bibinfo {volume} {11}},\ \bibinfo {pages} {101103} (\bibinfo {year} {2023})}\BibitemShut {NoStop}%
\bibitem [{\citenamefont {Bose}\ \emph {et~al.}(2022)\citenamefont {Bose}, \citenamefont {Schreiber}, \citenamefont {Jain}, \citenamefont {Shao}, \citenamefont {Nair}, \citenamefont {Sun}, \citenamefont {Zhang}, \citenamefont {Muller}, \citenamefont {Tsymbal}, \citenamefont {Schlom},\ and\ \citenamefont {Ralph}}]{Bose-Ralph:2022}%
  \BibitemOpen
  \bibfield  {author} {\bibinfo {author} {\bibfnamefont {A.}~\bibnamefont {Bose}}, \bibinfo {author} {\bibfnamefont {N.~J.}\ \bibnamefont {Schreiber}}, \bibinfo {author} {\bibfnamefont {R.}~\bibnamefont {Jain}}, \bibinfo {author} {\bibfnamefont {D.-F.}\ \bibnamefont {Shao}}, \bibinfo {author} {\bibfnamefont {H.~P.}\ \bibnamefont {Nair}}, \bibinfo {author} {\bibfnamefont {J.}~\bibnamefont {Sun}}, \bibinfo {author} {\bibfnamefont {X.~S.}\ \bibnamefont {Zhang}}, \bibinfo {author} {\bibfnamefont {D.~A.}\ \bibnamefont {Muller}}, \bibinfo {author} {\bibfnamefont {E.~Y.}\ \bibnamefont {Tsymbal}}, \bibinfo {author} {\bibfnamefont {D.~G.}\ \bibnamefont {Schlom}},\ and\ \bibinfo {author} {\bibfnamefont {D.~C.}\ \bibnamefont {Ralph}},\ }\bibfield  {title} {\bibinfo {title} {{Tilted spin current generated by the collinear antiferromagnet ruthenium dioxide}},\ }\href {https://doi.org/10.1038/s41928-022-00744-8} {\bibfield  {journal} {\bibinfo  {journal} {Nat. Electron.}\ }\textbf {\bibinfo {volume} {5}},\ \bibinfo {pages}
  {267} (\bibinfo {year} {2022})},\ \Eprint {https://arxiv.org/abs/2108.09150} {arXiv:2108.09150} \BibitemShut {NoStop}%
\bibitem [{\citenamefont {Bai}\ \emph {et~al.}(2022)\citenamefont {Bai}, \citenamefont {Han}, \citenamefont {Feng}, \citenamefont {Zhou}, \citenamefont {Su}, \citenamefont {Wang}, \citenamefont {Liao}, \citenamefont {Zhu}, \citenamefont {Chen}, \citenamefont {Pan}, \citenamefont {Fan},\ and\ \citenamefont {Song}}]{Bai-Song:2022}%
  \BibitemOpen
  \bibfield  {author} {\bibinfo {author} {\bibfnamefont {H.}~\bibnamefont {Bai}}, \bibinfo {author} {\bibfnamefont {L.}~\bibnamefont {Han}}, \bibinfo {author} {\bibfnamefont {X.~Y.}\ \bibnamefont {Feng}}, \bibinfo {author} {\bibfnamefont {Y.~J.}\ \bibnamefont {Zhou}}, \bibinfo {author} {\bibfnamefont {R.~X.}\ \bibnamefont {Su}}, \bibinfo {author} {\bibfnamefont {Q.}~\bibnamefont {Wang}}, \bibinfo {author} {\bibfnamefont {L.~Y.}\ \bibnamefont {Liao}}, \bibinfo {author} {\bibfnamefont {W.~X.}\ \bibnamefont {Zhu}}, \bibinfo {author} {\bibfnamefont {X.~Z.}\ \bibnamefont {Chen}}, \bibinfo {author} {\bibfnamefont {F.}~\bibnamefont {Pan}}, \bibinfo {author} {\bibfnamefont {X.~L.}\ \bibnamefont {Fan}},\ and\ \bibinfo {author} {\bibfnamefont {C.}~\bibnamefont {Song}},\ }\bibfield  {title} {\bibinfo {title} {{Observation of Spin Splitting Torque in a Collinear Antiferromagnet RuO$_2$}},\ }\href {https://doi.org/10.1103/PhysRevLett.128.197202} {\bibfield  {journal} {\bibinfo  {journal} {Phys. Rev. Lett.}\ }\textbf
  {\bibinfo {volume} {128}},\ \bibinfo {pages} {197202} (\bibinfo {year} {2022})},\ \Eprint {https://arxiv.org/abs/2109.05933} {arXiv:2109.05933} \BibitemShut {NoStop}%
\bibitem [{\citenamefont {Karube}\ \emph {et~al.}(2022)\citenamefont {Karube}, \citenamefont {Tanaka}, \citenamefont {Sugawara}, \citenamefont {Kadoguchi}, \citenamefont {Kohda},\ and\ \citenamefont {Nitta}}]{Karube-Nitta:2022}%
  \BibitemOpen
  \bibfield  {author} {\bibinfo {author} {\bibfnamefont {S.}~\bibnamefont {Karube}}, \bibinfo {author} {\bibfnamefont {T.}~\bibnamefont {Tanaka}}, \bibinfo {author} {\bibfnamefont {D.}~\bibnamefont {Sugawara}}, \bibinfo {author} {\bibfnamefont {N.}~\bibnamefont {Kadoguchi}}, \bibinfo {author} {\bibfnamefont {M.}~\bibnamefont {Kohda}},\ and\ \bibinfo {author} {\bibfnamefont {J.}~\bibnamefont {Nitta}},\ }\bibfield  {title} {\bibinfo {title} {{Observation of Spin-Splitter Torque in Collinear Antiferromagnetic RuO$_2$}},\ }\href {https://doi.org/10.1103/PhysRevLett.129.137201} {\bibfield  {journal} {\bibinfo  {journal} {Phys. Rev. Lett.}\ }\textbf {\bibinfo {volume} {129}},\ \bibinfo {pages} {137201} (\bibinfo {year} {2022})},\ \Eprint {https://arxiv.org/abs/2111.07487} {arXiv:2111.07487} \BibitemShut {NoStop}%
\bibitem [{\citenamefont {Wu}\ \emph {et~al.}(2007)\citenamefont {Wu}, \citenamefont {Sun}, \citenamefont {Fradkin},\ and\ \citenamefont {Zhang}}]{wu_prb_07}%
  \BibitemOpen
  \bibfield  {author} {\bibinfo {author} {\bibfnamefont {C.}~\bibnamefont {Wu}}, \bibinfo {author} {\bibfnamefont {K.}~\bibnamefont {Sun}}, \bibinfo {author} {\bibfnamefont {E.}~\bibnamefont {Fradkin}},\ and\ \bibinfo {author} {\bibfnamefont {S.-C.}\ \bibnamefont {Zhang}},\ }\bibfield  {title} {\bibinfo {title} {Fermi liquid instabilities in the spin channel},\ }\href {https://doi.org/10.1103/PhysRevB.75.115103} {\bibfield  {journal} {\bibinfo  {journal} {Phys. Rev. B}\ }\textbf {\bibinfo {volume} {75}},\ \bibinfo {pages} {115103} (\bibinfo {year} {2007})}\BibitemShut {NoStop}%
\bibitem [{\citenamefont {Kiselev}\ \emph {et~al.}(2017)\citenamefont {Kiselev}, \citenamefont {Scheurer}, \citenamefont {W\"olfle},\ and\ \citenamefont {Schmalian}}]{kiselev_prb_17}%
  \BibitemOpen
  \bibfield  {author} {\bibinfo {author} {\bibfnamefont {E.~I.}\ \bibnamefont {Kiselev}}, \bibinfo {author} {\bibfnamefont {M.~S.}\ \bibnamefont {Scheurer}}, \bibinfo {author} {\bibfnamefont {P.}~\bibnamefont {W\"olfle}},\ and\ \bibinfo {author} {\bibfnamefont {J.}~\bibnamefont {Schmalian}},\ }\bibfield  {title} {\bibinfo {title} {Limits on dynamically generated spin-orbit coupling: Absence of $l=1$ pomeranchuk instabilities in metals},\ }\href {https://doi.org/10.1103/PhysRevB.95.125122} {\bibfield  {journal} {\bibinfo  {journal} {Phys. Rev. B}\ }\textbf {\bibinfo {volume} {95}},\ \bibinfo {pages} {125122} (\bibinfo {year} {2017})}\BibitemShut {NoStop}%
\bibitem [{\citenamefont {Naka}\ \emph {et~al.}(2019)\citenamefont {Naka}, \citenamefont {Hayami}, \citenamefont {Kusunose}, \citenamefont {Yanagi}, \citenamefont {Motome},\ and\ \citenamefont {Seo}}]{Naka-Seo-SpinCurrentGeneration-2019}%
  \BibitemOpen
  \bibfield  {author} {\bibinfo {author} {\bibfnamefont {M.}~\bibnamefont {Naka}}, \bibinfo {author} {\bibfnamefont {S.}~\bibnamefont {Hayami}}, \bibinfo {author} {\bibfnamefont {H.}~\bibnamefont {Kusunose}}, \bibinfo {author} {\bibfnamefont {Y.}~\bibnamefont {Yanagi}}, \bibinfo {author} {\bibfnamefont {Y.}~\bibnamefont {Motome}},\ and\ \bibinfo {author} {\bibfnamefont {H.}~\bibnamefont {Seo}},\ }\bibfield  {title} {\bibinfo {title} {Spin current generation in organic antiferromagnets},\ }\href {https://doi.org/10.1038/s41467-019-12229-y} {\bibfield  {journal} {\bibinfo  {journal} {Nature Communications}\ }\textbf {\bibinfo {volume} {10}},\ \bibinfo {pages} {4305} (\bibinfo {year} {2019})},\ \Eprint {https://arxiv.org/abs/1902.02506} {arxiv:1902.02506 [cond-mat]} \BibitemShut {NoStop}%
\bibitem [{\citenamefont {Shao}\ \emph {et~al.}(2021)\citenamefont {Shao}, \citenamefont {Zhang}, \citenamefont {Li}, \citenamefont {Eom},\ and\ \citenamefont {Tsymbal}}]{Shao-Tsymbal:2021}%
  \BibitemOpen
  \bibfield  {author} {\bibinfo {author} {\bibfnamefont {D.-F.}\ \bibnamefont {Shao}}, \bibinfo {author} {\bibfnamefont {S.-H.}\ \bibnamefont {Zhang}}, \bibinfo {author} {\bibfnamefont {M.}~\bibnamefont {Li}}, \bibinfo {author} {\bibfnamefont {C.-B.}\ \bibnamefont {Eom}},\ and\ \bibinfo {author} {\bibfnamefont {E.~Y.}\ \bibnamefont {Tsymbal}},\ }\bibfield  {title} {\bibinfo {title} {Spin-neutral currents for spintronics},\ }\href {https://doi.org/10.1038/s41467-021-26915-3} {\bibfield  {journal} {\bibinfo  {journal} {Nature Communications}\ }\textbf {\bibinfo {volume} {12}},\ \bibinfo {pages} {7061} (\bibinfo {year} {2021})},\ \Eprint {https://arxiv.org/abs/2103.09219} {arxiv:2103.09219} \BibitemShut {NoStop}%
\bibitem [{\citenamefont {{\v S}mejkal}\ \emph {et~al.}(2022)\citenamefont {{\v S}mejkal}, \citenamefont {Hellenes}, \citenamefont {{Gonz{\'a}lez-Hern{\'a}ndez}}, \citenamefont {Sinova},\ and\ \citenamefont {Jungwirth}}]{Smejkal-Jungwirth:2022-TMR}%
  \BibitemOpen
  \bibfield  {author} {\bibinfo {author} {\bibfnamefont {L.}~\bibnamefont {{\v S}mejkal}}, \bibinfo {author} {\bibfnamefont {A.~B.}\ \bibnamefont {Hellenes}}, \bibinfo {author} {\bibfnamefont {R.}~\bibnamefont {{Gonz{\'a}lez-Hern{\'a}ndez}}}, \bibinfo {author} {\bibfnamefont {J.}~\bibnamefont {Sinova}},\ and\ \bibinfo {author} {\bibfnamefont {T.}~\bibnamefont {Jungwirth}},\ }\bibfield  {title} {\bibinfo {title} {Giant and {{Tunneling Magnetoresistance}} in {{Unconventional Collinear Antiferromagnets}} with {{Nonrelativistic Spin-Momentum Coupling}}},\ }\href {https://doi.org/10.1103/PhysRevX.12.011028} {\bibfield  {journal} {\bibinfo  {journal} {Physical Review X}\ }\textbf {\bibinfo {volume} {12}},\ \bibinfo {pages} {011028} (\bibinfo {year} {2022})},\ \Eprint {https://arxiv.org/abs/2103.12664} {arxiv:2103.12664} \BibitemShut {NoStop}%
\bibitem [{\citenamefont {Das}\ \emph {et~al.}(2023)\citenamefont {Das}, \citenamefont {Suri},\ and\ \citenamefont {Soori}}]{Das-Soori:2023}%
  \BibitemOpen
  \bibfield  {author} {\bibinfo {author} {\bibfnamefont {S.}~\bibnamefont {Das}}, \bibinfo {author} {\bibfnamefont {D.}~\bibnamefont {Suri}},\ and\ \bibinfo {author} {\bibfnamefont {A.}~\bibnamefont {Soori}},\ }\bibfield  {title} {\bibinfo {title} {Transport across junctions of altermagnets with normal metals and ferromagnets},\ }\href {https://doi.org/10.1088/1361-648X/acea12} {\bibfield  {journal} {\bibinfo  {journal} {J. Phys. Condens. Matter}\ }\textbf {\bibinfo {volume} {35}},\ \bibinfo {pages} {435302} (\bibinfo {year} {2023})},\ \Eprint {https://arxiv.org/abs/2305.06680} {arxiv:2305.06680} \BibitemShut {NoStop}%
\bibitem [{\citenamefont {Sun}\ and\ \citenamefont {Linder}(2023)}]{sun_prb_23}%
  \BibitemOpen
  \bibfield  {author} {\bibinfo {author} {\bibfnamefont {C.}~\bibnamefont {Sun}}\ and\ \bibinfo {author} {\bibfnamefont {J.}~\bibnamefont {Linder}},\ }\bibfield  {title} {\bibinfo {title} {Spin pumping from a ferromagnetic insulator into an altermagnet},\ }\href {https://doi.org/10.1103/PhysRevB.108.L140408} {\bibfield  {journal} {\bibinfo  {journal} {Phys. Rev. B}\ }\textbf {\bibinfo {volume} {108}},\ \bibinfo {pages} {L140408} (\bibinfo {year} {2023})}\BibitemShut {NoStop}%
\bibitem [{\citenamefont {Hodt}\ and\ \citenamefont {Linder}(2024)}]{hodt_prb_24}%
  \BibitemOpen
  \bibfield  {author} {\bibinfo {author} {\bibfnamefont {E.~W.}\ \bibnamefont {Hodt}}\ and\ \bibinfo {author} {\bibfnamefont {J.}~\bibnamefont {Linder}},\ }\bibfield  {title} {\bibinfo {title} {Spin pumping in an altermagnet/normal-metal bilayer},\ }\href {https://doi.org/10.1103/PhysRevB.109.174438} {\bibfield  {journal} {\bibinfo  {journal} {Phys. Rev. B}\ }\textbf {\bibinfo {volume} {109}},\ \bibinfo {pages} {174438} (\bibinfo {year} {2024})}\BibitemShut {NoStop}%
\bibitem [{\citenamefont {Andreev}(1996)}]{Andreev:1996}%
  \BibitemOpen
  \bibfield  {author} {\bibinfo {author} {\bibfnamefont {A.~F.}\ \bibnamefont {Andreev}},\ }\bibfield  {title} {\bibinfo {title} {Macroscopic magnetic fields of antiferromagnets},\ }\href {https://doi.org/10.1134/1.566978} {\bibfield  {journal} {\bibinfo  {journal} {J. Exp. Theor. Phys. Lett.}\ }\textbf {\bibinfo {volume} {63}},\ \bibinfo {pages} {758} (\bibinfo {year} {1996})}\BibitemShut {NoStop}%
\bibitem [{\citenamefont {Belashchenko}(2010)}]{Belashchenko-Belashchenko-EquilibriumMagnetizationBoundary-2010}%
  \BibitemOpen
  \bibfield  {author} {\bibinfo {author} {\bibfnamefont {K.~D.}\ \bibnamefont {Belashchenko}},\ }\bibfield  {title} {\bibinfo {title} {Equilibrium {{Magnetization}} at the {{Boundary}} of a {{Magnetoelectric Antiferromagnet}}},\ }\href {https://doi.org/10.1103/PhysRevLett.105.147204} {\bibfield  {journal} {\bibinfo  {journal} {Phys. Rev. Lett.}\ }\textbf {\bibinfo {volume} {105}},\ \bibinfo {pages} {147204} (\bibinfo {year} {2010})},\ \Eprint {https://arxiv.org/abs/1004.3974} {arxiv:1004.3974 [cond-mat.mtrl-sci]} \BibitemShut {NoStop}%
\bibitem [{\citenamefont {Mitsui}\ \emph {et~al.}(2020)\citenamefont {Mitsui}, \citenamefont {Sakai}, \citenamefont {Li}, \citenamefont {Ueno}, \citenamefont {Watanuki}, \citenamefont {Kobayashi}, \citenamefont {Masuda}, \citenamefont {Seto},\ and\ \citenamefont {Akai}}]{Mitsui-Akai-MagneticFriedelOscillation-2020}%
  \BibitemOpen
  \bibfield  {author} {\bibinfo {author} {\bibfnamefont {T.}~\bibnamefont {Mitsui}}, \bibinfo {author} {\bibfnamefont {S.}~\bibnamefont {Sakai}}, \bibinfo {author} {\bibfnamefont {S.}~\bibnamefont {Li}}, \bibinfo {author} {\bibfnamefont {T.}~\bibnamefont {Ueno}}, \bibinfo {author} {\bibfnamefont {T.}~\bibnamefont {Watanuki}}, \bibinfo {author} {\bibfnamefont {Y.}~\bibnamefont {Kobayashi}}, \bibinfo {author} {\bibfnamefont {R.}~\bibnamefont {Masuda}}, \bibinfo {author} {\bibfnamefont {M.}~\bibnamefont {Seto}},\ and\ \bibinfo {author} {\bibfnamefont {H.}~\bibnamefont {Akai}},\ }\bibfield  {title} {\bibinfo {title} {Magnetic {{Friedel Oscillation}} at the {{Fe}}(001) {{Surface}}: {{Direct Observation}} by {{Atomic-Layer-Resolved Synchrotron Radiation Fe}} 57 {{M{\"o}ssbauer Spectroscopy}}},\ }\href {https://doi.org/10.1103/PhysRevLett.125.236806} {\bibfield  {journal} {\bibinfo  {journal} {Phys. Rev. Lett.}\ }\textbf {\bibinfo {volume} {125}},\ \bibinfo {pages} {236806} (\bibinfo {year} {2020})}\BibitemShut
  {NoStop}%
\bibitem [{\citenamefont {Lapa}\ \emph {et~al.}(2020)\citenamefont {Lapa}, \citenamefont {Lee}, \citenamefont {Roshchin}, \citenamefont {Belashchenko},\ and\ \citenamefont {Schuller}}]{Lapa-Schuller-DetectionUncompensatedMagnetization-2020}%
  \BibitemOpen
  \bibfield  {author} {\bibinfo {author} {\bibfnamefont {P.~N.}\ \bibnamefont {Lapa}}, \bibinfo {author} {\bibfnamefont {M.-H.}\ \bibnamefont {Lee}}, \bibinfo {author} {\bibfnamefont {I.~V.}\ \bibnamefont {Roshchin}}, \bibinfo {author} {\bibfnamefont {K.~D.}\ \bibnamefont {Belashchenko}},\ and\ \bibinfo {author} {\bibfnamefont {I.~K.}\ \bibnamefont {Schuller}},\ }\bibfield  {title} {\bibinfo {title} {Detection of uncompensated magnetization at the interface of an epitaxial antiferromagnetic insulator},\ }\href {https://doi.org/10.1103/PhysRevB.102.174406} {\bibfield  {journal} {\bibinfo  {journal} {Phys. Rev. B}\ }\textbf {\bibinfo {volume} {102}},\ \bibinfo {pages} {174406} (\bibinfo {year} {2020})}\BibitemShut {NoStop}%
\bibitem [{\citenamefont {Weber}\ \emph {et~al.}(2024)\citenamefont {Weber}, \citenamefont {Urru}, \citenamefont {Bhowal}, \citenamefont {Ederer},\ and\ \citenamefont {Spaldin}}]{Weber-Spaldin-SurfaceMagnetizationAntiferromagnets-2023}%
  \BibitemOpen
  \bibfield  {author} {\bibinfo {author} {\bibfnamefont {S.~F.}\ \bibnamefont {Weber}}, \bibinfo {author} {\bibfnamefont {A.}~\bibnamefont {Urru}}, \bibinfo {author} {\bibfnamefont {S.}~\bibnamefont {Bhowal}}, \bibinfo {author} {\bibfnamefont {C.}~\bibnamefont {Ederer}},\ and\ \bibinfo {author} {\bibfnamefont {N.~A.}\ \bibnamefont {Spaldin}},\ }\bibfield  {title} {\bibinfo {title} {Surface {{Magnetization}} in {{Antiferromagnets}}: {{Classification}}, example materials, and relation to magnetoelectric responses},\ }\href {https://doi.org/10.1103/PhysRevX.14.021033} {\bibfield  {journal} {\bibinfo  {journal} {Phys. Rev. X}\ }\textbf {\bibinfo {volume} {14}},\ \bibinfo {pages} {021033} (\bibinfo {year} {2024})},\ \Eprint {https://arxiv.org/abs/2306.06631} {arxiv:2306.06631 [cond-mat.mtrl-sci]} \BibitemShut {NoStop}%
\bibitem [{\citenamefont {Pylypovskyi}\ \emph {et~al.}(2024)\citenamefont {Pylypovskyi}, \citenamefont {Weber}, \citenamefont {Makushko}, \citenamefont {Veremchuk}, \citenamefont {Spaldin},\ and\ \citenamefont {Makarov}}]{Pylypovskyi-Makarov-SurfacesymmetrydrivenDzyaloshinskiiMoriya-2023}%
  \BibitemOpen
  \bibfield  {author} {\bibinfo {author} {\bibfnamefont {O.~V.}\ \bibnamefont {Pylypovskyi}}, \bibinfo {author} {\bibfnamefont {S.~F.}\ \bibnamefont {Weber}}, \bibinfo {author} {\bibfnamefont {P.}~\bibnamefont {Makushko}}, \bibinfo {author} {\bibfnamefont {I.}~\bibnamefont {Veremchuk}}, \bibinfo {author} {\bibfnamefont {N.~A.}\ \bibnamefont {Spaldin}},\ and\ \bibinfo {author} {\bibfnamefont {D.}~\bibnamefont {Makarov}},\ }\bibfield  {title} {\bibinfo {title} {Surface-symmetry-driven {{Dzyaloshinskii--Moriya}} interaction and canted ferrimagnetism in collinear magnetoelectric antiferromagnet {{Cr}}{$_2$}{{O}}{$_3$}},\ }\href {https://doi.org/10.1103/PhysRevLett.132.226702} {\bibfield  {journal} {\bibinfo  {journal} {Phys. Rev. Lett.}\ }\textbf {\bibinfo {volume} {132}},\ \bibinfo {pages} {226702} (\bibinfo {year} {2024})},\ \Eprint {https://arxiv.org/abs/2310.13438} {arxiv:2310.13438 [cond-mat]} \BibitemShut {NoStop}%
\bibitem [{\citenamefont {Yershov}\ \emph {et~al.}(2024)\citenamefont {Yershov}, \citenamefont {Gomonay}, \citenamefont {Sinova}, \citenamefont {van~den Brink},\ and\ \citenamefont {Kravchuk}}]{Yershov-Kravchuk-CurvatureInducedMagnetization-2024}%
  \BibitemOpen
  \bibfield  {author} {\bibinfo {author} {\bibfnamefont {K.~V.}\ \bibnamefont {Yershov}}, \bibinfo {author} {\bibfnamefont {O.}~\bibnamefont {Gomonay}}, \bibinfo {author} {\bibfnamefont {J.}~\bibnamefont {Sinova}}, \bibinfo {author} {\bibfnamefont {J.}~\bibnamefont {van~den Brink}},\ and\ \bibinfo {author} {\bibfnamefont {V.~P.}\ \bibnamefont {Kravchuk}},\ }\href {http://arxiv.org/abs/2406.05103} {\bibinfo {title} {Curvature induced magnetization of altermagnetic films}} (\bibinfo {year} {2024}),\ \Eprint {https://arxiv.org/abs/2406.05103} {arxiv:2406.05103 [cond-mat]} \BibitemShut {NoStop}%
\bibitem [{\citenamefont {Wortmann}\ \emph {et~al.}(2001)\citenamefont {Wortmann}, \citenamefont {Heinze}, \citenamefont {Kurz}, \citenamefont {Bihlmayer},\ and\ \citenamefont {Bl{\"u}gel}}]{Wortmann-Blugel-ResolvingComplexAtomicScale-2001}%
  \BibitemOpen
  \bibfield  {author} {\bibinfo {author} {\bibfnamefont {D.}~\bibnamefont {Wortmann}}, \bibinfo {author} {\bibfnamefont {S.}~\bibnamefont {Heinze}}, \bibinfo {author} {\bibfnamefont {{\relax Ph}.}~\bibnamefont {Kurz}}, \bibinfo {author} {\bibfnamefont {G.}~\bibnamefont {Bihlmayer}},\ and\ \bibinfo {author} {\bibfnamefont {S.}~\bibnamefont {Bl{\"u}gel}},\ }\bibfield  {title} {\bibinfo {title} {Resolving {{Complex Atomic-Scale Spin Structures}} by {{Spin-Polarized Scanning Tunneling Microscopy}}},\ }\href {https://doi.org/10.1103/PhysRevLett.86.4132} {\bibfield  {journal} {\bibinfo  {journal} {Phys. Rev. Lett.}\ }\textbf {\bibinfo {volume} {86}},\ \bibinfo {pages} {4132} (\bibinfo {year} {2001})}\BibitemShut {NoStop}%
\bibitem [{\citenamefont {Wiesendanger}(2009)}]{Wiesendanger:review-2009}%
  \BibitemOpen
  \bibfield  {author} {\bibinfo {author} {\bibfnamefont {R.}~\bibnamefont {Wiesendanger}},\ }\bibfield  {title} {\bibinfo {title} {Spin mapping at the nanoscale and atomic scale},\ }\href {https://doi.org/10.1103/RevModPhys.81.1495} {\bibfield  {journal} {\bibinfo  {journal} {Reviews of Modern Physics}\ }\textbf {\bibinfo {volume} {81}},\ \bibinfo {pages} {1495} (\bibinfo {year} {2009})}\BibitemShut {NoStop}%
\bibitem [{\citenamefont {Schlenhoff}\ \emph {et~al.}(2020)\citenamefont {Schlenhoff}, \citenamefont {Kovarik}, \citenamefont {Krause},\ and\ \citenamefont {Wiesendanger}}]{Schlenhoff-Wiesendanger-RealspaceImagingAtomicscale-2020}%
  \BibitemOpen
  \bibfield  {author} {\bibinfo {author} {\bibfnamefont {A.}~\bibnamefont {Schlenhoff}}, \bibinfo {author} {\bibfnamefont {S.}~\bibnamefont {Kovarik}}, \bibinfo {author} {\bibfnamefont {S.}~\bibnamefont {Krause}},\ and\ \bibinfo {author} {\bibfnamefont {R.}~\bibnamefont {Wiesendanger}},\ }\bibfield  {title} {\bibinfo {title} {Real-space imaging of atomic-scale spin textures at nanometer distances},\ }\href {https://doi.org/10.1063/1.5145363} {\bibfield  {journal} {\bibinfo  {journal} {Appl. Phys. Lett.}\ }\textbf {\bibinfo {volume} {116}},\ \bibinfo {pages} {122406} (\bibinfo {year} {2020})}\BibitemShut {NoStop}%
\bibitem [{\citenamefont {Casola}\ \emph {et~al.}(2018)\citenamefont {Casola}, \citenamefont {Van Der~Sar},\ and\ \citenamefont {Yacoby}}]{Casola-Yacoby-ProbingCondensedMatter-2018}%
  \BibitemOpen
  \bibfield  {author} {\bibinfo {author} {\bibfnamefont {F.}~\bibnamefont {Casola}}, \bibinfo {author} {\bibfnamefont {T.}~\bibnamefont {Van Der~Sar}},\ and\ \bibinfo {author} {\bibfnamefont {A.}~\bibnamefont {Yacoby}},\ }\bibfield  {title} {\bibinfo {title} {Probing condensed matter physics with magnetometry based on nitrogen-vacancy centres in diamond},\ }\href {https://doi.org/10.1038/natrevmats.2017.88} {\bibfield  {journal} {\bibinfo  {journal} {Nat Rev Mater}\ }\textbf {\bibinfo {volume} {3}},\ \bibinfo {pages} {17088} (\bibinfo {year} {2018})}\BibitemShut {NoStop}%
\bibitem [{\citenamefont {Hedrich}\ \emph {et~al.}(2021)\citenamefont {Hedrich}, \citenamefont {Wagner}, \citenamefont {Pylypovskyi}, \citenamefont {Shields}, \citenamefont {Kosub}, \citenamefont {Sheka}, \citenamefont {Makarov},\ and\ \citenamefont {Maletinsky}}]{Hedrich-Maletinsky-NanoscaleMechanicsAntiferromagnetic-2021}%
  \BibitemOpen
  \bibfield  {author} {\bibinfo {author} {\bibfnamefont {N.}~\bibnamefont {Hedrich}}, \bibinfo {author} {\bibfnamefont {K.}~\bibnamefont {Wagner}}, \bibinfo {author} {\bibfnamefont {O.~V.}\ \bibnamefont {Pylypovskyi}}, \bibinfo {author} {\bibfnamefont {B.~J.}\ \bibnamefont {Shields}}, \bibinfo {author} {\bibfnamefont {T.}~\bibnamefont {Kosub}}, \bibinfo {author} {\bibfnamefont {D.~D.}\ \bibnamefont {Sheka}}, \bibinfo {author} {\bibfnamefont {D.}~\bibnamefont {Makarov}},\ and\ \bibinfo {author} {\bibfnamefont {P.}~\bibnamefont {Maletinsky}},\ }\bibfield  {title} {\bibinfo {title} {Nanoscale mechanics of antiferromagnetic domain walls},\ }\href {https://doi.org/10.1038/s41567-020-01157-0} {\bibfield  {journal} {\bibinfo  {journal} {Nat. Phys.}\ }\textbf {\bibinfo {volume} {17}},\ \bibinfo {pages} {574} (\bibinfo {year} {2021})},\ \Eprint {https://arxiv.org/abs/2009.08986} {arxiv:2009.08986} \BibitemShut {NoStop}%
\bibitem [{\citenamefont {Huxter}\ \emph {et~al.}(2022)\citenamefont {Huxter}, \citenamefont {Palm}, \citenamefont {Davis}, \citenamefont {Welter}, \citenamefont {Lambert}, \citenamefont {Trassin},\ and\ \citenamefont {Degen}}]{Huxter-Degen-ScanningGradiometrySingle-2022}%
  \BibitemOpen
  \bibfield  {author} {\bibinfo {author} {\bibfnamefont {W.~S.}\ \bibnamefont {Huxter}}, \bibinfo {author} {\bibfnamefont {M.~L.}\ \bibnamefont {Palm}}, \bibinfo {author} {\bibfnamefont {M.~L.}\ \bibnamefont {Davis}}, \bibinfo {author} {\bibfnamefont {P.}~\bibnamefont {Welter}}, \bibinfo {author} {\bibfnamefont {C.-H.}\ \bibnamefont {Lambert}}, \bibinfo {author} {\bibfnamefont {M.}~\bibnamefont {Trassin}},\ and\ \bibinfo {author} {\bibfnamefont {C.~L.}\ \bibnamefont {Degen}},\ }\bibfield  {title} {\bibinfo {title} {Scanning gradiometry with a single spin quantum magnetometer},\ }\href {https://doi.org/10.1038/s41467-022-31454-6} {\bibfield  {journal} {\bibinfo  {journal} {Nat Commun}\ }\textbf {\bibinfo {volume} {13}},\ \bibinfo {pages} {3761} (\bibinfo {year} {2022})},\ \Eprint {https://arxiv.org/abs/2202.09130} {arxiv:2202.09130 [cond-mat.mes-hall]} \BibitemShut {NoStop}%
\bibitem [{\citenamefont {Vasyukov}\ \emph {et~al.}(2013)\citenamefont {Vasyukov}, \citenamefont {Anahory}, \citenamefont {Embon}, \citenamefont {Halbertal}, \citenamefont {Cuppens}, \citenamefont {Neeman}, \citenamefont {Finkler}, \citenamefont {Segev}, \citenamefont {Myasoedov}, \citenamefont {Rappaport}, \citenamefont {Huber},\ and\ \citenamefont {Zeldov}}]{Vasyukov-Zeldov-ScanningSuperconductingQuantum-2013}%
  \BibitemOpen
  \bibfield  {author} {\bibinfo {author} {\bibfnamefont {D.}~\bibnamefont {Vasyukov}}, \bibinfo {author} {\bibfnamefont {Y.}~\bibnamefont {Anahory}}, \bibinfo {author} {\bibfnamefont {L.}~\bibnamefont {Embon}}, \bibinfo {author} {\bibfnamefont {D.}~\bibnamefont {Halbertal}}, \bibinfo {author} {\bibfnamefont {J.}~\bibnamefont {Cuppens}}, \bibinfo {author} {\bibfnamefont {L.}~\bibnamefont {Neeman}}, \bibinfo {author} {\bibfnamefont {A.}~\bibnamefont {Finkler}}, \bibinfo {author} {\bibfnamefont {Y.}~\bibnamefont {Segev}}, \bibinfo {author} {\bibfnamefont {Y.}~\bibnamefont {Myasoedov}}, \bibinfo {author} {\bibfnamefont {M.~L.}\ \bibnamefont {Rappaport}}, \bibinfo {author} {\bibfnamefont {M.~E.}\ \bibnamefont {Huber}},\ and\ \bibinfo {author} {\bibfnamefont {E.}~\bibnamefont {Zeldov}},\ }\bibfield  {title} {\bibinfo {title} {A scanning superconducting quantum interference device with single electron spin sensitivity},\ }\href {https://doi.org/10.1038/nnano.2013.169} {\bibfield  {journal} {\bibinfo  {journal}
  {Nature Nanotech}\ }\textbf {\bibinfo {volume} {8}},\ \bibinfo {pages} {639} (\bibinfo {year} {2013})}\BibitemShut {NoStop}%
\bibitem [{\citenamefont {Persky}\ \emph {et~al.}(2022)\citenamefont {Persky}, \citenamefont {Sochnikov},\ and\ \citenamefont {Kalisky}}]{Persky-Kalisky-StudyingQuantumMaterials-2022}%
  \BibitemOpen
  \bibfield  {author} {\bibinfo {author} {\bibfnamefont {E.}~\bibnamefont {Persky}}, \bibinfo {author} {\bibfnamefont {I.}~\bibnamefont {Sochnikov}},\ and\ \bibinfo {author} {\bibfnamefont {B.}~\bibnamefont {Kalisky}},\ }\bibfield  {title} {\bibinfo {title} {Studying {{Quantum Materials}} with {{Scanning SQUID Microscopy}}},\ }\href {https://doi.org/10.1146/annurev-conmatphys-031620-104226} {\bibfield  {journal} {\bibinfo  {journal} {Annu. Rev. Condens. Matter Phys.}\ }\textbf {\bibinfo {volume} {13}},\ \bibinfo {pages} {385} (\bibinfo {year} {2022})}\BibitemShut {NoStop}%
\bibitem [{\citenamefont {Wang}\ and\ \citenamefont {Freeman}(1981)}]{Wang-Freeman-SurfaceStatesSurface-1981}%
  \BibitemOpen
  \bibfield  {author} {\bibinfo {author} {\bibfnamefont {C.~S.}\ \bibnamefont {Wang}}\ and\ \bibinfo {author} {\bibfnamefont {A.~J.}\ \bibnamefont {Freeman}},\ }\bibfield  {title} {\bibinfo {title} {Surface states, surface magnetization, and electron spin polarization: {{Fe}}(001)},\ }\href {https://doi.org/10.1103/PhysRevB.24.4364} {\bibfield  {journal} {\bibinfo  {journal} {Phys. Rev. B}\ }\textbf {\bibinfo {volume} {24}},\ \bibinfo {pages} {4364} (\bibinfo {year} {1981})}\BibitemShut {NoStop}%
\bibitem [{\citenamefont {Freeman}\ \emph {et~al.}(1982)\citenamefont {Freeman}, \citenamefont {Krakauer}, \citenamefont {Ohnishi}, \citenamefont {Wang}, \citenamefont {Weinert},\ and\ \citenamefont {Wimmer}}]{Freeman-Wimmer-MagnetismSurfacesInterfaces-1982}%
  \BibitemOpen
  \bibfield  {author} {\bibinfo {author} {\bibfnamefont {A.~J.}\ \bibnamefont {Freeman}}, \bibinfo {author} {\bibfnamefont {H.}~\bibnamefont {Krakauer}}, \bibinfo {author} {\bibfnamefont {S.}~\bibnamefont {Ohnishi}}, \bibinfo {author} {\bibfnamefont {D.-s.}\ \bibnamefont {Wang}}, \bibinfo {author} {\bibfnamefont {M.}~\bibnamefont {Weinert}},\ and\ \bibinfo {author} {\bibfnamefont {E.}~\bibnamefont {Wimmer}},\ }\bibfield  {title} {\bibinfo {title} {Magnetism at {{Surfaces}} and {{Interfaces}}},\ }\href {https://doi.org/10.1051/jphyscol:1982725} {\bibfield  {journal} {\bibinfo  {journal} {J. Phys. Colloques}\ }\textbf {\bibinfo {volume} {43}},\ \bibinfo {pages} {C7} (\bibinfo {year} {1982})}\BibitemShut {NoStop}%
\bibitem [{\citenamefont {Harrison}(2012)}]{Harrison:book}%
  \BibitemOpen
  \bibfield  {author} {\bibinfo {author} {\bibfnamefont {W.~A.}\ \bibnamefont {Harrison}},\ }\href@noop {} {\emph {\bibinfo {title} {Solid {{State Theory}}}}}\ (\bibinfo  {publisher} {Dover Publications},\ \bibinfo {year} {2012})\BibitemShut {NoStop}%
\bibitem [{\citenamefont {Mahan}(2000)}]{Mahan:book-2013}%
  \BibitemOpen
  \bibfield  {author} {\bibinfo {author} {\bibfnamefont {G.~D.}\ \bibnamefont {Mahan}},\ }\href {https://doi.org/10.1007/978-1-4757-5714-9} {\emph {\bibinfo {title} {Many-{{Particle Physics}}}}}\ (\bibinfo  {publisher} {Springer New York},\ \bibinfo {address} {New York},\ \bibinfo {year} {2000})\BibitemShut {NoStop}%
\bibitem [{\citenamefont {{\v{S}}mejkal}\ \emph {et~al.}(2022)\citenamefont {{\v{S}}mejkal}, \citenamefont {Sinova},\ and\ \citenamefont {Jungwirth}}]{Smejkal-Jungwirth:2022}%
  \BibitemOpen
  \bibfield  {author} {\bibinfo {author} {\bibfnamefont {L.}~\bibnamefont {{\v{S}}mejkal}}, \bibinfo {author} {\bibfnamefont {J.}~\bibnamefont {Sinova}},\ and\ \bibinfo {author} {\bibfnamefont {T.}~\bibnamefont {Jungwirth}},\ }\bibfield  {title} {\bibinfo {title} {{Emerging Research Landscape of Altermagnetism}},\ }\href {https://doi.org/10.1103/PhysRevX.12.040501} {\bibfield  {journal} {\bibinfo  {journal} {Phys. Rev. X}\ }\textbf {\bibinfo {volume} {12}},\ \bibinfo {pages} {040501} (\bibinfo {year} {2022})},\ \Eprint {https://arxiv.org/abs/2204.10844} {arXiv:2204.10844} \BibitemShut {NoStop}%
\bibitem [{\citenamefont {Granata}\ and\ \citenamefont {Vettoliere}(2016)}]{Granata-Vettoliere-NanoSuperconductingQuantum-2016}%
  \BibitemOpen
  \bibfield  {author} {\bibinfo {author} {\bibfnamefont {C.}~\bibnamefont {Granata}}\ and\ \bibinfo {author} {\bibfnamefont {A.}~\bibnamefont {Vettoliere}},\ }\bibfield  {title} {\bibinfo {title} {Nano {{Superconducting Quantum Interference}} device: {{A}} powerful tool for nanoscale investigations},\ }\href {https://doi.org/10.1016/j.physrep.2015.12.001} {\bibfield  {journal} {\bibinfo  {journal} {Physics Reports}\ }\textbf {\bibinfo {volume} {614}},\ \bibinfo {pages} {1} (\bibinfo {year} {2016})},\ \Eprint {https://arxiv.org/abs/1505.06887} {arXiv:1505.06887 [cond-mat.supr-con]} \BibitemShut {NoStop}%
\bibitem [{\citenamefont {Buchner}\ \emph {et~al.}(2018)\citenamefont {Buchner}, \citenamefont {H{\"o}fler}, \citenamefont {Henne}, \citenamefont {Ney},\ and\ \citenamefont {Ney}}]{Buchner-Ney-TutorialBasicPrinciples-2018}%
  \BibitemOpen
  \bibfield  {author} {\bibinfo {author} {\bibfnamefont {M.}~\bibnamefont {Buchner}}, \bibinfo {author} {\bibfnamefont {K.}~\bibnamefont {H{\"o}fler}}, \bibinfo {author} {\bibfnamefont {B.}~\bibnamefont {Henne}}, \bibinfo {author} {\bibfnamefont {V.}~\bibnamefont {Ney}},\ and\ \bibinfo {author} {\bibfnamefont {A.}~\bibnamefont {Ney}},\ }\bibfield  {title} {\bibinfo {title} {Tutorial: {{Basic}} principles, limits of detection, and pitfalls of highly sensitive {{SQUID}} magnetometry for nanomagnetism and spintronics},\ }\href {https://doi.org/10.1063/1.5045299} {\bibfield  {journal} {\bibinfo  {journal} {J. Appl. Phys.}\ }\textbf {\bibinfo {volume} {124}},\ \bibinfo {pages} {161101} (\bibinfo {year} {2018})}\BibitemShut {NoStop}%
\bibitem [{\citenamefont {Paulsen}\ \emph {et~al.}(2023)\citenamefont {Paulsen}, \citenamefont {Lindner}, \citenamefont {Klemke}, \citenamefont {Beyer}, \citenamefont {Fechner}, \citenamefont {Meier},\ and\ \citenamefont {Kiefer}}]{Paulsen-Kiefer-UltralowFieldSQUID-2022}%
  \BibitemOpen
  \bibfield  {author} {\bibinfo {author} {\bibfnamefont {M.}~\bibnamefont {Paulsen}}, \bibinfo {author} {\bibfnamefont {J.}~\bibnamefont {Lindner}}, \bibinfo {author} {\bibfnamefont {B.}~\bibnamefont {Klemke}}, \bibinfo {author} {\bibfnamefont {J.}~\bibnamefont {Beyer}}, \bibinfo {author} {\bibfnamefont {M.}~\bibnamefont {Fechner}}, \bibinfo {author} {\bibfnamefont {D.}~\bibnamefont {Meier}},\ and\ \bibinfo {author} {\bibfnamefont {K.}~\bibnamefont {Kiefer}},\ }\bibfield  {title} {\bibinfo {title} {{An ultra-low field SQUID magnetometer for measuring antiferromagnetic and weakly remanent magnetic materials at low temperatures}},\ }\href {https://doi.org/10.1063/5.0135877} {\bibfield  {journal} {\bibinfo  {journal} {Review of Scientific Instruments}\ }\textbf {\bibinfo {volume} {94}},\ \bibinfo {pages} {103904} (\bibinfo {year} {2023})}\BibitemShut {NoStop}%
\bibitem [{\citenamefont {Brekke}\ \emph {et~al.}(2023)\citenamefont {Brekke}, \citenamefont {Brataas},\ and\ \citenamefont {Sudb{\o}}}]{Brekke-Sudbo:2023}%
  \BibitemOpen
  \bibfield  {author} {\bibinfo {author} {\bibfnamefont {B.}~\bibnamefont {Brekke}}, \bibinfo {author} {\bibfnamefont {A.}~\bibnamefont {Brataas}},\ and\ \bibinfo {author} {\bibfnamefont {A.}~\bibnamefont {Sudb{\o}}},\ }\bibfield  {title} {\bibinfo {title} {Two-dimensional altermagnets: {{Superconductivity}} in a minimal microscopic model},\ }\href {https://doi.org/10.1103/PhysRevB.108.224421} {\bibfield  {journal} {\bibinfo  {journal} {Phys. Rev. B}\ }\textbf {\bibinfo {volume} {108}},\ \bibinfo {pages} {224421} (\bibinfo {year} {2023})},\ \Eprint {https://arxiv.org/abs/2308.08606} {arxiv:2308.08606} \BibitemShut {NoStop}%
\bibitem [{\citenamefont {Lee}(2023)}]{lee_arxiv_23}%
  \BibitemOpen
  \bibfield  {author} {\bibinfo {author} {\bibfnamefont {Y.-l.}\ \bibnamefont {Lee}},\ }\href {http://arxiv.org/abs/2312.15733} {\bibinfo {title} {Magnetic impurities in an altermagnetic metal}} (\bibinfo {year} {2023}),\ \Eprint {https://arxiv.org/abs/2312.15733} {arXiv:2312.15733} \BibitemShut {NoStop}%
\bibitem [{\citenamefont {Amundsen}\ \emph {et~al.}(2024)\citenamefont {Amundsen}, \citenamefont {Brataas},\ and\ \citenamefont {Linder}}]{amundsen_prb_24}%
  \BibitemOpen
  \bibfield  {author} {\bibinfo {author} {\bibfnamefont {M.}~\bibnamefont {Amundsen}}, \bibinfo {author} {\bibfnamefont {A.}~\bibnamefont {Brataas}},\ and\ \bibinfo {author} {\bibfnamefont {J.}~\bibnamefont {Linder}},\ }\bibfield  {title} {\bibinfo {title} {Rkky interaction in rashba altermagnets},\ }\href {https://doi.org/10.1103/PhysRevB.110.054427} {\bibfield  {journal} {\bibinfo  {journal} {Phys. Rev. B}\ }\textbf {\bibinfo {volume} {110}},\ \bibinfo {pages} {054427} (\bibinfo {year} {2024})}\BibitemShut {NoStop}%
\bibitem [{\citenamefont {Chilcote}\ \emph {et~al.}(2024)\citenamefont {Chilcote}, \citenamefont {Mazza}, \citenamefont {Lu}, \citenamefont {Gray}, \citenamefont {Tian}, \citenamefont {Deng}, \citenamefont {Moseley}, \citenamefont {Chen}, \citenamefont {Lapano}, \citenamefont {Gardner}, \citenamefont {Eres}, \citenamefont {Ward}, \citenamefont {Feng}, \citenamefont {Cao}, \citenamefont {Lauter}, \citenamefont {McGuire}, \citenamefont {Hermann}, \citenamefont {Parker}, \citenamefont {Han}, \citenamefont {Kayani}, \citenamefont {Rimal}, \citenamefont {Wu}, \citenamefont {Charlton}, \citenamefont {Moore},\ and\ \citenamefont {Brahlek}}]{Chilcote-Brahlek-StoichiometryinducedFerromagnetismAltermagnetic-2024}%
  \BibitemOpen
  \bibfield  {author} {\bibinfo {author} {\bibfnamefont {M.}~\bibnamefont {Chilcote}}, \bibinfo {author} {\bibfnamefont {A.~R.}\ \bibnamefont {Mazza}}, \bibinfo {author} {\bibfnamefont {Q.}~\bibnamefont {Lu}}, \bibinfo {author} {\bibfnamefont {I.}~\bibnamefont {Gray}}, \bibinfo {author} {\bibfnamefont {Q.}~\bibnamefont {Tian}}, \bibinfo {author} {\bibfnamefont {Q.}~\bibnamefont {Deng}}, \bibinfo {author} {\bibfnamefont {D.}~\bibnamefont {Moseley}}, \bibinfo {author} {\bibfnamefont {A.-H.}\ \bibnamefont {Chen}}, \bibinfo {author} {\bibfnamefont {J.}~\bibnamefont {Lapano}}, \bibinfo {author} {\bibfnamefont {J.~S.}\ \bibnamefont {Gardner}}, \bibinfo {author} {\bibfnamefont {G.}~\bibnamefont {Eres}}, \bibinfo {author} {\bibfnamefont {T.~Z.}\ \bibnamefont {Ward}}, \bibinfo {author} {\bibfnamefont {E.}~\bibnamefont {Feng}}, \bibinfo {author} {\bibfnamefont {H.}~\bibnamefont {Cao}}, \bibinfo {author} {\bibfnamefont {V.}~\bibnamefont {Lauter}}, \bibinfo {author} {\bibfnamefont {M.~A.}\ \bibnamefont {McGuire}},
  \bibinfo {author} {\bibfnamefont {R.}~\bibnamefont {Hermann}}, \bibinfo {author} {\bibfnamefont {D.}~\bibnamefont {Parker}}, \bibinfo {author} {\bibfnamefont {M.-G.}\ \bibnamefont {Han}}, \bibinfo {author} {\bibfnamefont {A.}~\bibnamefont {Kayani}}, \bibinfo {author} {\bibfnamefont {G.}~\bibnamefont {Rimal}}, \bibinfo {author} {\bibfnamefont {L.}~\bibnamefont {Wu}}, \bibinfo {author} {\bibfnamefont {T.~R.}\ \bibnamefont {Charlton}}, \bibinfo {author} {\bibfnamefont {R.~G.}\ \bibnamefont {Moore}},\ and\ \bibinfo {author} {\bibfnamefont {M.}~\bibnamefont {Brahlek}},\ }\bibfield  {title} {\bibinfo {title} {Stoichiometry-{{Induced Ferromagnetism}} in {{Altermagnetic Candidate MnTe}}},\ }\href {https://doi.org/10.1002/adfm.202405829} {\bibfield  {journal} {\bibinfo  {journal} {Adv Funct Materials}\ ,\ \bibinfo {pages} {2405829}} (\bibinfo {year} {2024})},\ \Eprint {https://arxiv.org/abs/2406.04474} {arXiv:2406.04474 [cond-mat]} \BibitemShut {NoStop}%
\end{thebibliography}%

\begin{widetext}

\appendix

\section{Derivation of continuum models} 
\label{App:tight-binding}

The continuum models of antiferromagnets given in Eq.~(\ref{analytics-H-AF}) and altermagnets given in Eq.~(\ref{ribbon-H-AM}) can be derived both by using the symmetry arguments and directly from the lattice model presented in Sec.~\ref{sec:numerics}.

In the case of antiferromagnets, we use the full lattice model~\footnote{A simplified model of an antiferromagnet with only two magnetic sites in the unit-cell and next-nearest-neighbor hopping results in the same continuum model. However, this model does not allow one to tune between antiferromagnetic and altermagnetic phases.} described in Sec.~\ref{sec:numerics-model} whose geometry is shown in Fig.~\ref{fig:latticemodel}(b). Performing the Fourier transformation, we obtain 
\begin{equation}
\label{app-tb-afm}
H_{\rm AFM}(\mathbf{k}) = \begin{pmatrix}
-s J_{sd} -\mu & 0 & 2t \cos{(k_ya)} & 2t \cos{(k_x a)}\\
2t \cos{(k_y a)} & \epsilon_3-\mu & 2t \cos{(k_y a)} & 0\\
0 & 2t \cos{(k_x a)} & s J_{sd} -\mu & 2t \cos{(k_y a)}\\
2t \cos{(k_x a)} & 0 & 2t \cos{(k_y a)} & \epsilon_2-\mu
\end{pmatrix}.
\end{equation}
The above Hamiltonian contains the spin projection $s$ and is written in the site basis $\left\{1,3,4,2\right\}$ and describes four energy bands with energies
\begin{eqnarray}
\label{app-tb-afm-eps}
\epsilon_{\rho_1,\rho_2} = -\mu +\rho_1 \sqrt{\frac{J_{\rm sd}^2}{4} +2t^2\left[\cos^2{(k_xa)}+\cos^2{(k_ya)}\right] +\rho_2 \sqrt{\frac{J_{\rm sd}^2}{4} +2t^2\cos^2{(k_xa)}} \sqrt{\frac{J_{\rm sd}^2}{4} +2t^2\cos^2{(k_ya)}}},
\end{eqnarray}
where $\rho_1,\rho_2=\pm$ and, for simplicity, we set $\epsilon_3=\epsilon_4=0$.

The effective low-energy Hamiltonian is derived by considering only the lowest-energy band and projecting out the three higher-energy ones. Then, expanding the expression (\ref{app-tb-afm-eps}) for the lowest-energy band ($\rho_1=-$ and $\rho_2=+$) around the $\Gamma$ point, we derive the following low-energy continuum model:
\begin{equation}
\label{app-tb-afm-eff-fin}
H_{\rm AFM, eff}(\mathbf{k}) = -\mu -\sqrt{16t^2 +J_{\rm sd}^2} +\frac{4t^2 (ka)^2}{\sqrt{16t^2+J_{\rm sd}^2}},
\end{equation}
cf. with Eq.~(\ref{analytics-H-AF}).

The continuum model for altermagnets can be derived in a similar way. Removing one nonmagnetic lattice site, e.g., at the position $4$ in Fig.~\ref{fig:latticemodel}(b), and performing the Fourier transformation, we obtain the following tight-binding model:
\begin{equation}
\label{app-tb-am}
H_{\rm AM}(\mathbf{k}) = \begin{pmatrix}
-s J_{\rm sd} -\mu & 2t \cos{(k_ya)} & 0 \\
2t \cos{(k_y a)} & \epsilon_3-\mu & 2t \cos{(k_x a)} \\
0 & 2t \cos{(k_x a)} & s J_{\rm sd} -\mu
\end{pmatrix}
\end{equation}
whose energy dispersion at $\epsilon_3=0$ follows from the following cubic equation:
\begin{equation}
\label{app-tb-am-eps-eq}
\epsilon\left(J_{\rm sd}^2 -\epsilon^2\right) +4\left[\left(J_{\rm sd} +\epsilon\right)\cos^2{(k_ya)} -\left(J_{\rm sd} -\epsilon\right)\cos^2{(k_xa)}\right]=0.
\end{equation}
Expanding its roots up to the second order in $ka$ around the $\Gamma$ point, we obtain
\begin{eqnarray}
\label{app-tb-am-eps-pm}
\epsilon_{\pm} &=& -\mu \pm \sqrt{8t^2 -J_{\rm sd}^2} \mp \frac{2 (ka)^2}{\sqrt{8t^2+J_{\rm sd}^2}} -\frac{2s J_{\rm sd}  t^2 a^2}{8t^2+J_{\rm sd}^2} \left(k_x^2 - k_y^2\right),\\
\label{app-tb-am-eps-0}
\epsilon_0 &=& -\mu + \frac{4J_{\rm sd} t^2 a^2}{8t^2+J_{\rm sd}^2} \left(k_x^2-k_y^2\right).
\end{eqnarray}
The effective Hamiltonian is derived by considering only the lowest-energy band $\epsilon_{-}$ and projecting out the two higher-energy ones. Then, the corresponding low-energy continuum Hamiltonian is 
\begin{equation}
\label{app-tb-am-eff-fin}
H_{\rm AM, eff}(\mathbf{k}) = -\mu -\sqrt{8t^2 +J_{\rm sd}^2} +\frac{2 t^2(ak)^2}{\sqrt{8t^2+J_{\rm sd}^2}} -\frac{2s J_{\rm sd} t^2 a^2}{8t^2+J_{\rm sd}^2} \left(k_x^2 - k_y^2\right).
\end{equation}
Comparing this model with the continuum model (\ref{ribbon-H-AM}) derived from the symmetry principles, we identify both the kinetic energy term (the third term in Eq.~(\ref{app-tb-am-eff-fin})) and the altermagnetic splitting term (the last term in Eq.~(\ref{app-tb-am-eff-fin})).

\section{Spectral function in the lattice model 
\label{App: A}}

In this appendix, we define the density of states and the spectral function in the lattice model with translational invariance in both \textit{x} and \textit{y} (i.e., the bulk model). The unitary matrix diagonalizing a particular $(k_x, k_y)$- and $\sigma$-block of the Hamiltonian defined in Eq. (\ref{eqn: total hamiltonian}) with Fourier-transformed operators also in \textit{x}, has the general structure 
\begin{equation}
    U_{\boldsymbol{k},\sigma} = \begin{bmatrix}
        \phi_{\boldsymbol{k},\sigma, 1} & \phi_{\boldsymbol{k},\sigma, 2} & \phi_{\boldsymbol{k},\sigma, 3} & \phi_{\boldsymbol{k}, \sigma, 4}
    \end{bmatrix},
\end{equation}
where each column is an eigenvector 
\begin{equation}
    \phi_{\boldsymbol{k},\sigma, n} = \begin{pmatrix}
        u_{\boldsymbol{k},\sigma, n} & v_{\boldsymbol{k},\sigma, n} & w_{\boldsymbol{k},\sigma, n} & x_{\boldsymbol{k},\sigma, n}
    \end{pmatrix}^T
\end{equation}
with corresponding eigenvalue $E_{\boldsymbol{k},\sigma, n}$  for $n=1,2,3,4$.

The operators in the original fermion basis $\tilde{c}_{\boldsymbol{k},\sigma}= \begin{pmatrix}
    \ca[]{\boldsymbol{k},1,\sigma} & \ca[]{\boldsymbol{k}, 2, \sigma} & \ca[]{\boldsymbol{k}, 3, \sigma} & \ca[]{\boldsymbol{k}, 4, \sigma}
\end{pmatrix}$ is related to the new operators ${\gamma}_{\boldsymbol{k}, \sigma, n}$ through the coefficients of the eigenvectors, 
\begin{align}
    {c}_{\boldsymbol{k}, 1, \sigma} & = \sum_{n = 1}^{4} u_{\boldsymbol{k}, \sigma, n} \gamma_{k,\sigma, n} \label{eqn: u}, \qquad
    {c}_{\boldsymbol{k}, 2, \sigma}  = \sum_{n = 1}^{4} v_{\boldsymbol{k}, \sigma, n} \gamma_{k,\sigma, n}, \\
    {c}_{\boldsymbol{k}, 3, \sigma} & = \sum_{n = 1}^{4} w_{\boldsymbol{k},\sigma, n} \gamma_{k,\sigma, n}, \qquad
    {c}_{\boldsymbol{k}, 4, \sigma}  = \sum_{n = 1}^{4} x_{\boldsymbol{k},\sigma, n} \gamma_{k,\sigma, n}.
    \label{eqn: v}
\end{align}

The number of electrons in the translationally invariant system with spin $\sigma$ summed over all momentum modes and unit-cell sites is given by 
\begin{equation}
    \rho_{\sigma}=\sum_{\boldsymbol{k}}\sum_{\alpha}\left\langle \cc[]{\boldsymbol{k},\alpha,\sigma} \ca[]{\boldsymbol{k}, \alpha, \sigma} \right\rangle = \int_{-\infty}^\infty d\epsilon D_{\sigma}(\epsilon)n_F(\epsilon),
\end{equation}
where $D_{\sigma}(\epsilon)$ is the spin-dependent density of states. By inserting the expressions for the operators in the new basis, see Eqs.~(\ref{eqn: u}) and (\ref{eqn: v}), we rewrite the above expressions as
\begin{equation}
    \int_{-\infty}^{\infty}d\epsilon D_{\sigma}(\epsilon)n_F(\epsilon) = \sum_{n, \boldsymbol{k}}\big(|u_{\boldsymbol{k},\sigma,n}|^2 + |v_{\boldsymbol{k},\sigma,n}|^2 + |w_{\boldsymbol{k},\sigma,n}|^2 +|x_{\boldsymbol{k},\sigma,n}|^2 \big) n_F(E_{\boldsymbol{k},\sigma, n}),
\end{equation}
where we have used that $\langle \gamma_{\boldsymbol{k},\sigma,n}^\dagger \gamma_{\boldsymbol{k}, \sigma, n'}^{\vphantom{\dagger}} \rangle = \delta_{n,n'}n_F(E_{\boldsymbol{k},\sigma,n})$. We can now identify the density of states as
\begin{equation}
    D_{\sigma}(\epsilon) = \sum_{n, \boldsymbol{k}}\big(|u_{\boldsymbol{k},\sigma,n}|^2 + |v_{\boldsymbol{k},\sigma,n}|^2 + |w_{\boldsymbol{k},\sigma,n}|^2 +|x_{\boldsymbol{k},\sigma,n}|^2 \big) \delta(\epsilon - E_{\boldsymbol{k},\sigma,n})=\sum_{\boldsymbol{k}} A_{\sigma}(\epsilon, \boldsymbol{k}).
\end{equation}

The spectral function reads
\begin{equation}
    A_{\sigma}(\epsilon, \boldsymbol{k})=\sum_{n}\big(|u_{\boldsymbol{k},\sigma,n}|^2 + |v_{\boldsymbol{k},\sigma,n}|^2 + |w_{\boldsymbol{k},\sigma,n}|^2 +|x_{\boldsymbol{k},\sigma,n}|^2 \big) \delta(\epsilon - E_{\boldsymbol{k},\sigma, n}).
\end{equation}

\color{black}

\end{widetext}

\end{document}